\documentclass[12pt,a4paper]{article}
\usepackage{amsmath}
\usepackage{amssymb}
\usepackage{graphicx,epsfig}
\usepackage[mathscr]{eucal}
\usepackage[dvipsnames]{xcolor}
\usepackage{import}
\usepackage{mathtools}
\usepackage{braket}
\usepackage{comment}
\usepackage[numbers,sort&compress]{natbib}
\usepackage[english]{babel}

\usepackage{tabularx}

\usepackage{setspace}
\usepackage{xspace}

\usepackage{multirow}

\def\msbar{{\overline {\rm MS}}}

\usepackage{scalefnt,ulem,pstricks}

\newcommand\as{\alpha_{\mathrm{S}}}

\def\pik  {(p_i\cdot k)}
\def\pjk  {(p_j\cdot k)}
\def\puk  {(p_1\cdot k)}
\def\pdk  {(p_2\cdot k)}
\def\ptk  {(p_3\cdot k)}
\def\pqk  {(p_4\cdot k)}

\def\piqu  {(p_i\cdot k_1)}
\def\piqd { (p_i\cdot k_2)}
\def\pjqu  {(p_j\cdot k_1)}
\def\pjqd {(p_j\cdot k_2)}
\def\piq  {(p_i\cdot(k_1 + k_2))}
\def\pjq  {(p_j\cdot(k_1 + k_2))}
\def\piq  {(p_i\cdot k)}
\def\pjq  {(p_j\cdot k)}
\def\quqd  {(k_1\cdot k_2)}
\def\pipj  {(p_i\cdot p_j)}

\def\pupd  {(p_1\cdot p_2)}
\def\pupdk  {(p_1+p_2)\cdot k}

\def\beeq{\begin{eqnarray}}
\def\eeeq{\end{eqnarray}}

\def\nini  {\vec n_i^2}
\def\njnj  {\vec n_j^2}

\def\ninj  {\vec n_i \cdot \vec n_j}

\def \Li{{\rm Li}}

\def\liD{\text{Li}_2}
\def\liT{\text{Li}_3}

\def \lni#1{\ln^{#1}\Big(\frac{\alpha_3}{v_+}\Big)}
\def \lnip#1{\ln^{#1}\Big(1-\frac{\alpha_3}{v_+}\Big)}
\def \lnj#1{\ln^{#1}\Big(\frac{\alpha_4}{v_+}\Big)}
\def \lnjp#1{\ln^{#1}\Big(1-\frac{\alpha_4}{v_+}\Big)}
\def \lix#1{{\Li_#1\left(\frac{v_-}{v_+}\right)}}
\def \lixp#1{{\Li_#1\left(1-\frac{v_-}{v_+}\right)}}
\def \lii#1{{\Li_#1\Big(\frac{\alpha_3}{v_+}\Big)}}
\def \lij#1{{\Li_#1\Big(\frac{\alpha_4}{v_+}\Big)}}

\usepackage{array}
\newcolumntype{L}[1]{>{\raggedright\let\newline\\\arraybackslash\hspace{0pt}}m{#1}}
\newcolumntype{C}[1]{>{\centering\let\newline\\\arraybackslash\hspace{0pt}}m{#1}}
\newcolumntype{R}[1]{>{\raggedleft\let\newline\\\arraybackslash\hspace{0pt}}m{#1}}

\usepackage{pstricks}
\usepackage[colorlinks=true,allcolors={black}]{hyperref}
\usepackage{cleveref}

\usepackage{footnotebackref}

\usepackage[section]{placeins}

\usepackage{etoolbox}
\def\appendixname{Appendix}
\appto\appendix{%
  \addtocontents{toc}{\patch@l@section}
  \appto\appendixname{ }
}
\protected\def\patch@l@section{%
  \patchcmd{\l@section}{1.5em}{\widthof{\appendixname\space}+2.5em}{}{}%
}

\begin{document}
\begin{titlepage}
\begin{flushright}
TIF-UNIMI-2023-2\\
ZU-TH 04/23\\
PSI-PR-23-1
\end{flushright}

\renewcommand{\thefootnote}{\fnsymbol{footnote}}
\vspace*{0.5cm}

\begin{center}
  {\Large \bf Soft-parton contributions to heavy-quark production\\[2ex]at low transverse momentum}
\end{center}

\par \vspace{2mm}
\begin{center}
  {\bf Stefano Catani${}^{(a)}$}, {\bf Simone Devoto${}^{(b)}$},\\[0.2cm]
  {\bf Massimiliano Grazzini${}^{(c)}$} and {\bf Javier Mazzitelli${}^{(d)}$}

\vspace{5mm}

${}^{(a)}$INFN, Sezione di Firenze and
Dipartimento di Fisica e Astronomia,\\
Universit\`a di Firenze,
50019 Sesto Fiorentino, Firenze, Italy\\[0.25cm]

${}^{(b)}$Dipartimento di Fisica ``Aldo Pontremoli'', University of Milano and INFN, Sezione di Milano, I-20133 Milano, Italy\\[0.25cm]

${}^{(c)}$Physik Institut, Universit\"at Z\"urich, 8057 Z\"urich, Switzerland\\[0.25cm]

${}^{(d)}$Paul Scherrer Institut, CH-5232 Villigen PSI, Switzerland

\vspace{5mm}

\end{center}

\par \vspace{2mm}
\begin{center} {\large \bf Abstract}

\end{center}
\begin{quote}
\pretolerance 10000

We consider QCD radiative corrections to the production of a heavy-quark pair in hadronic collisions.
We present the computation of the soft-parton contributions at low transverse momentum of the heavy-quark pair up to second order in the QCD coupling $\as$.
These results complete the evaluation at the next-to-next-to-leading order (NNLO) of the transverse-momentum resummation formula for this process.
Moreover, they give all the ingredients that are needed
for the NNLO implementation of the $q_T$ subtraction formalism for heavy-quark production.
We discuss the details of the computation and we provide a code that can be used to obtain the relevant results in numerical form.

\end{quote}

\vspace*{\fill}
\begin{flushleft}
January 2023
\end{flushleft}
\end{titlepage}

\tableofcontents


\section{Introduction}
\label{sec:introduction}

Heavy-quark pair production is one of the classic hard-scattering processes at hadron colliders.
For a sufficiently-heavy quark, the cross section is
perturbatively computable as an expansion in the QCD coupling $\as(\mu^2_R)$ where the renormalisation scale $\mu_R$ is of the order of the mass $m$ of the heavy quark. A large variety of QCD studies of heavy-quark hadroproduction have been carried out over the years.
In this context top-quark pair production plays a special role: being the heaviest particle in the Standard Model,
the top quark couples strongly to the Higgs boson and is therefore particularly relevant for the mechanism of electroweak symmetry breaking.
As such, top-quark pair production is especially relevant in searches for physics beyond the Standard Model,
it constitutes a possible window on new physics and, at the same time, a crucial background in many analyses.
Bottom and charm quark production have also been extensively studied at hadron colliders, and allow us to probe QCD at smaller energy scales.

For the above reasons, the study of the hadroproduction of a heavy-quark pair
has attracted the attention and the efforts of the theoretical community for decades.
Next-to-leading order (NLO) QCD corrections to this process have been available since a long time,
both for the total cross section and for differential distributions \cite{Nason:1987xz, Beenakker:1988bq, Beenakker:1990maa, Nason:1989zy,Mangano:1991jk}.
Nevertheless, because of the challenging complications arising at the next order in the perturbative expansion, more than 20 years passed before next-to-next-to-leading order (NNLO) QCD corrections for top-quark pair production were also computed~\cite{Barnreuther:2012wtj,Czakon:2012zr,Czakon:2012pz,Czakon:2013goa,Czakon:2015owf,Czakon:2016ckf,Catani:2019iny,Catani:2019hip}.
Further progress regards the combination of QCD and EW corrections~\cite{Czakon:2017wor} and the inclusion of top-quark decays~\cite{Behring:2019iiv}.
Results by using the $\overline{\text{MS}}$ scheme for the renormalisation of the top-quark mass are also available~\cite{Dowling:2013baa,Catani:2020tko}.
More recently, the NNLO calculation of Refs.~\cite{Catani:2019iny,Catani:2019hip} has been extended to bottom-quark pair production~\cite{Catani:2020kkl}.

One of the two available NNLO computations for heavy-quark production~\cite{Catani:2019iny, Catani:2019hip,Catani:2020tko,Catani:2020kkl} is based on the $q_T$ subtraction formalism \cite{Catani:2007vq}.
The $q_T$ subtraction formalism is a method to handle and cancel the IR divergences in QCD computations at NLO, NNLO and beyond. The method uses IR subtraction counterterms that are constructed by evaluating the $q_T$ distribution of the produced final-state system in the limit $q_T\to 0$. If the produced final-state system is composed of colourless particles (such as vector bosons, Higgs bosons, and so forth),
the behaviour of the $q_T$ distribution in the limit $q_T\to 0$ has a universal structure that is explicitly known to the next-to-next-to-next-to leading order (N$^3$LO) through the formalism of transverse-momentum resummation \cite{Li:2016ctv,Vladimirov:2016dll,Luo:2019szz,Ebert:2020yqt,Luo:2020epw}.
The resummation formalism can be extended to the production of final states containing a heavy-quark pair \cite{Zhu:2012ts,Li:2013mia,Catani:2014qha,Catani:2018mei}. The heavy quarks do not lead to additional collinear singularities (which are absent because of the finite heavy-quark mass) but, being coloured, they lead to additional soft singularities that need to be properly taken into account.
The NNLO computations of Refs.~\cite{Catani:2019iny, Catani:2019hip,Catani:2020tko,Catani:2020kkl}
rely on the explicit evaluation of 
such soft-parton contributions due to the coloured massive quarks.

The purpose of this paper is to report on the details of the computation of such soft-parton terms.
The final numerical results can be obtained by using 
the program attached to the arXiv submission of this paper.\footnote{The soft-parton contributions evaluated in this work also enter the {\sc MiNNLO}$_{\rm PS}$ formalism for the matching of NNLO calculations to parton showers for heavy-quark production \cite{Mazzitelli:2020jio,Mazzitelli:2021mmm,minnlo_bb}.} 
Our calculation is performed within the transverse-momentum resummation formalism of Ref.~\cite{Catani:2014qha}.
A similar computation, carried out within the framework of
Soft Collinear Effective Theory (SCET) used in Refs.~\cite{Zhu:2012ts,Li:2013mia},
has been presented in Ref.~\cite{Angeles-Martinez:2018mqh}.

We note that our formalism can be extended to the production of a heavy-quark pair accompanied by colourless particles~\cite{Catani:2021cbl}. Such extension has been recently applied to the evaluation of NNLO corrections to $t{\bar t}H$~\cite{Catani:2022mfv} and $Wb{\bar b}$~\cite{Buonocore:2022pqq} production. In this paper, however, we will limit ourselves to the case of heavy-quark production, i.e., with no additional colourless particles. The soft-parton contributions relevant for the production of a heavy-quark pair and a colourless system will be documented elsewhere.

The paper is organised as follows. In Sect.~\ref{sec:resum+soft} we review the resummation formalism for heavy-quark production
and we discuss the soft-parton contributions we want to compute. In Sect.~\ref{sec:details} we illustrate our calculation, starting from single-gluon emission at tree level in Sect.~\ref{sub:nlo}, then going to single-gluon emission at one loop in Sect.~\ref{sub:1L}, soft $q{\bar q}$ emission in Sect.~\ref{sub:qq} and double-gluon emission in Sect.~\ref{sub:gg}. Our numerical implementation and final results are presented in Sect.~\ref{sec:code}. In Sect.~\ref{sec:summa} we summarise our findings.

\section{Heavy-quark production at low transverse momentum}
\label{sec:resum+soft}

\subsection{Resummation formalism for heavy-quark production}
\label{sec:resummation}

We consider the inclusive hard-scattering process
\begin{equation}
  \label{eq:process}
h_1(P_1)+h_2(P_2)\to Q(p_3)+{\bar Q}(p_4)+X
    \end{equation}
    where the collision of the two hadrons $h_1$ and $h_2$ with momenta $P_1$ and $P_2$ produces the heavy-quark pair $Q{\bar Q}$, and $X$ denotes the accompanying final-state radiation.
    The heavy quarks have four-momenta $p_3$ and $p_4$, total momentum $q=p_3+p_4$, invariant mass $M^2=q^2$ and total transverse momentum ${\vec q}_T={\vec p}_{3,T}+{\vec p}_{4,T}$.
    The rapidity of the $Q{\bar Q}$ pair is $y=1/2 \ln (q\cdot P_2/q\cdot P_1)$.
    The knowledge of $M$, $y$ and ${\vec q}_T$ completely specifies the total momentum $q$ of the heavy-quark pair.
    The kinematics of the observed heavy quarks is fully determined by $q$ and two additional independent kinematical variables, that we denote by ${\vec \Omega}$.
    For example, we can choose ${\vec \Omega}=\{y_3,\phi_3\}$, where $y_3$ and $\phi_3$ are the rapidity and the azimuthal angle of the heavy quark $Q$.

    The hadronic cross section corresponding to Eq.~(\ref{eq:process}) can be computed by convoluting partonic cross sections with parton distribution functions $f_{a/h}(x,\mu_F^2)$ ($a=q,{\bar q},g$ denotes the massless partons) of the colliding hadrons. The partonic cross sections can be computed in QCD perturbation theory. At the leading order (LO) only two partonic processes contribute: quark-antiquark annihilation $q{\bar q}\to Q{\bar Q}$ and gluon fusion $gg\to Q{\bar Q}$.
    For both processes the ${\vec q}_T$ dependence of the cross section at LO is simply proportional to $\delta^{(2)}({\vec q}_T)$, since no radiation is emitted at this perturbative order.
    At higher perturbative orders the partonic cross section in the limit $q_T\to 0$ receives large logarithmic contributions of the form $\as^{n+2}\frac{1}{q_T^2} \ln^k (M^2/q_T^2)$ ($k\leq 2n-1$)
    that need be resummed to all orders.
    The resummation is customarily carried out in impact parameter (${\vec b}$) space, to factorise the kinematics of multiple parton emission.
    
The all-order structure of the logarithmically enhanced contributions can be written as~\cite{Catani:2014qha}
\begin{align}
\label{eq:resumm}
    &\frac{d\sigma(P_1,P_2;\, \vec q_T, M, y, \vec \Omega)}{d^2\vec q_T\, dM^2\, dy\, d\vec\Omega}=\frac{M^2}{2P_1\cdot P_2}\sum_{c=q, \bar q, g}\left[ d\sigma^{(0)}_{c\bar c}\right]\int\frac{d^2\vec b}{(2\pi)^2}e^{i\vec b\cdot \vec q_T}S_c(M,b)
    \nonumber\\
    &\times \sum_{a_1, a_2}\int_{x_1}^{1} \frac{dz_1}{z_1}\int_{x_2}^1\frac{dz_2}{z_2} [(\mathbf{H\Delta})C_1C_2]_{c\bar c;a_1a_2} f_{a_1/h_1}(x_1/z_1,b_0^2/b^2)f_{a_2/h_2}(x_2/z_2,b_0^2/b^2)\;,
\end{align}
where $b_0=2e^{-\gamma_E}$ ($\gamma_E=0.5772....$ is the Euler number) and the kinematic variables $x_1$ and $x_2$ are defined as
\begin{equation}
    x_{1,2}=\frac M{\sqrt{2P_1\cdot P_2}}\, e^{\pm y}\;.
\end{equation}
The symbol $\left[ d\sigma^{(0)}_{c\bar c}\right]$ is related to the LO cross section $d{\hat \sigma}_{c{\bar c}\to Q{\bar Q}}^{(0)}$
for the partonic process
\begin{equation}
  \label{eq:bornpro}
  c(p_1)+{\bar c}(p_2)\to Q(p_3) +{\bar Q}(p_4),~~~~~~~~~~~~~~c=q,{\bar q},g 
\end{equation}
with $p_i=x_i P_i$ ($i=1,2$), and we have
\begin{equation}
  \left[ d\sigma^{(0)}_{c\bar c}\right]=\as^2(M^2)\frac{d{\hat \sigma}^{(0)}_{c{\bar c}\to Q{\bar Q}}}{M^2d{\vec \Omega}}\, .
  \end{equation}
We briefly recall the perturbative ingredients entering the resummation formula in Eq.~(\ref{eq:resumm}) (more details can be found in Ref.~\cite{Catani:2014qha}). The formula contains process-dependent and process-independent contributions.
The functions $C_i$ include the contribution of radiation collinear to the initial-state partons at small momentum scales $q\lesssim 1/b$, while the Sudakov form factor $S_c$
accounts for soft and flavour-conserving collinear emissions at scales $1/b\lesssim q\lesssim M$.
Since they are originated by the spin- and $q_T$-dependent collinear splitting kernels, the functions $C_{i}$ feature also a dependence on the azimuthal degree of freedom of $\vec b$.
All the information on the process-dependent corrections is embodied in the term $\mathbf{H\Delta}$, while the collinear functions $C_{i}$ and the Sudakov form factor $S_c$ are universal.
The radiative factor $\mathbf{\Delta}$ is specific of heavy-quark pair production and is due to soft
radiation from the $Q \bar Q$ final state and from the initial-state and final-state interference.
It depends on the invariant mass $M^2$, on the kinematics of the partonic process in Eq.~(\ref{eq:bornpro}) and on the impact parameter $\vec b$. The azimuthal dependence can be specified through the angle $\phi=\phi_3-\phi_b$, where $\phi_3$ and $\phi_b$ are the azimuthal angles
of ${\vec p}_{3,T}$ and $\vec b$, respectively.
The hard-virtual term $\mathbf{H}$, which embodies virtual contributions at scale $q\sim M$, depends on the all-loop scattering amplitude ${\cal M}_{c{\bar c}\to Q{\bar Q}}$ for the partonic process $c{\bar c}\to Q{\bar Q}$.

The explicit form of the term  $\mathbf{H\Delta}$ is
\begin{equation}
\label{eq:HDelta}
(\mathbf{H\Delta})_{c{\bar c}}=\frac{\langle\widetilde{\cal M}_{c{\bar c}\to Q{\bar Q}}|\mathbf{\Delta}|\widetilde{\cal M}_{c{\bar c}\to Q{\bar Q}}\rangle}{\as^2(M^2)\left|{\cal M}_{c{\bar c}\to Q{\bar Q}}^{(0)}\right|^2}\;.
\end{equation}
The symbol ${\cal M}^{(0)}_{c{\bar c}\to Q{\bar Q}}$ denotes the Born-level amplitude, while $\widetilde{\cal M}_{c{\bar c}\to Q{\bar Q}}$ represents the all-loop renormalised amplitude after subtraction of the IR singularities (see Eq.~(\ref{eq:Mtilde})).
The amplitude $|\widetilde{\cal M}_{c{\bar c}\to Q{\bar Q}}\rangle$ is a vector in the colour space of $\{c,{\bar c},Q,{\bar Q}\}$ and $\mathbf{\Delta}$ is a colour-space operator. In the gluon fusion channel ($c=g$), the Lorentz (spin) indeces of $\widetilde{\cal M}$ in Eq.~(\ref{eq:HDelta}) are properly summed with the corresponding indeces of the gluon collinear functions $C_i$ (see Eqs.~(11) and (13) in Ref.~\cite{Catani:2014qha}).

The action of the colour factor $\mathbf{\Delta}$ is expressed in terms of the operators\footnote{Here and in the following, the additional dependence on the rapidity difference $y_{34}=y_3-y_4$ is left understood.} $\mathbf{D}$ and $\mathbf{V}$ \cite{Catani:2014qha}
\begin{align}
\label{eq:VDV}
    \mathbf{\Delta}(\vec b, M)=\mathbf{V}^\dagger(b,M)\mathbf{D}(\phi,\as(b_0^2/b^2))\mathbf{V}(b,M)\;.
\end{align}
The evolution factor $\mathbf{V}$ resums logarithmic terms $\as^n(M^2)\ln^k(M^2 b^2)$ (with $k\leq n$).
It is obtained by the exponentiation of the integral of the anomalous dimension matrix $\mathbf{\Gamma}_t$, which is specific of transverse-momentum resummation for $Q\overline{Q}$ production
\begin{align}
\label{eq:def V}
    \mathbf{V}(b,M)=\overline{P}_q\exp\left(-\int_{b_0^2/b^2}^{M^2}\frac{dq^2}{q^2}\mathbf{\Gamma}_t(\as(q^2))\right)\;.
\end{align}
The symbol $\overline{P}_q$ in Eq.~(\ref{eq:def V}) denotes
the anti path-ordering of the exponential matrix with respect to the integration variable $q^2$.

The soft-parton factor $\mathbf{D}$ in Eq.~(\ref{eq:VDV}) embodies the azimuthal correlations produced by the soft radiation and
it is defined \cite{Catani:2014qha} in such a way that it gives a trivial contribution after integration over the azimuthal angle. We have
\begin{align}
\label{eq:averageD}
    \braket{\mathbf{D}(\phi,\as)}_{\rm av.}=1\;,
\end{align}
where the symbol $\langle ...\rangle_{\rm av.}$ denotes the average with respect to the azimuthal angle $\phi$.

The explicit expressions of the factor $\mathbf{H\Delta}$ up to ${\cal O}(\as)$ and of the anomalous dimension $\mathbf{\Gamma}_t$ up to ${\cal O}(\as^2)$ are given in Ref.~\cite{Catani:2014qha}.\footnote{The explicit expressions of the corresponding resummation coefficients for the production of an arbitrary number of heavy quarks accompanied by a colourless system is reported in Ref.~\cite{Catani:2021cbl}. Note that the expression of the first-order contribution $\mathbf{D}^{(1)}$ to $\mathbf{D}$ therein is mistyped.
The correct expression is obtained by replacing 
${\widehat {\bf b}} \to - {\widehat {\bf b}}$
in Eqs.~(25) and (26).} 
In this paper we present a general discussion of the resummation factor $\mathbf{H\Delta}$
and of its detailed origin and dependence on soft-parton contributions. Moreover we explicitly compute $\mathbf{H\Delta}$ up to ${\cal O}(\as^2)$. This ${\cal O}(\as^2)$ result is also relevant in the context of the QCD computation of heavy-quark production at NNLO.
Indeed, as recalled below, it permits the NNLO implementation of the $q_T$ subtraction formalism for this production process.

Within the $q_T$-subtraction formalism, the NNLO differential cross section $d\sigma^{Q\overline{Q}}_{\rm NNLO}$ of the process in Eq.~(\ref{eq:process}) is split into a part with $q_T=0$ and one with $q_T\neq 0$	
\begin{equation}
  d\sigma^{Q\overline{Q}}_{\rm NNLO}=d\sigma^{Q\overline{Q}}_{\rm NNLO}\big|_{q_T=0}+d\sigma^{Q\overline{Q}}_{\rm NNLO}\big|_{q_T\neq 0}\;.
\end{equation}
Since at the Born level the final state ${Q\overline{Q}}$ has $q_T=0$,
the NNLO contributions at $q_T\neq 0$ are actually given by NLO contributions for the final state ${Q\overline{Q}}$+jets
\begin{align}
    d\sigma^{Q\overline{Q}}_{\rm NNLO}\big|_{q_T\neq0}=d\sigma^{{Q\overline{Q}}+\text{jets}}_{\rm NLO}\;.
\end{align}
At NNLO, we can hence handle the IR divergences of the $q_T\neq0$ part with the available NLO techniques.
By doing so, we are nevertheless left with additional singularities of purely NNLO origin connected to the limit $q_T\to0$, for which we need an additional subtraction.
Following this strategy, we write the cross section as~\cite{Catani:2007vq}
\begin{equation}
  \label{eq:qt_sub_QQ}
  d\sigma^{Q\overline{Q}}_{\rm NNLO}=\mathcal H^{Q\overline{Q}}_{\rm NNLO}\otimes d\sigma_{\rm LO}^{Q\overline{Q}}+\left[d\sigma_{\rm NLO}^{{Q\overline{Q}}+\text{jets}}-d\sigma^{\rm CT}_{\rm NNLO}\right]\;.
\end{equation}
The cancellation of the extra singularities of NNLO type is performed by introducing the counterterm $d\sigma^{\rm CT}_{\rm NNLO}$, while the coefficient $\mathcal H^{Q\overline Q}_{\rm NNLO}$ embodies the information on the virtual corrections to the process and contains the $q_T=0$ contribution.

The counterterm $d\sigma^{\rm CT}_{\rm NNLO}$ needs to capture the singular behaviour of the amplitude in the limit $q_T\to0$ and it can been derived by using the knowledge on the low transverse-momentum spectrum.
In particular, it can be obtained from the NNLO perturbative expansion of  the logarithmically-enhanced contributions of the resummation formula in Eq.~(\ref{eq:resumm}). 
It depends \cite{Bonciani:2015sha} on the resummation coefficients that already appear in the case of a colourless final state,
on the additional $Q{\bar Q}$ resummation coefficients at ${\cal O}(\as)$ and on the anomalous dimension $\mathbf{\Gamma}_t$ at ${\cal O}(\as^2)$.

The coefficient $\mathcal H^{Q\overline Q}_{\rm NNLO}$ contains the virtual corrections to the process in Eq.~(\ref{eq:bornpro}) and contributions that compensate for the subtraction of the counterterm $d\sigma^{\rm CT}_{\rm NNLO}$.
It is defined as the NNLO truncation of the following perturbative series
\begin{equation}
	    \label{eq:expansion H}
	    \mathcal H^{Q\overline{Q}}=1+\frac{\as}\pi \mathcal H^{{Q\overline{Q}}(1)}+\left(\frac{\as}\pi\right)^2 \mathcal H^{{Q\overline{Q}}(2)}+\dots
\end{equation}
where $\mathcal H^{Q\overline{Q}}$ can be expressed \cite{Bonciani:2015sha,Catani:2019iny,Catani:2021cbl} in terms of the functions that we just introduced in the context of $q_T$ resummation. We have
\begin{align}
\label{eq:HQQ}
\mathcal H^{Q\overline{Q}}= \braket{(\mathbf{H{\bf D}})C_1 C_2}_{\rm av.}\;,
\end{align}
where the average is over the azimuthal angle $\phi$, which appears \cite{Catani:2014qha} both in the factor
$\mathbf{D}$ and through the functions $C_i$ in the gluon channel.
Analogously to Eq.~(\ref{eq:HDelta}), the explicit form of the term $\mathbf{HD}$ reads
\begin{equation}
\label{eq:HD}
(\mathbf{HD})_{c{\bar c}}=\frac{\langle\widetilde{\cal M}_{c{\bar c}\to Q{\bar Q}}|\mathbf{D}|\widetilde{\cal M}_{c{\bar c}\to Q{\bar Q}}\rangle}{\as^2(M^2)\left|{\cal M}_{c{\bar c}\to Q{\bar Q}}^{(0)}\right|^2}\;.
\end{equation}
The second-order coefficient $\mathcal{H}^{Q\overline{Q}(2)}$ can be computed with the results presented in this paper.

\subsection{Soft contributions}
\label{sec:soft}

In our computation we regularise both ultraviolet and IR divergences by using conventional dimensional regularisation
in $D=4-2\epsilon$ space-time dimensions (see, e.g., Ref.~\cite{Baernreuther:2013caa}).
The $SU(N_c)$ QCD colour factors are $C_F=(N^2_c-1)/(2N_c)$, $C_A=N_c$, $T_R=1/2$ and we use $C_c=C_F$ if $c=q$ and $C_c=C_A$ if $c=g$.
We consider $n_f$ flavours of massless quarks in addition to the heavy quark $Q$.
The QCD running coupling $\as(\mu^2_R)=\as^{(n_f)}(\mu^2_R)$ is introduced through $\msbar$ renormalisation at the
scale $\mu_R$ and decoupling of the heavy quark \cite{Baernreuther:2013caa}.

We start our discussion by considering the finite part $\widetilde{\cal M}_{c{\bar c}\to Q{\bar Q}}$ of the all-order virtual
amplitude ${\cal M}_{c{\bar c}\to Q{\bar Q}}$, which is defined through the relation \cite{Catani:2014qha}
\begin{equation}
\label{eq:Mtilde}
|\widetilde{\cal M}_{c{\bar c}\to Q{\bar Q}}\rangle=\left[ 1-\widetilde{ \bf I}_{c{\bar c}\to Q{\bar Q}}\right]|{\cal M}_{c{\bar c}\to Q{\bar Q}}\rangle\;\;,
\end{equation}
where the subtraction operator $\widetilde{ \bf I}_{c{\bar c}\to Q{\bar Q}}$ in colour space has the following expansion
\begin{align}
\widetilde{\bf I}_{c{\bar c}\to Q{\bar Q}}(\as(M^2), \epsilon;\{p_i\}) =
\sum_{n=1}^{\infty}
\left(\frac{\as(\mu_R^2)}{2\pi} \right)^{\!\!n}
\;\widetilde{\bf I}^{(n)}_{c{\bar c}\to Q{\bar Q}}(\epsilon,M^2/\mu_R^2;\{p_i\})\, .
\label{eq:itilall}
\end{align}
It is useful to introduce the subtraction operator in the simpler case in which a colourless system $F$ with invariant mass $M$ is produced.
In this case we can write \cite{Catani:2013tia}
\begin{equation}
\label{eq:Mtildecolourless}
|\widetilde{\cal M}_{c{\bar c}\to F}\rangle=\left[ 1-\widetilde{I}_{c}\right]|{\cal M}_{c{\bar c}\to F}\rangle\;\;,
\end{equation}
where the subtraction operator $\widetilde{I}_c(\as(M^2),\epsilon)$ is now a c-number, and it can be perturbatively expanded as in Eq.~(\ref{eq:itilall}).
The explicit expression of the first two perturbative coefficients $\widetilde{I}^{(1)}_c(\epsilon,M^2/\mu_R^2)$ and $\widetilde{I}^{(2)}_c(\epsilon,M^2/\mu_R^2)$ can be found in Ref.~\cite{Catani:2013tia}.
We note that $\widetilde{I}_c$ depends on the initial-state parton $c$, but it is completely independent of the produced
colourless system $F$.
For later convenience we also define $V_c$ as follows
\begin{equation}
  \label{eq:Vc}
  V_c=\ln(1-\widetilde{I}_c)\, ,
\end{equation}
and we write its decomposition in IR divergent and IR finite components
\begin{equation}
  \label{eq:Vcsplit}
  V_c=V_c^{\rm sing}+V_c^{\rm fin}\, .
\end{equation}
The term $V_c^{\rm sing}$, which includes the complete IR divergent contributions to $V_c$,
is a perturbative series in powers of $\as(M^2)$ and the corresponding perturbative coefficients are proportional
to $\epsilon$ poles, with no additional $\epsilon$ dependence. The remaining $\epsilon$ dependence of $V_c$ is entirely embodied
in $V_c^{\rm fin}$, which is finite in the limit $\epsilon\to 0$.
The all-order virtual amplitude ${\cal M}_{c{\bar c}\to F}$ has IR divergent contributions that are cancelled by $\widetilde{I}_c$,
and $\widetilde{\cal M}_{c{\bar c}\to F}$ in Eq.~(\ref{eq:Mtildecolourless}) is IR finite in the limit $\epsilon\to 0$.

Comparing the transverse-momentum resummation formula in Eq.~(\ref{eq:resumm}) with the corresponding formula for the production
of a colourless system $F$ \cite{Catani:2013tia}, we recall \cite{Catani:2014qha} that the factor
$\langle\widetilde{\cal M}_{c{\bar c}\to Q{\bar Q}}|\mathbf{\Delta}|\widetilde{\cal M}_{c{\bar c}\to Q{\bar Q}}\rangle$ in Eq.~(\ref{eq:HDelta})
is analogous to the factor $|\widetilde{\cal M}_{c{\bar c}\to F}|^2$ for $F$ production and, therefore,
we can introduce the following master formula\footnote{For convenience, here and in the following the amplitudes ${\cal M}_{c{\bar c}\to Q{\bar Q}}$ and $\widetilde{{\cal M}}_{c{\bar c}\to Q{\bar Q}}$ are denoted as ${\cal M}$ and $\widetilde{\cal M}$, by removing the subscript $c{\bar c}\to Q{\bar Q}$.}
\begin{align}
\label{eq:HDelta master}
   \braket{\widetilde {\cal M}|\mathbf{\Delta}|\widetilde {\cal M}}=\left[\bra{{\cal M}}e^{V_c^*}e^{2\mathbf{F}_{\rm ex}(\vec{b})}e^{V_c}\ket{{\cal M}}\right]_{\epsilon=0}\;,
\end{align}
where we have written $1-\widetilde{I}_c=e^{V_c}$, according to Eq.~(\ref{eq:Vc}).
The term $\widetilde{I}_c$ in Eq.~(\ref{eq:Mtildecolourless}) is due to real emission contributions to the underlying partonic process
$c{\bar c}\to F$. More precisely, $\widetilde{I}_c$ is produced by radiation of final-state partons that are either soft or collinear to the colliding partons $c$ and ${\bar c}$ \cite{Catani:2013tia}.
In the case of $Q{\bar Q}$ production, the underlying partonic process is $c{\bar c}\to Q{\bar Q}$,
and the produced $Q$ and ${\bar Q}$ act as extra source of soft-parton radiation, while the accompanying initial-state collinear radiation is the same as for $F$ production. The amount of extra soft radiation due to $Q$ and ${\bar Q}$ is embodied by the factor
$e^{2\mathbf{F}_{\rm ex}}$ in the right-hand side of Eq.~(\ref{eq:HDelta master}).
This factor is the result of the integration of the soft-emission contributions after factorisation of the initial-state emission,
which is taken into account by the factor $e^{V_c}$. We note that $\mathbf{F}_{\rm ex}$ is a colour space operator, which depends on
the colour charges of the partons $c,{\bar c},Q,{\bar Q}$.

The real-emission factor $e^{V_c^*}e^{2\mathbf{F}_{\rm ex}(\vec{b})}e^{V_c}$ in Eq.~(\ref{eq:HDelta master}) is IR divergent, and it cancels
the IR divergences of the virtual amplitude ${\cal M}$. This cancellation mechanism and the ensuing structure of the IR-finite terms
$\mathbf{\Delta}$ and $\widetilde {\cal M}$ are discussed in the remaining part of this Section.

The structure of the IR singular contributions in QCD amplitudes with massive partons is discussed in Refs.~\cite{Catani:2000ef,Mitov:2009sv,Mitov:2010xw,Ferroglia:2009ep,Ferroglia:2009ii}.
The IR singularities of the amplitude ${\cal M}$ in Eq.~(\ref{eq:HDelta master}) are factorised
in the IR divergent operator $\mathbf{Z}$ \cite{Ferroglia:2009ii} that permits to write
the IR-finite remainder ${\cal M}_{\text{fin}}$ of the amplitude as follows
\begin{align}
\label{eq:Mfin}
|{\cal M}_{\text{fin}}(\mu_{\rm IR})\rangle=\mathbf{Z}^{-1}(\mu_{\rm IR})\ket{\cal M}\;.
\end{align}
Both $\mathbf{Z}(\mu_{\rm IR})$ and ${\cal M}_{\text{fin}}(\mu_{\rm IR})$ depend on the arbitrary subtraction scale $\mu_{\rm IR}$.
The operator $\mathbf{Z}(\mu_{\rm IR})$ is a perturbative series in powers of $\as(\mu^2_{\rm IR})$ and the corresponding perturbative
coefficients are proportional to $\epsilon$ poles, with no additional $\epsilon$ dependence.
Unless otherwise stated we will use $\mu_{\rm IR}=M$.

We write the operator $\mathbf{Z}$ as follows
\begin{align}
  \label{eq:ZM}
    \mathbf{Z}(M)=\mathbf{Z}_{\rm ex}Z_c(M)\;,
\end{align}
where the factor $Z_c(M)$ embodies the IR divergences due to the initial-state partons $c$ and ${\bar c}$,
while $\mathbf{Z}_{\rm ex}$ includes the additional IR divergences due to soft wide-angle radiation from the colour-charged
heavy quarks. Therefore $Z_c(M)$ is the IR divergent operator of the amplitude ${\cal M}_{c{\bar c}\to F}$ in Eq.~(\ref{eq:Mtildecolourless}), and we also have
\begin{equation}
  \label{eq:Zc}
  Z_c(M)=e^{-V_c^{\rm sing}}\, ,
\end{equation}
since the real-emission factor $e^{V_c}$ in Eq.~(\ref{eq:Mtildecolourless}) cancels the virtual IR divergences of ${\cal M}_{c{\bar c}\to F}$.
The operator $\mathbf{Z}_{\rm ex}$ can be obtained by exponentiation of the integral of the subtracted soft anomalous dimension $\mathbf{\Gamma}_\text{sub}$ introduced in Ref.~\cite{Catani:2014qha}. We have
\begin{equation}
  \label{eq:Zex}
    \mathbf{Z}_{\rm ex}(M)=\overline{P}_q \exp\left\{-\frac{1}{2}\int_0^{M^2} \frac{dq^2}{q^2}\mathbf{\Gamma}_{\rm sub}(\as(q^2))\right\}\;\;,
\end{equation}
where $\as(q^2)$ is the renormalised QCD coupling in $D=4-2\epsilon$ dimensions and the perturbative expansion of $\mathbf{\Gamma}_\text{sub}$ is
\begin{align}
\label{eq:subtractedGamma}
    \mathbf{\Gamma}_{\text{sub}}=\frac{\as}{2\pi}\mathbf{\Gamma}^{(1)}_{\text{sub}}
    +\left(\frac{\as}{2\pi}\right)^2\mathbf{\Gamma}^{(2)}_{\text{sub}}+\mathcal O(\as^3)\;.
\end{align}
The explicit scale dependence of $\as$ is
\begin{equation}
  \label{eq:asrun}
  \as(q^2)=\as(\mu^2)\left(\frac{\mu^2}{q^2}\right)^\epsilon\left[1-\frac{\beta_0}{\epsilon}\as(\mu^2)\left(1-\left(\frac{\mu^2}{q^2}\right)^\epsilon\right)+{\cal O}(\as^2)\right]\,
\end{equation}
where $\beta_0$ is the first coefficient of the QCD beta function
\begin{equation}
  \label{eq:beta0}
  12\pi \beta_0=11 C_A -2 n_f.
\end{equation}
Using Eq.~(\ref{eq:asrun}), the operator $\mathbf{Z}_{\rm ex}$ in Eq.~(\ref{eq:Zex}) can be written as
\begin{equation}
  \label{eq:ZexM}
  \mathbf{Z}_{\rm ex}(M)=e^{-\mathbf{V}_{\rm ex}(M^2)}\, ,
  \end{equation}
where the explicit expression of $\mathbf{V}_{\rm ex}$ up to ${\cal O}(\as^2)$ reads
\begin{align}
  \label{eq:VexM2}
    \mathbf{V}_{\rm ex}(M^2)=
    \frac{\as(M^2)}{2\pi}\left(-\frac 1{2\epsilon}\mathbf{\Gamma}^{(1)}_{\text{sub}}\right)
    +\left(\frac{\as(M^2)}{2\pi}\right)^2\left(\frac 1{\epsilon^2}\frac{\pi\beta_0}{2}\mathbf{\Gamma}^{(1)}_{\text{sub}}
    -\frac 1{\epsilon}\frac14\mathbf{\Gamma}^{(2)}_{\text{sub}}\right)
+{\cal O}(\as^3)    \;.
\end{align}
We note that the anti-path ordered operator ${\bar P}_q$ in Eq.~(\ref{eq:Zex}) is irrelevant
to evaluate $\mathbf{Z}_{\rm ex}$ 
up to ${\cal O}(\as^2)$.

Using Eqs.~(\ref{eq:Mfin})--(\ref{eq:Zc}) in the right-hand side of Eq.~(\ref{eq:HDelta master})  
we see that the colourless subtraction operator $V_c$ only cancels the IR singularities of ${\cal M}$ that originate
from the initial-state emission factor $Z_c(M)$.
The virtual IR divergences in $\mathbf{Z}_{\rm ex}$ are removed by the IR divergences in $\mathbf{F}_{\rm ex}$, as discussed in the following.
The perturbative expansion of $\mathbf{F}_{\rm ex}(\vec{b})$ can be written as
\begin{align}
  \label{eq:Fex_exp}
    \mathbf{F}_{\rm ex}(\vec{b})=
    \frac{\alpha_0}{2\pi} S_\epsilon\left(\frac{b^2\mu_0^2}{b_0^2}\right)^{\epsilon}\mathbf{F}_{{\rm ex},1}\left(\phi\right)
    +\left(\frac{\alpha_0}{2\pi}S_\epsilon\right)^2\left(\frac{b^2\mu_0^2}{b_0^2}\right)^{2\epsilon}\mathbf{F}_{{\rm ex},2}\left(\phi\right)+\mathcal O(\alpha_0^3)\;,
\end{align}
where $\alpha_0$ denotes the unrenormalised QCD coupling. 
In our calculation of $\mathbf{F}_{\rm ex}(\vec{b})$, the renormalisation of the coupling constant is taken into account by using the $\msbar$  scheme: the running coupling $\as$ is related to the bare coupling $\alpha_0$ via the relation
\begin{equation}
	\label{eq:renorm}
	\alpha_0\mu_0^{2\epsilon} S_\epsilon=\as(\mu_R^2) \mu_R^{2\epsilon}\left(1-\as(\mu_R^2) \frac{\beta_0}{\epsilon}
        +\mathcal O(\as^2)\right)\;,
\end{equation}
where\footnote{At the end of Sect.~\ref{sub:1L}
we comment on the contribution to $\mathbf{F}_{\rm ex}(\vec{b})$
of heavy-quark loops.} $\beta_0$ is given in Eq.~(\ref{eq:beta0}) and
\begin{equation}
S_\epsilon=(4\pi)^\epsilon e^{-\epsilon \gamma_E}
\end{equation}
is the customary $D$-dimensional spherical factor.
We note that the operator $\mathbf{F}_{\rm ex}(\vec b)$ fulfils the relation
$\mathbf{F}_{\rm ex}^\dagger(\vec b)=\mathbf{F}_{\rm ex}(-\vec b)$.
We also point out that, while the function $\mathbf{F}_{\rm ex}(\vec{b})$ depends on the vector $\vec b$, the dependence on $b$ in Eq.~(\ref{eq:Fex_exp})
is fully embodied in the prefactors $b^{2n\epsilon}$ and, therefore, the
perturbative coefficients $\mathbf{F}_{{\rm ex},n}(\phi)$ only depend on the azimuthal degree of freedom $\phi$ of $\vec b$.
In the following, this dependence is left understood.
Each perturbative coefficient can also be expanded in $\epsilon$ as follows
\begin{align}
\label{eq: Fex1 epsilon}
    &\mathbf{F}_{{\rm ex},1}=\frac 1\epsilon\, \mathbf{F}_{{\rm ex},1}^{(-1)}+\mathbf{F}_{{\rm ex},1}^{(0)}+\epsilon\, \mathbf{F}_{{\rm ex},1}^{(1)}+\dots\, ,\\
    \label{eq: Fex2 epsilon}
    &\mathbf{F}_{{\rm ex},2}=\frac 1{\epsilon^2}\, \mathbf{F}_{{\rm ex},2}^{(-2)}+\frac 1\epsilon\, \mathbf{F}_{{\rm ex},2}^{(-1)}+\mathbf{F}_{{\rm ex},2}^{(0)}+\dots\, .
\end{align}
The poles in Eqs.~(\ref{eq: Fex1 epsilon}) and (\ref{eq: Fex2 epsilon}) are due to the soft singularities of the real-emission contributions and,
as previously mentioned, they have to cancel the virtual IR divergences due to the factor $\mathbf{Z}_{\rm ex}$ in Eq.~(\ref{eq:HDelta master}).
The cancellation of IR divergences leads to relations between the coefficients $\mathbf{F}_{{\rm ex},n}^{(k)}$ in Eqs.~(\ref{eq: Fex1 epsilon}), (\ref{eq: Fex2 epsilon}) and the coefficients $\mathbf{\Gamma}_{\rm sub}^{(n)}$ of the $\epsilon$ pole contributions in Eqs.~(\ref{eq:ZexM}),(\ref{eq:VexM2}).
We find the following relations
\begin{align}
\label{eq: Fex1 check}
    &\mathbf{F}_{{\rm ex},1}^{(-1)}=-\frac 14 \left(\mathbf{\Gamma}^{(1)}_{\text{sub}}+{\rm h.c.}\right)\;,\\
\label{eq: Fex22 check}
    &\mathbf{F}_{{\rm ex},2}^{(-2)}=\pi\beta_0\mathbf{F}_{{\rm ex},1}^{(-1)}+\frac 18 \left[\left(\mathbf{\Gamma}_{\rm sub}^{(1)}-{\rm h.c.}\right),\mathbf{F}_{{\rm ex},1}^{(-1)}\right]\;,\\
\label{eq: Fex21 check}
    &\mathbf{F}_{{\rm ex},2}^{(-1)}=-\frac 18
    \left(\mathbf{\Gamma}^{(2)}_{\text{sub}}+{\rm h.c.}\right)
    +2\pi\beta_0\mathbf{F}_{{\rm ex},1}^{(0)}+ 
    \frac{1}{4}\left[\left(\mathbf{\Gamma}_{\rm sub}^{(1)}-{\rm h.c.}\right),\mathbf{F}_{{\rm ex},1}^{(0)}\right]\;.
\end{align}
The explicit calculation of $\mathbf{F}_{\rm ex}$ is presented in Sect.~\ref{sec:details}.
We have verified that Eqs.~(\ref{eq: Fex1 check})--(\ref{eq: Fex21 check}) are fulfilled by our final result for $\mathbf{F}_{\rm ex}$, which is an important cross-check of our computation.

We can now consider the master formula in Eq.~(\ref{eq:HDelta master}), implement the cancellation of the real and virtual IR divergences and
derive the expressions of $\mathbf{\Delta}$ and $\widetilde{\cal M}$.
We write
\begin{align}
\label{eq:MDeltaM intermediate}
\braket{\widetilde {\cal M}|\mathbf{\Delta}|\widetilde {\cal M}}
=&
       \left[\bra{{\cal M}_{\text{fin}}}e^{-\mathbf{V}_{\rm ex}^\dagger(M^2)}e^{-V_c^{\text{sing}*}}e^{V_c^*}e^{2\mathbf{F}_{\rm ex}(\vec{b})}e^{V_c}e^{-V_c^{\text{sing}}}e^{-\mathbf{V}_{\rm ex}(M^2)}\ket{{\cal M}_{\text{fin}}}\right]_{\epsilon=0}\nonumber\\\nonumber
       =&\left[\bra{{\cal M}_{\text{fin}}}e^{V_c^{\text{fin}*}}e^{-\mathbf{V}_{\rm ex}^\dagger(M^2)}e^{2\mathbf{F}_{\rm ex}(\vec{b})}e^{-\mathbf{V}_{\rm ex}(M^2)}e^{V_c^{\text{fin}}}\ket{{\cal M}_{\text{fin}}}\right]_{\epsilon=0}\\
       =&
    \left[\bra{{\cal M}_{\text{fin}}}e^{V_c^{\text{fin}*}}\mathbf{V}^\dagger_{\text{sub}}(b,M)e^{-\mathbf{V}_{\rm ex}^\dagger(b_0^2/b^2)}e^{2\mathbf{F}_{\rm ex}(\vec{b})}e^{-\mathbf{V}_{\rm ex}(b_0^2/b^2)}\mathbf{V}_{\text{sub}}(b,M)e^{V_c^{\text{fin}}}\ket{{\cal M}_{\text{fin}}}\right]_{\epsilon=0}\;.
\end{align}
In the first line of Eq.~(\ref{eq:MDeltaM intermediate}) we have used Eqs.~(\ref{eq:Mfin}), (\ref{eq:ZM}), (\ref{eq:Zc}) and (\ref{eq:ZexM}). In the second line we have used Eq.~(\ref{eq:Vcsplit}), and the fact that $V_c$ is a c-number that commutes with the other operators in colour space.
In the third line we have introduced the evolution operator $\mathbf{V}_{\text{sub}}$, defined by the following relation:
\begin{align}
  \label{eq:defvsub}
    e^{-\mathbf{V}_{\rm ex}(M^2)}&=e^{-\mathbf{V}_{\rm ex}\left(b_0^2/b^2\right)}{\bar P}_{q}\exp\left(-\frac 12 \int_{b_0^2/b^2}^{M^2}\frac{dq^2}{q^2}\mathbf{\Gamma_\text{sub}}(\as(q^2))\right)\nonumber\\
    &\equiv e^{-\mathbf{V}_{\rm ex}(b_0^2/b^2)}\mathbf{V}_{\text{sub}}(b,M)\;.    
\end{align}
The IR poles in the third line of Eq.~(\ref{eq:MDeltaM intermediate}) are fully contained in the
individual factors of the operator $e^{-\mathbf{V}_{\rm ex}^{\dagger}(b_0^2/b^2)}e^{2\mathbf{F}_{\rm ex}(\vec{b})}e^{-\mathbf{V}_{\rm ex}(b_0^2/b^2)}$.
Their cancellation takes place at the operator level after combining the exponential functions together, and it is guaranteed by the relations between $\mathbf{F}_{\rm ex}$ and $\mathbf{\Gamma}_{\text{sub}}$ that are reported in Eqs.~(\ref{eq: Fex1 check}) and (\ref{eq: Fex21 check}).
Therefore we can safely perform the limit $\epsilon\to0$ and we obtain a finite reminder that, for later convenience, we define as follows
\begin{align}
\label{eq:KK}
\lim_{\epsilon\to 0}\left(e^{-\mathbf{V}_{\rm ex}^{\dagger}(b_0^2/b^2)}e^{2\mathbf{F}_{\rm ex}(\vec{b})}e^{-\mathbf{V}_{\rm ex}(b_0^2/b^2)}\right)=
\mathbf{K}^\dagger(-\vec b)\mathbf{K}(\vec b)\;,
\end{align}
where we also used the relation $\mathbf{F}_{\rm ex}^\dagger(\vec b)=\mathbf{F}_{\rm ex}(-\vec b)$.

To recast Eqs.~(\ref{eq:MDeltaM intermediate}) and (\ref{eq:KK}) in the form of Eqs.~(\ref{eq:HDelta}) and (\ref{eq:VDV}) we isolate
the azimuthal dependence of $\mathbf{K}^\dagger(-\vec b)\mathbf{K}(\vec b)$ in a factor with azimuthal average equal to unity,
thus identifying the operator $\mathbf{D}(\phi,\as)$.
We write
\begin{align}
  \label{eq:KKD}
     \mathbf{K}^\dagger(-\vec b)\mathbf{K}(\vec b)=\mathbf{h}(\as(b_0^2/b^2))\mathbf{D}(\phi,\as)\mathbf{h}(\as(b_0^2/b^2))\;,
\end{align}
with
\begin{align}
    &\mathbf{h}^\dagger(\as(b_0^2/b^2))=\mathbf{h}(\as(b_0^2/b^2))\;,\\
    &\braket{\mathbf{D}(\phi,\as)}_{\rm av.}=1\;.
\end{align}
The expressions for the colour operators $\mathbf{h}$ and $\mathbf{D}$ can be trivially obtained from $\mathbf{K}$ as follows
\begin{align}
    &(\mathbf{h}(\as(b_0^2/b^2))^2=\braket{\mathbf{K}^\dagger(-\vec b)\mathbf{K}(\vec b)}_{\rm av.}\;,\\
    &\mathbf{D}(\phi,\as(b_0^2/b^2))=\mathbf{h}^{-1}(\as(b_0^2/b^2))\left(\mathbf{K}^\dagger(-\vec b)\mathbf{K}(\vec b)\right)\mathbf{h}^{-1}(\as(b_0^2/b^2))\;.
\end{align}
In terms of $\mathbf{F}_{\rm ex}$ and $\mathbf{\Gamma}_{\text{sub}}$ they read
\begin{align}
    \label{eq:h}
    &\mathbf{h}(\as)=
    1+\frac{\as}{2\pi}\braket{\mathbf{F}_{{\rm ex},1}^{(0)}}_{\rm av.}
    +\left(\frac{\as}{2\pi}\right)^2\Big\{\braket{(\mathbf{F}_{{\rm ex},1}^{(0)})^2}_{\rm av.}-\frac 12 \left(\braket{\mathbf{F}_{{\rm ex},1}^{(0)}}_{\rm av.}\right)^2
    \nonumber\\&\phantom{\mathbf{h}(\as)=}
    +\braket{\mathbf{F}_{{\rm ex},2}^{(0)}}_{\rm av.}
    -2\pi\beta_0\braket{\mathbf{F}_{{\rm ex},1}^{(1)}}_{\rm av.}
    -\frac 14\left[\left(\mathbf{\Gamma}^{(1)}_{\text{sub}}-{\rm h.c.}\right),\langle\mathbf{F}_{{\rm ex},1}^{(1)}\rangle_{\rm av.}\right]   
    \Big\}+\mathcal{O}(\as^3)\;,\\
    \label{eq:D}
    &\mathbf{D}(\phi,\as)=1+
    2\,\frac{\as}{2\pi}\left(\mathbf{F}_{{\rm ex},1}^{(0)}\right)_{\text{cor}}
    \nonumber\\&\phantom{\mathbf{D}(\phi,\as)=}
    +2\left(\frac{\as}{2\pi}\right)^2\left\{
    \left(\mathbf{F}_{{\rm ex},2}^{(0)} -2\pi\beta_0\mathbf{F}_{{\rm ex},1}^{(1)}-\frac 14\left[\left(\mathbf{\Gamma}^{(1)}_{\text{sub}}-{\rm h.c.}\right),\mathbf{F}_{{\rm ex},1}^{(1)}\right]
    \right)_{\text{cor}}
    \right.\nonumber\\&\phantom{\mathbf{D}(\phi,\as)=}\left.
    +\left((\mathbf{F}_{{\rm ex},1}^{(0)})^2\right)_{\text{cor}}
    -\braket{\mathbf{F}_{{\rm ex},1}^{(0)}}_{\rm av.}\left(\mathbf{F}_{{\rm ex},1}^{(0)}\right)_{\text{cor}}
    -\left(\mathbf{F}_{{\rm ex},1}^{(0)}\right)_{\text{cor}}\braket{\mathbf{F}_{{\rm ex},1}^{(0)}}_{\rm av.}
    \right\}+\mathcal{O}(\as^3)\;,
\end{align}
where, to keep the notation compact, we have defined the azimuthal correlation $(f)_{\rm cor}$ of an operator $f$ as
\begin{align}
    (f)_{\text{cor}}=f-\braket{f}_{\rm av.}\;.
\end{align}
In the operator $\mathbf{h}$ of Eq.~(\ref{eq:KKD}) the scale of $\as$ is $b_0^2/b^2$. The scale in $\mathbf{h}$ can be evolved up to the
hard scale $M^2$ by using the operator $\mathbf{V}_{\rm sub}$ of Eq.~(\ref{eq:defvsub}) and by introducing the operator $\mathbf{V}$ of Eq.~(\ref{eq:def V})
through the following relation
\begin{align}
\label{eq:hVh}
    \mathbf{V}(b,M)=\mathbf{h}(\as(b_0^2/b^2))\mathbf{V}_{\text{sub}}(b,M)\mathbf{h}^{-1}(\as(M^2))\;.
\end{align}
From here we also obtain the relation between the anomalous dimensions ${\bf \Gamma}_t$ and ${\bf \Gamma}_{\rm sub}$ in Eqs.~(\ref{eq:def V}) and (\ref{eq:defvsub}). Computing the logarithmic derivative of Eq.~(\ref{eq:hVh}) with respect to $M^2$ we find
\begin{equation}
  \label{eq:GtGsub}
  {\bf \Gamma}_t(\as)=\frac{1}{2}\mathbf{h}(\as){\bf \Gamma}_{\rm sub}(\as)\mathbf{h}^{-1}(\as)+\beta(\as)\frac{d\mathbf{h}(\as)}{d\ln \as} \mathbf{h}^{-1}(\as)\, ,
\end{equation}
where we have introduced the QCD $\beta$ function
\begin{equation}
  \beta(\as(q^2))=\frac{d\ln\as(q^2)}{d\ln q^2}=-\sum_{k=1}^\infty \beta_{k-1}\as^k(q^2)\, ,
\end{equation}
with $\beta_0$ given in Eq.~(\ref{eq:beta0}).
At ${\cal O}(\as^2)$ Eq.~(\ref{eq:GtGsub}) is Eq.~(40) of Ref.~\cite{Catani:2014qha} with the identification ${\bf F}_t^{(1)}=2\braket{\mathbf{F}_{{\rm ex},1}^{(0)}}_{\rm av.}$.

We can collect all the results of our discussion by inserting Eqs.~(\ref{eq:KK}), (\ref{eq:KKD}) and (\ref{eq:hVh})
in the third line of Eq.~(\ref{eq:MDeltaM intermediate}), and we obtain
\begin{align}
\label{eq:final mDelta}
    \bra{\widetilde {\cal M}}\mathbf{\Delta}\ket{\widetilde {\cal M}}=\bra{\widetilde {\cal M}}\mathbf{V}^\dagger(b,M)\mathbf{D}(\phi,\as(b_0^2/b^2))\mathbf{V}(b,M)\ket{\widetilde {\cal M}}\;,
\end{align}
where the IR finite matrix element $\ket{\widetilde{\cal M}}$ can be expressed as
\begin{align}
\label{eq:mtilde new}
    \ket{\widetilde {\cal M}}=\lim_{\epsilon\to 0}\left(
    \mathbf{h}(\as(M^2))e^{V_c^{\text{fin}}}\mathbf{Z}^{-1}\ket{{\cal M}}
    \right)\;,
\end{align}
and $\mathbf{Z}^{-1}\ket{{\cal M}}=\ket{{\cal M}_{\text{fin}}}$ can be obtained from Ref.~\cite{Baernreuther:2013caa}.\footnote{To be precise the numerical expression of the two-loop amplitude in Ref.~\cite{Baernreuther:2013caa} is presented by using $\mu_{\rm IR}=m$ as IR subtraction scale, while in Eq.~(\ref{eq:mtilde new}) the operator $\mathbf{Z}$ is defined at the IR subtraction scale $\mu_{\rm IR}=M$. Therefore the implementation of the results of Ref.~\cite{Baernreuther:2013caa} in Eq.~(\ref{eq:mtilde new}) requires the evolution of the numerical result presented in Ref.~\cite{Baernreuther:2013caa} from the scale $m$ to the scale $M$. We also note that a fully analytic result for the two-loop amplitude in the $q{\bar q}\to Q{\bar Q}$ channel became available recently \cite{Mandal:2022vju}.}
For convenience, we report the explicit expression of $V_c^{\rm fin}$ in the limit $\epsilon\to 0$ 
\begin{align}
\label{eq:Vcfin}
  V_c^{\rm fin}=&C_c\Big\{-\frac{\pi^2}{12}\left(\frac{\as(M^2)}{2\pi}\right)
  +\left(\frac{\as(M^2)}{2\pi}\right)^2\Big[\left(\frac{607}{162}-\frac{67}{144}\pi^2+\frac{\pi^4}{72}-\frac{77}{36}\right)C_A\nonumber\\
  &+\left(-\frac{41}{81}+\frac{5}{72}\pi^2+\frac{7}{18}\zeta_3\right)n_f-i\frac{\pi^4}{6}\beta_0\Big]+{\cal O}(\as^3)\Big\}\, .
\end{align}
In this Section we have discussed how the factors $\mathbf{\Delta}$ and $\widetilde{\cal M}$ that appear in
the transverse-momentum resummation formalism of Sect.~\ref{sec:resummation} are related to the soft-radiation contribution $\mathbf{F}_{\rm ex}(\vec b)$.
The first order resummation coefficients that were presented in Ref.~\cite{Catani:2014qha} depend on the first-order term $\mathbf{F}_{{\rm ex},1}$ in Eq.~(\ref{eq:Fex_exp}). In the following sections we illustrate the explicit computation of the first- and second-order terms $\mathbf{F}_{{\rm ex},1}$ and $\mathbf{F}_{{\rm ex},2}$.
In particular, $\mathbf{F}_{{\rm ex},2}$, controls the NNLO contribution to the operator $\mathbf{h}$ (see Eq.~(\ref{eq:h})) and, through Eq.~(\ref{eq:mtilde new}),
it allows us to evaluate the NNLO subtracted amplitude $\widetilde{\cal M}$.

\section{Details of the calculation}
\label{sec:details}
\subsection{The subtracted integrals}
The evaluation of the operator $\mathbf{F_{\rm ex}}(\vec b)$ introduced in the previous section requires the integration of the soft-parton contributions
after subtraction of the corresponding contribution of initial-state emission.
We can write this symbolically as
\begin{align}
\label{eq:Fex}
    \mathbf{F}_{\rm ex}(\vec{b})=\frac 12 \left(\mathcal{F}_{Q\bar Q}-\mathcal F_{\text{colourless}}\right)\equiv \frac 12 \mathcal{F}_{\rm sub}\;.
\end{align}
At NLO we just have to consider the emission of one soft gluon, which can be described by the customary tree-level eikonal factorisation formula, after subtraction of initial-state emission.
The relevant contribution is
\begin{equation}
  \label{eq:I0g}
  \mathbf{I}^{(0)}_g(\vec{b})=-\int \frac{d^Dk}{(2\pi)^{D-1}}\delta_+(k^2)\left|\mathbf{J}^{(0)}_g(k)\right|^2_{\rm sub} e^{i \vec{b}\cdot \vec k_T}\;,
\end{equation}
  where the subtracted squared current $\left|\mathbf{J}^{(0)}_g(k)\right|^2_{\rm sub}$ is defined in Eq.~(\ref{eq:single_gluon_NLO}).
At NNLO we need to consider contributions from:
\begin{itemize}
  	\item single-gluon emission at one-loop order (see Sect.~\ref{sub:1L});
	\item emission of a soft quark-antiquark pair (see Sect.~\ref{sub:qq});
	\item emission of two soft gluons (see Sect.~\ref{sub:gg}).
\end{itemize}
The corresponding integrals read
\begin{align}
  \mathbf{I}^{(1)}_g(\vec{b})&=-\int \frac{d^Dk}{(2\pi)^{D-1}}\delta_+(k^2)\left(\mathbf{J}^{(0)\dagger}_g(k)\mathbf{J}^{(1)}_g(k)+{\rm c.c.}\right)_{\rm sub} e^{i \vec{b}\cdot \vec k_T} \label{eq:Ibom1g}\;,\\
  \mathbf{I}^{(0)}_{q\bar{q}}(\vec{b})&=\int \frac{d^Dk_1}{(2\pi)^{D-1}} \frac{d^Dk_2}{(2\pi)^{D-1}}\delta_+(k_1^2)\delta_+(k_2^2)
  {\boldsymbol I}^{(0)}_{q\bar{q}}(k_1,k_2)\big|_{\rm sub} e^{i \vec{b}\cdot (\vec k_{T1}+k_{T2})}\label{eq:Ibomqq}\;,\\
   \mathbf{I}^{(0)}_{gg}(\vec{b})&=\frac{1}{2}\int \frac{d^Dk_1}{(2\pi)^{D-1}} \frac{d^Dk_2}{(2\pi)^{D-1}}\delta_+(k_1^2)\delta_+(k_2^2)  \mathbf{W}^{(0)}_{gg}(k_1,k_2)\big|_{\rm sub} e^{i \vec{b}\cdot (\vec k_{T1}+k_{T2})}\label{eq:Ibomgg}\;,
  \end{align}
where the soft factors $\left(\mathbf{J}^{(0)\dagger}_g(k)\mathbf{J}^{(1)}_g(k)+{\rm c.c.}\right)$, $\boldsymbol{I}^{(0)}_{q\bar{q}}(k_1,k_2)$ and $\mathbf{W}^{(0)}_{gg}(k_1,k_2)$
are explicitly given in Eq.~(\ref{eq:J0J1pcc}), (\ref{eq:taskqq_2}) and (\ref{eq:ggW}), respectively.
As in Eq.~(\ref{eq:I0g}), the label ``sub'' in Eqs.~(\ref{eq:Ibom1g})--(\ref{eq:Ibomgg}) denotes the subtraction procedure that removes the initial-state emission contributions.
Details of this procedure are given in Sects.~\ref{sub:1L}, \ref{sub:qq} and \ref{sub:gg}.
All the integrals are computed by using dimensional regularisation with $D=4-2\epsilon$ dimensions.
The relations with the perturbative coefficients $\mathbf{F}_{{\rm ex},1}$ and $\mathbf{F}_{{\rm ex},2}$ in Eq.~(\ref{eq:Fex_exp}) are
\begin{align}
 \label{eq:Fex1I} 
  2\times\frac{S_\epsilon}{8\pi^2}\left(\frac{b^2}{b_0^2}\right)^\epsilon \mathbf{F}_{{\rm ex},1}(\phi)&=\mathbf{I}_g^{(0)}(\vec{b})\, ,\\
  2\times\frac{S_\epsilon^2}{(8\pi^2)^2}\left(\frac{b^2}{b_0^2}\right)^{2\epsilon} \mathbf{F}_{{\rm ex},2}(\phi)&=\mathbf{I}_g^{(1)}(\vec{b})+\mathbf{I}_{q\bar{q}}^{(0)}(\vec{b})+\mathbf{I}_{gg}^{(0)}(\vec{b})\, .
  \label{eq:Fex2I}
\end{align}
We observe that the expression for $\mathbf{F}_{{\rm ex},2}$ (see Eq.~(\ref{eq: Fex2 epsilon})) has up to double poles in $\epsilon$. This is however not the case for the different contributions defined here: in particular terms $1/\epsilon^3$ are separately present in $\mathbf{I}^{(1)}_g$ and $\mathbf{I}^{(0)}_{gg}$, but they cancel in Eq.~(\ref{eq:Fex2I}).

All the integrals presented so far have been written in $b$-space, that is, in the space of the impact parameter $b$, connected to the ordinary space ($q_T$-space) by a Fourier transform.
The transformation from a $b$-space integral in a $q_T$-space one is hence obtained with the formal substitution
\begin{equation}
\label{eq:phasespace}
\delta^{(D-2)}\left(\vec q_T+ \vec k_{T1}+\vec k_{T2}\right)~~~\longrightarrow~~~e^{i \vec{b}\cdot\left(\vec k_{T1}+\vec k_{T2}\right)}\, .
\end{equation}

For the computation of the function $\mathbf h$ in Eq.~(\ref{eq:h}), azimuthal averages are required, which are denoted as $\braket{...}_{\rm av.}$. 
We compute the $D$-dimensional azimuthal average of a function $F(\phi)$ as
\begin{equation}
	\label{eq:azimuthal average}
	\braket{F(\phi)}_{\rm av.}=\frac1{B\left(\frac 12 , \frac 12-\epsilon\right)}\int_{-1}^1d \cos\phi\,(1-\cos^2\phi)^{-\frac 12-\epsilon}F(\phi)\;,
\end{equation}
where $B(x,y)$ is the Euler beta function.

We note that, when considering the azimuthally averaged result, the step from $b$-space to $q_T$-space is straightforward, and is determined by the overall dependence on $b$ of the integral under consideration.
Given a $b$-space function $I({\vec b})$ we introduce the corresponding $q_T$-space transform as
\begin{equation}
\label{eq:qttransform}
 {\tilde I}({\vec q_T})=\frac{1}{(2\pi)^{D-2}}\int d^{D-2}{\vec b}\, I({\vec b})\,e^{-i{\vec b}\cdot {\vec q_T}}\, . 
\end{equation}
Performing the azimuthal average in $q_T$ space of Eq.~(\ref{eq:qttransform}) we obtain
\begin{equation}
\label{eq:avequiv}
\langle{\tilde I}({\vec q_T})\rangle_{\rm av.}=\frac{1}{(2\pi)^{D-2}}\int d^{D-2}{\vec b}\, \langle I({\vec b})\rangle_{\rm av.}\,e^{-i{\vec b}\cdot {\vec q_T}}\, . 
\end{equation}
If the $b$-space function has the factorised form
\begin{equation}
I({\vec b})=f(b^2) {\bar I}({\hat b})  \;,
\end{equation}
where the function ${\bar I}({\hat b})$ depends only on the azimuthal angle of $\vec b$, Eq.~(\ref{eq:avequiv}) gives
\begin{equation}
  \label{eq:bspgen}
\langle {\tilde I}({\vec q_T})\rangle_{\rm av.}=\langle {\bar I}({\hat b})\rangle_{\rm av.}\frac{1}{(2\pi)^{D-2}}\int d^{D-2}{\vec b}\, f(b^2)\,e^{-i{\vec b}\cdot {\vec q_T}}\, .
\end{equation}  
By inspection of the structure of Eq.~(\ref{eq:Fex_exp}), we see that the soft integrals to be evaluated at N$^n$LO are of the form
\begin{equation}
  I({\vec b})=b^{2n\epsilon}{\bar I}(\hat b)\, .
\end{equation}
We can then use Eq.~(\ref{eq:bspgen}) with $f(b^2)=b^{2n\epsilon}$ to obtain
\begin{equation}
\label{eq:space_conversion}
\langle {\tilde I}({\vec q_T})\rangle_{\rm av.}= 4^{n\epsilon} \pi^{-1+\epsilon}\left(\frac{1}{q_T^2}\right)^{1+(n-1)\epsilon}\frac{\Gamma(1+(n-1)\epsilon)}{\Gamma(-n\epsilon)}\, \langle {\bar I}({\hat b})\rangle_{\rm av.}
\end{equation}

We conclude this section by specifying the kinematical variables for the Born level process in Eq.~(\ref{eq:bornpro}).
The polar angle $\theta$ is defined as the angle between the beam axis and the momentum of the final-state heavy quark in the centre-of-mass frame of the colliding partons. The variable $\beta$ is defined as
\begin{equation}
\label{eq:var_beta}
    \beta=\sqrt{1-\tau}\;,
\end{equation}
with $0<\tau<1$
\begin{equation}
\label{eq:var_tau}
    \tau = \frac{4 m^2}s\;,
\end{equation}
where $s=(p_1+p_2)^2=(p_3+p_4)^2$.
We also introduce the following auxiliary variables, that will be useful in order to write our partial results in a more compact form
	\begin{align}
		\label{eq:var_B}
		&B=\frac{p_{T,3}^2}{m^2}=\frac{p_{T,4}^2}{m^2}= \frac{\beta^2}{1-\beta^2}\,\sin^2 \theta\;\;,\\
		\label{eq:var_r}
		&r=\sqrt{1+B}\;,\\
		\label{eq:var_v}
		&  v =  \sqrt{1-\left( \frac{2m^2}{s-2m^2}\right)^2}=\frac{2\beta}{1+\beta^2}\;,\\
	    \label{eq:var_c}
		&  c= \frac{1-\beta}{1+\beta}\;,\\
		\label{eq:var_ct}
		&  c_T =\frac{1-\sqrt{1-r^2 \,\tau }}{1+\sqrt{1-r^2\, \tau }}\;.
	\end{align}

\subsection{Single gluon emission at tree level}
\label{sub:nlo}

The evaluation of the soft-gluon contributions at NLO has already been performed in Ref.~\cite{Catani:2014qha}.
In the following, we describe the strategy adopted to carry out the calculation.
We note that, for the extension to NNLO, we need to obtain the NLO result up to ${\cal O}(\epsilon)$ (see Eq.~(\ref{eq:h})).

The integral $\mathbf{I}_g^{(0)}(\vec{b})$ in Eq.~(\ref{eq:I0g}) is obtained from the subtracted current $\left|\mathbf{J}^{(0)}_g(k)\right|^2_{\rm sub}$, which is constructed as follows. 
We start from the tree-level eikonal current $\mathbf{J}^{(0)}_g(k)$ describing the emission of a soft gluon with momentum $k$ from the $c(p_1){\bar c}(p_2)\to Q(p_3){\bar Q}(p_4)$ Born level amplitude
\begin{equation}
    \label{eq:nlo current}
  	\mathbf{J}^{(0)}_{g,\mu}(k)=\sum_{i=1}^4 \textbf{T}_i\frac {p_{i\mu}}{\pik}\;.
\end{equation}
The corresponding factorisation formula reads\footnote{Here and in the following the unrenormalised scattering amplitudes are denoted as
${\cal M}^u=\alpha_0 \mu_0^{2\epsilon}\left({\cal M}^{(0)}+{\cal M}^{(1)}+....\right)$ where ${\cal M}^{(0)}$ is the tree-level contribution, ${\cal M}^{(1)}$ is the one-loop virtual correction and so forth.}
\begin{align}
    |{\cal M}^{(0)}_{c{\bar c}\to Q{\bar Q}g}|^2&\sim (g_0\mu_0^\epsilon)^2 \langle {\cal M}^{(0)}_{c{\bar c}\to Q{\bar Q}}|
\mathbf{J}^{(0)\dagger}_{g,\mu}(k)\,d^{\mu\nu}(k)\,\mathbf{J}^{(0)}_{g,\nu}(k)|{\cal M}^{(0)}_{c{\bar c}\to Q{\bar Q}}\rangle\nonumber\\
&=-(g_0\mu_0^\epsilon)^2 \langle {\cal M}^{(0)}_{c{\bar c}\to Q{\bar Q}}|
\left|\mathbf{J}^{(0)}_{g}(k)\right|^2|{\cal M}^{(0)}_{c{\bar c}\to Q{\bar Q}}\rangle\;,
\end{align}
where $g_0$ is the bare coupling ($g_0^2=4\pi\alpha_0$),
\begin{equation}
d^{\mu\nu}(k)=-g^{\mu\nu}+{\rm gauge~terms}
\end{equation}
is the spin-polarisation tensor of the soft gluon and the gauge terms give vanishing contribution due to current conservation.
The square of the current can be written in the form
\begin{align}
  \left|\mathbf{J}^{(0)}_g(k)\right|^2=&
  \sum_{j=3,4}\Bigg[\frac{p_j^2}{(p_j\cdot k)^2}\textbf{T}_j^2
  +\sum_{i=1,2}\frac{2p_i\cdot p_j}{(p_i\cdot k)(p_j\cdot k)}\textbf{T}_i\cdot\textbf{T}_j\Bigg]
  \nonumber\\&
  +\frac {2p_3\cdot p_4}{(p_3\cdot k) (p_4\cdot k)}\textbf{T}_3\cdot\textbf{T}_4 +\frac {2p_1\cdot p_2}{(p_1\cdot k) (p_2\cdot k)}\textbf{T}_1\cdot\textbf{T}_2
  \;.
\end{align}
From this expression we need to subtract the initial-state contribution, which is relevant for the production of a colourless system.
It reads
\begin{align}
  \label{eq:colourless square current NLO}
    \left|\mathbf{J}^{(0)}_g(k)\right|^2_{\text{colourless}}&=\frac{2p_1\cdot p_2}{\puk\pdk}\textbf{T}_1\cdot\textbf{T}_2
    =-\,\frac{\pupd}{\puk\pdk}\left(\textbf{T}_1^2+\textbf{T}_2^2\right)
    \nonumber\\&
     =-\,\left(\frac{\pupd}{\puk(p_1+p_2)\cdot k}+\frac{\pupd}{\pdk(p_1+p_2)\cdot k}\right)\left(\textbf{T}_1^2+\textbf{T}_2^2\right)\;,
\end{align}
where we used the colour conservation relation $\mathbf{T}_1+\mathbf{T}_2=0$ for the corresponding production process.
The subtracted squared current that appears in Eq.~(\ref{eq:I0g}) is defined as
\begin{align}
  \label{eq:single_gluon_NLO}
  \left|\mathbf{J}_g^{(0)}(k)\right|^2_{\text{sub}}=& \left|\mathbf{J}_g^{(0)}(k)\right|^2- \left|\mathbf{J}_g^{(0)}(k)\right|^2_{\rm colourless}\nonumber\\
  =&\sum_{j=3,4}\Bigg[\frac{m^2}{(p_j\cdot k)^2}\textbf{T}_j^2
  +2\sum_{i=1,2}\left(\frac{p_i\cdot p_j}{p_j\cdot k}-\frac{p_1\cdot p_2}{(p_1+p_2)k}\right)\frac{\textbf{T}_i\cdot\textbf{T}_j}{p_i\cdot k}\Bigg]
  \nonumber\\&
  +\frac {2p_3\cdot p_4}{(p_3\cdot k) (p_4\cdot k)}\textbf{T}_3\cdot\textbf{T}_4
  \;,
\end{align}
We emphasise that each of the three colour contributions in Eq.~(\ref{eq:single_gluon_NLO}) is separately collinear safe.

Using Eq.~(\ref{eq:single_gluon_NLO}) the evaluation of Eq.~(\ref{eq:I0g}) is reduced to the computation of the following integrals
\begin{align}
\label{eq:ijjnlo}
&I_{jj}(\vec{b}) =
\int d^D k \, \delta_+(k^2) \, \frac{m^2}{(p_j \cdot k)^2} \,e^{i \vec{b}\cdot \vec k_T}\;, \\
\label{eq:iijnlo}
&I_{ij}(\vec{b}) = \int d^D k  \, \delta_+(k^2)  \,
\frac{1}{p_i \cdot k}
\left[
\frac{p_i \cdot p_j}{p_j \cdot k} - \frac{p_1 \cdot p_2}{(p_1+p_2) \cdot k}
\right]\,e^{i \vec{b}\cdot \vec k_T}\;, \\
\label{eq:i34nlo}
&I_{34}(\vec{b}) = \int d^D k \, \delta_+(k^2)\,
\frac{p_3\cdot p_4}{(p_3 \cdot k) (p_4 \cdot k)}\,e^{i \vec{b}\cdot \vec k_T}
\;,
\end{align}
where $i=1,2$ labels an initial-state parton, while $j=3,4$ labels one of the final-state massive particles.
In terms of these definitions, Eq.~(\ref{eq:I0g}) reads
\begin{equation}
	\mathbf{I}^{(0)}_g(\vec{b})
  =
  -\frac1{(2\pi)^{D-1}}
  \Bigg\{\sum_{j=3,4}\left[ \,I_{jj}(\vec{b})\,\textbf{T}_j^2
  +2\sum_{i=1,2} I_{ij}(\vec{b}) \,\textbf{T}_i\cdot\textbf{T}_j\right]
  +2\, I_{34}(\vec{b})\, \textbf{T}_3\cdot\textbf{T}_4 \Bigg\}
  \;.
\end{equation}
The simplest contribution is the one where only a massive final-state particle is involved, $I_{jj}$ defined in Eq.~(\ref{eq:ijjnlo}).
To perform its computation, we introduce light-cone coordinates
	\begin{equation}
		\label{eq:lc}
		p_\pm=\frac{p_0\pm p_z}{\sqrt2}\;,~~~~~~~~~~~p^\mu k_\mu=p_+ k_-+ p_- k_+ - \vec{p}_T\cdot \vec{k}_T\;,
	\end{equation}
and Eq.~(\ref{eq:ijjnlo}) becomes
\begin{equation}
    I_{jj}({\vec b})=\int dk_+\,dk_-\,d^{D-2}\vec{k}_T\,\delta(2k_+k_--\vec{k}_T^2)\,
    \frac{m^2\, e^{i \vec{b}\cdot \vec k_T}}{\left(p_{j,+}k_-+p_{j,-}k_+-\vec{p}_{j,T}\cdot\vec{k}_T\right)^2}\;.
\end{equation}
The delta function can now be used to perform the integral over $k_+$
\begin{equation}
    I_{jj}({\vec b})=\int dk_-\,d^{D-2}\vec{k}_T\,
    \frac{2\,m^2\,k_-\,e^{i \vec{b}\cdot \vec k_T}}{\left(p_{j,-}k_T^2+2p_{j,+}k_-^2-2k_-\vec{p}_{j,T}\cdot\vec{k}_T\right)^2}\;.
\end{equation}
The leftover angular integral can be simplified by removing the angular dependence from the denominator
with an appropriate shift of the $\vec{k}_T$ variable.
We obtain
\begin{equation}
    I_{jj}({\vec b})=(2\pi)^{1-\epsilon}b^\epsilon m^2\,\int dk_-\,\frac{2k_-}{p_{j,-}^2}
    e^{i\frac{k_-}{p_{j,-}}\vec b \cdot \vec {p}_{j,T}}
    \int dk_T\,\frac{k_T^{1-\epsilon} J_{-\epsilon}(b k_T)}{\left(k_T^2+m^2\frac{k_-^2}{p_{j,-}^2}\right)^2}\;,
\end{equation}
where $J_n(x)$ is the Bessel function of the first kind.
Including the azimuthal average, we are now left with a three-fold integral that can be performed via standard techniques to all orders in $\epsilon$.

A similar strategy can be followed to compute $I_{ij}({\vec b})$ in Eq.~(\ref{eq:iijnlo}), but this requires some additional care.
The term $I_{ij}({\vec b})$ involves the contribution of the initial-state
emitter that is massless, and this may lead to a collinear singularity 
in the region $p_i \cdot k \to 0$. The collinear singularity is absent
in the complete integrand of $I_{ij}({\vec b})$, but it is present in the two 
separate contributions that correspond to the two terms in the square bracket
of Eq.~(\ref{eq:iijnlo}).
To apply the same integration procedure used for $I_{jj}({\vec b})$, the two contributions must be computed separately and, therefore,
a regulator for the collinear singularity needs be introduced.
We thus multiply the integrand by the factor 
\cite{Smirnov:1993ta,Becher:2011dz}
\begin{equation}
    \label{eq: lambda regulator}
    \left(\frac{p_i \cdot k}{m^2}\right)^{2\lambda}\;,
\end{equation}
where $\lambda$ is a small, positive coefficient and the mass scale $m$ has been chosen equal to the heavy-quark mass, but it is in principle arbitrary.
With the inclusion of this additional factor, the collinear singularity is regularised, and after integration it leads to poles in $\lambda$, which cancel 
with each other once the results from the two contributions are combined.

The divergence that appears in the intermediate steps of the evaluation of $I_{ij}({\vec b})$ is just an artifact of the approximation used to compute the small-$q_T$ behavior. Similar divergences  arise in SCET computations
and are usually called {\it rapidity divergences} \cite{Collins:2008ht,Collins:2011zzd,Becher:2010tm,Echevarria:2011epo,Chiu:2012ir}. However, we point out that
the term $I_{ij}({\vec b})$ and our entire soft contributions in 
Eqs.~(\ref{eq:I0g})--(\ref{eq:Fex2I})
have no collinear or rapidity divergences. In our computation
the collinear singularities from initial-state emission
can only appear
due to technical reasons, since for practical purposes we split integrable integrands 
in several non-integrable terms that are evaluated separately.

We now focus on the final integral, $I_{34}({\vec b})$ in Eq.~(\ref{eq:i34nlo}).
It can be computed from $I_{jj}({\vec b})$ by using Feynman parametrisation
\begin{equation}
    I_{34}({\vec b})=\int_0^1\,dx\int d^D k \, \delta_+(k^2) \, \frac{p_3\cdot p_4}{(p(x) \cdot k)^2} \,e^{i \vec{b}\cdot \vec k_T}=\frac {p_3\cdot p_4}{m^2}\,\int_0^1\,dx\,I_{jj}({\vec b})\big|_{p_j=p(x)}\;,
\end{equation}
where we introduced the auxiliary momentum
\begin{equation}
    \label{eq:auxiliary momentum}
    p^\mu(x)=xp_3^\mu+(1-x)p_4^\mu\;.
\end{equation}
The integration over the Feynman parameter can be easily performed in terms of multiple polylogarithms after the azimuthal average
and an expansion to ${\cal O}(\epsilon)$.

We now present our results for the azimuthally averaged integrals $\langle I_{jj}(\vec{b}) \rangle_{\rm av.}$, $\langle I_{ij}(\vec{b})\rangle_{\rm av.}$ and $\langle I_{34}(\vec{b})\rangle_{\rm av.}$.
When the all order result is available, we show both the expression before and after the $\epsilon$ expansion.
We find
\begingroup\allowdisplaybreaks
\begin{align}
  \label{eq:int AA NLO1}
\langle I_{jj}(\vec{b}) \rangle_{\rm av.} =& \pi^{1-\epsilon}\,\Gamma(1-\epsilon)
\left(\frac{b^2}{4}\right)^\epsilon
\left[
-\frac{1}{\epsilon}\, {}_2 F_1\left(1,-\epsilon ; 1-\epsilon ; -B\right)
\right]
\nonumber\\
=& \pi^{1-\epsilon}\,\Gamma(1-\epsilon)
\left(\frac{b^2}{4}\right)^\epsilon \Bigg[
-\frac{1}{\epsilon} - \ln\left(1+B\right)
+ \epsilon \, \liD \left(-B\right)
+ {\cal O}(\epsilon^2) \Bigg]\;,
\\
  \label{eq:int AA NLO2}
\langle I_{ij}(\vec{b})\rangle_{\rm av.} =&\lim_{\lambda\to0}\,
\frac 12 \pi^{1-\epsilon}\,
\left(\frac{b^2}{4}\right)^\epsilon
\Gamma(\tfrac{\lambda}{2}-\epsilon)\Gamma(\tfrac{\lambda}{2})
\Bigg[
\left(
\frac{p_i\cdot p_j}{m}
\right)^\lambda
{}_2 F_1\left(
\frac{\lambda}{2} , \frac{\lambda}{2}-\epsilon ; 1-\epsilon ; -B
\right)
\nonumber\\
&
-
\left(
\frac{p_1\cdot p_2}{\sqrt s}
\right)^\lambda
{}_2 F_1\left(
\frac{\lambda}{2} , \frac{\lambda}{2}-\epsilon ; 1-\epsilon ; -\frac{1}{s}
\right)
\Bigg]\nonumber\\
=& \pi^{1-\epsilon}\frac {\Gamma(1-\epsilon) }2 \,
\left(\frac{b^2}{4}\right)^\epsilon
\Bigg[
-\frac{2}{\epsilon} \ln \left( \frac{2\,p_i\cdot p_j}{\sqrt s\, m } \right)
+\liD \left(
-B
\right)
+ \epsilon
\liT \left(
-B
\right)
+ {\cal O}(\epsilon^2)\Bigg]\;,
\\
\label{eq:int AA NLO3}
\langle I_{34}(\vec{b})\rangle_{\rm av.} =&
\pi^{1-\epsilon}\,\Gamma(1-\epsilon)
\left(\frac{b^2}{4}\right)^\epsilon
\frac{1+\beta^2}{2\beta}
\left[
-\frac{1}{\epsilon} L_0(\beta) - L_1(\beta, \theta)
+ \epsilon \, P_2(\beta,\theta)
+ {\cal O}(\epsilon^2) \right]\;,
\end{align}
\endgroup
where the coefficient $\lambda$ is the one introduced with the collinear regulator in Eq.~(\ref{eq: lambda regulator}) and the functions $L_n(\beta, \theta)$, $P_n(\beta,\theta)$ are defined as
\begin{align}
  \label{eq:ln}
&L_n (\beta, \theta)=(p_3 \cdot p_4)\,
\frac{2\beta\, }{1+\beta^2}
\int_0^1 \frac{d x}{p(x)^2} \ln^n\left(1+\frac{\vec p_T(x)^2}{p(x)^2}\right) \to
\int_{-\beta}^{\beta} \frac{dz}{1 - z^2} \ln^n\left(
\frac{1 - z^2 \cos\theta}{1 - z^2} \right)\;,
\\
\label{eq:pn}
&P_n(\beta) =(p_3 \cdot p_4)\,
\frac{2\beta}{1+\beta^2}
\int_0^1 \frac{d x}{p(x)^2} \text{Li}_n\left(-\frac{\vec p_T(x)^2}{p(x)^2}\right) \to
\int_{-\beta}^{\beta} \frac{dz}{1 - z^2} \text{Li}_n\left(
  \frac{z^2\sin^2\theta}{z^2-1}
\right)\;.
\end{align}
The momentum $p^\mu(x)$ is defined in Eq.~(\ref{eq:auxiliary momentum}), and
in the last step in Eqs.~(\ref{eq:ln}), (\ref{eq:pn}) we have used $\vec p_3=-\vec p_4$.
The explicit expressions of the functions $L_0(\beta)$, $L_1(\beta,\theta)$ and $P_2(\beta,\theta)$ read
\begingroup\allowdisplaybreaks
\begin{align}
\label{eq:L0_result}
&L_0 (\beta)=\,  \ln\left(
\frac{1+\beta}{1-\beta}
\right)\;,
\\
&L_1 (\beta, \theta)=\,
\ln\left(\frac{1+\beta}{1-\beta} \right) \ln\left(1+B\right)
- \liD \left(\frac{4\beta}{(1+\beta)^2}\right)
-\frac{1}{2} \ln^2\left(\frac{1+\beta}{1-\beta}\right) +  \liD(1 - c \, c_T) 
\nonumber\\&\phantom{L_1 (\beta, \theta)=}
 + \liD\left( 1- \frac{c}{c_T} \right)+ \frac{1}{2} \ln^2 c_T \;,\\
\label{eq:P2_result}
&P_2(\beta,\theta)=
G\left(0,0,\frac{\beta -1}{2 \beta },\sin ^2\left(\frac{\theta }{2}\right)\right)+G\left(0,1,\frac{\beta
   -1}{2 \beta },\sin ^2\left(\frac{\theta }{2}\right)\right)
\nonumber\\\nonumber&\phantom{P_2(\beta,\theta)=}
   +G\left(0,\frac{\beta -1}{2 \beta },0,\sin
   ^2\left(\frac{\theta }{2}\right)\right)
   +G\left(0,\frac{\beta -1}{2 \beta },1,\sin ^2\left(\frac{\theta
   }{2}\right)\right)
 \\\nonumber&\phantom{P_2(\beta,\theta)=}
   -G\left(1,0,\frac{\beta -1}{2 \beta },\sin ^2\left(\frac{\theta
   }{2}\right)\right)
   -G\left(1,1,\frac{\beta -1}{2 \beta },\sin ^2\left(\frac{\theta
   }{2}\right)\right)
  \\\nonumber&\phantom{P_2(\beta,\theta)=}  
   -G\left(1,\frac{\beta -1}{2 \beta },0,\sin ^2\left(\frac{\theta
   }{2}\right)\right)-2
   \ln (1-\beta ) G\left(1,0,\sin ^2\left(\frac{\theta }{2}\right)\right)
    \\\nonumber&\phantom{P_2(\beta,\theta)=}
   -2 \ln (1-\beta )
   G\left(1,1,\sin ^2\left(\frac{\theta }{2}\right)\right)
   -G\left(1,\frac{\beta -1}{2 \beta },1,\sin ^2\left(\frac{\theta }{2}\right)\right)
   \\\nonumber&\phantom{P_2(\beta,\theta)=}
-\ln \left(\frac{\sin ^2(\theta )}{4}\right)
   G\left(0,\frac{\beta -1}{2 \beta },\sin ^2\left(\frac{\theta }{2}\right)\right)
   +\ln \left(\frac{\sin
   ^2(\theta )}{4}\right) G\left(1,\frac{\beta -1}{2 \beta },\sin ^2\left(\frac{\theta }{2}\right)\right)
      \\&\phantom{P_2(\beta,\theta)=}
   +2
   \ln (1-\beta ) \ln \left(\frac{\sin ^2(\theta )}{4}\right) \ln \left(\cos ^2\left(\frac{\theta
   }{2}\right)\right)\;.
\end{align}
\endgroup
with $c$ and $c_T$ defined in Eq.~(\ref{eq:var_c}) and Eq.~(\ref{eq:var_ct}) respectively.

The function $P_2(\beta,\theta)$ is expressed in terms of multiple polylogarithmic functions $G$. We note that the same kind of integrals in Eqs.~(\ref{eq:ln}), (\ref{eq:pn}) will also appear at a later stage in the computation of the double gluon emission contribution (see Sect.~\ref{sub:gg}): in this case though we will need $L_n$ and $P_n$ up to $n=3$.

\subsection{Single gluon emission at one loop}
\label{sub:1L}

\def\RR{{\cal R}}
\def\RRt{R}
\def\rrt{r}

We now focus on the emission of a soft gluon at one loop order.
The corresponding factorisation formula reads \cite{Catani:2000pi,Bierenbaum:2011gg}
\begin{align}
  \label{eq:oneloopfact}
  \langle {\cal M}^{(0)}_{c\bar{c}\to Q{\bar Q}g}|{\cal M}^{(1)}_{c\bar{c}\to Q{\bar Q}g}\rangle+{\rm c.c.}
  \simeq &-(g_0\mu_0^\epsilon)^2
  \left[\langle {\cal M}^{(0)}_{c\bar{c}\to Q{\bar Q}}|\mathbf{J}_{g}^{(0)}(k)\cdot \mathbf{J}_g^{(0)}(k)|{\cal M}^{(1)}_{c\bar{c}\to Q{\bar Q}}\rangle+{\rm c.c.}\right]\nonumber\\
  & +(g_0\mu_0^\epsilon)^4\left[\langle {\cal M}^{(0)}_{c\bar{c}\to Q{\bar Q}}|\mathbf{J}_g^{(0)\dagger}(k)\cdot \mathbf{J}_g^{(1)}(k)|{\cal M}^{(0)}_{c\bar{c}\to Q{\bar Q}}\rangle+{\rm c.c.}\right]\,,
\end{align}
where $\mathbf{J}_{g}^{(1)}(k)$ is the one-loop correction to the soft-gluon current.
The first contribution in Eq.~(\ref{eq:oneloopfact}) factorises the tree-level squared current from the interference between the $c{\bar c}\to Q{\bar Q}$ Born and one-loop amplitudes.
Such term does not lead to new soft contributions to $\mathbf{F}_{\rm ex}$.
The product of the tree and loop soft currents can be written as
\begin{align}
  \label{eq:J0J1pcc}
  \mathbf{J}_g^{(0)\dagger}(k)\cdot \mathbf{J}_g^{(1)}(k)+{\rm c.c.}=&2C_A\sum_{i\neq j}\left(\frac{(p_i\cdot p_j)}{(p_i\cdot k)(p_j\cdot k)}-\frac{m^2}{(p_j\cdot k)^2}\right)\RR_{ij}\, \mathbf{T}_i\cdot \mathbf{T}_j\nonumber\\
    &-4\pi\sum_{i,j,k}{\vphantom{\sum}}'
    \frac{p_i\cdot p_j}{(p_i\cdot k)(p_j\cdot k)} {\cal I}_{ik} f^{abc} T_i^a T_k^bT^c_j\, ,
\end{align}
where $\sum_{i,j,k}^\prime$ denotes the sum over distinct indices ($i\neq j$, $j\neq k$, $k\neq i$).
The expansion in $\epsilon$ of the $\RR_{ij}$, ${\cal I}_{ij}$  functions can be found in Ref.~\cite{Bierenbaum:2011gg} and, in the case of two massive emitters, a simplified expression has been presented in Ref.~\cite{Czakon:2018iev}.

Since we limit ourselves to considering heavy-quark production, we only need to evaluate the contribution proportional to $\RR_{ij}$. In fact ${\cal I}_{ij}$ is proportional to the three-partons correlator $f^{abc} T_i^a T_k^bT^c_j$ that vanishes when
acting on the tree-level amplitudes of the process\footnote{Note that this is not generally the case for processes in which the heavy-quark pair is accompanied by particles with complex couplings (see the Note Added in Ref.~\cite{Catani:2011st}).} $c{\bar c}\to Q{\bar Q}$~\cite{Forshaw:2008cq,Seymour:2008xr,
Czakon:2013hxa}.

By using colour conservation, we can apply the same procedure employed in Sect.~\ref{sub:nlo} to isolate the initial-state radiation and thus replace the first contribution
on the right hand side of Eq.~(\ref{eq:J0J1pcc}) with
\begin{align}
  \label{eq:subtracted_1L}
\Big(\mathbf{J}_g^{(0)}(k)\cdot &\mathbf{J}_g^{(1)}(k)+{\rm c.c.}\Big)_{\rm sub}\equiv\nonumber\\
=&2C_A\sum_{\substack{i=1,2\\j=3,4}}
\left[\left(
\frac {2\pipj}{\piq\pjq} - \frac {m^2}{\pjq^2} \right) \RR_{ij}-\frac {2\pipj}{\piq (p_1+p_2)\cdot k} \RR_{12}
\right] \textbf{T}_i \cdot \textbf{T}_j\nonumber\\
&+ 2C_A\left(
\frac {2(p_3\cdot p_4)}{(p_3\cdot k)(p_4\cdot k)} - \frac {m^2}{(p_3\cdot k)^2} - \frac {m^2}{(p_4\cdot k)^2}
\right) \, \RR_{34} \, \textbf{T}_3 \cdot \textbf{T}_4+.... 
\end{align}
where the dots stand for the contributions proportional to ${\cal I}_{ij}$ that will eventually vanish when evaluated onto tree-level amplitudes.
We now need to expand $\RR_{ij}$ in powers of $\epsilon$.  In order to match the normalisation used in Ref.~\cite{Bierenbaum:2011gg} we write
\begin{equation}
\RR_{ij}= \left(\frac{\pipj}{2 \piq\pjq}\right)^\epsilon \RRt_{ij}\;\;,
\end{equation}
where
\begin{equation}
\label{eq:Rijexp}
\RRt_{ij}=\sum_{n=-2}^\infty \RRt^{(n)}_{ij} \epsilon^n\;,
\end{equation}
and $\RRt^{(n)}_{ij}$ with $n\leq 2$ are given in Sect.~2 of Ref.~\cite{Bierenbaum:2011gg}.
The integral of the one-loop squared current in Eq.~(\ref{eq:Ibom1g}) 
can be organised into a massless-massive and a massive-massive contribution based on their colour factor
\begin{align}
 \mathbf{I}_g^{(1)}({\vec b}) =-\frac{2C_A}{(2\pi)^{D-1}}\left\{\sum_{\substack{j=1,2\\i=3,4}}I_{ij}^{(1)}({\vec b})\,\textbf{T}_i \cdot \textbf{T}_j+I_{34}^{(1)}({\vec b})\,\textbf{T}_3 \cdot \textbf{T}_4\right\}\;,
\end{align}
where the massless-massive contribution reads
\begin{equation}
\label{eq: Iij1L}
I_{ij}^{(1)}({\vec b})= \int d^Dk\,  \delta_+(k^2)\left[\left( \frac {2\pipj}{\piq\pjq} - \frac {m^2}{\pjq^2} \right) \RR_{ij}- \frac {2\pipj}{\piq (p_1+p_2)\cdot k} \RR_{12}\right]\, e^{i\vec{b}\cdot \vec{k}_T} \;,
\end{equation}
while the massive-massive contribution is
\begin{equation}
\label{eq: I341L}
I_{34}^{(1)}({\vec b})= \int d^Dk\,\delta_+(k^2) \left( \frac {2(p_3\cdot p_4)}{(p_3\cdot k)(p_4\cdot k)} - \frac {m^2}{(p_3\cdot k)^2} - \frac {m^2}{(p_4\cdot k)^2}\right) \, \RR_{34}\,  e^{i\vec{b}\cdot \vec{k}_T}  \;.
\end{equation}

\subsubsection{Massive-massless contribution: \texorpdfstring{$I_{ij}^{(1)}$}{Iij 1L}}
\label{sub:1L_ij}

By inspecting Eq.~(\ref{eq: Iij1L})
we can identify three different contributions proportional to $\frac \pipj{\piq\pjq}$, $\frac {m^2}{\pjq^2}$ and $\frac {\pipj}{\piq (p_1+p_2)\cdot k}$, respectively.
We therefore define the three auxiliary integrals
\begin{align}
\label{eq: Iijij1L}
&I_{ij,ij}^{(1)}({\vec b})=\int d^Dk\, \delta_+(k^2)\; \frac \pipj{\piq\pjq}\, \RRt_{ij} \left(\frac{\pipj}{2 \pik\pjk}\right)^\epsilon e^{i \vec b\cdot \vec k_T}\;,\\
\label{eq: Iijjj1L}
&I_{ij,jj}^{(1)}({\vec b})=\int d^Dk\, \delta_+(k^2)\; \frac {m^2}{\pjq^2}\,\RRt_{ij} \left(\frac{\pipj}{2 \pik\pjk}\right)^\epsilon e^{i \vec b\cdot \vec k_T}\;,\\
\label{eq: Iiji121L}
&I_{ij,i(12)}^{(1)}({\vec b})=\int d^Dk\, \delta_+(k^2)\; \frac {\pipj}{\piq (p_1+p_2)\cdot k}\, \RRt_{12} \left(\frac{\pupd}{2 \puk\pdk}\right)^\epsilon e^{i \vec b\cdot \vec k_T}\;.
\end{align}
In terms of these auxiliary integrals, $I_{ij}^{(1)}$ reads
\begin{equation}
\label{eq: Iij1L_2}
I_{ij}^{(1)}({\vec b})=2\, I_{ij,ij}^{(1)}({\vec b})-I_{ij,jj}^{(1)}({\vec b})-2I_{ij,i(12)}^{(1)}({\vec b})\;.
\end{equation}

We start from $I_{ij,ij}^{(1)}$. In this case we have a collinear singularity associated with the radiation from the initial-state massless particle which is due to the factor $\pik$ in the denominator.
To take care of it, we can introduce a $\lambda$ regulator similarly to what was done in the case of the NLO contribution in Eq.~(\ref{eq: lambda regulator})
\begin{equation}
\left(\frac{p_i\cdot k}{m^2}\right)^{2\lambda}\;,
\end{equation}
with $\lambda$ being positive. The collinear singularity will then be translated into poles in $\lambda$, which will cancel with analogous poles in the massless-massless contribution $I_{ij,i(12)}^{(1)}$.

From this stage, we can closely follow the procedure used in Sect.~\ref{sub:nlo} to perform the integral over the phase space of the emitted gluon.
We obtain
\begin{align}
\label{eq:Iay}
I_{ij,ij}^{(1)}({\vec b})=& (m^2)^{-3\lambda}\pi^{1-\epsilon} \pipj^{2\lambda}\left(\frac{b^2}4\right)^{2\epsilon-\lambda}\frac{\Gamma(-2\epsilon+\lambda)}{2\Gamma(1+\epsilon-\lambda)}\int_0^{\infty}\,dw\,\frac{\RRt_{ij}(w)}{w(1+w)^{1+\epsilon}}\nonumber\\\nonumber
&\times\left\{w^\lambda+\left[{}_2F_1\left(-2\epsilon,-\epsilon;\frac 12;\left(\frac{\vec b \cdot \vec p_{T,j}}{b\,m}\right)^2 w\right)-1-4\epsilon \frac{i\vec b \cdot \vec p_{T,j}}{b\,m}
\right.\right.
\\&
\left.\left.
 \times\sqrt w\, \frac{\Gamma(\frac 12-2\epsilon)\Gamma(1+\epsilon)}{\Gamma(1-2\epsilon)\Gamma(\frac 12+\epsilon)}{}_2F_1\left(\frac 12-2\epsilon,\frac 12 -\epsilon;\frac 32;\left(\frac{\vec b \cdot \vec p_{T,j}}{b\,m}\right)^2 w\right)\right]\right\}\;,
\end{align}
which after azimuthal average becomes
\begin{align}
\label{eq:Iay_aa}
\braket{I_{ij,ij}^{(1)}({\vec b})}_{\rm av.}= & \; (m^2)^{-3\lambda}\pi^{1-\epsilon} \pipj^{2\lambda}\left(\frac{b^2}4\right)^{2\epsilon-\lambda}\frac{\Gamma(-2\epsilon+\lambda)}{2\Gamma(1+\epsilon-\lambda)}\int_0^{\infty}\,dw\,\frac{\RRt_{ij}(w)}{w(1+w)^{1+\epsilon}}\nonumber\\
 &\times\left\{w^\lambda+
\text{Re}\left\{\left[{}_2F_1\left(-2\epsilon, -\epsilon; 1-\epsilon;w B\right)-1\right]\left(1+i\cot(\pi\epsilon)\right)\right\}
\right\}
\;.
\end{align}
Expanding in $\epsilon$ the expression in the curly bracket of Eq.~(\ref{eq:Iay_aa}) we obtain
\begin{align}
\label{eq:IayAA}
\text{Re}\left\{\left[{}_2F_1\left(-2\epsilon, -\epsilon; 1-\epsilon;w B\right)-1\right]\left(1+i\cot(\pi\epsilon)\right)\right\}
=
&
-\frac {2\epsilon}{\pi}\text{Im}\left[\text{Li}_2\left(w B\right)\right]+\mathcal O(\epsilon^2)\nonumber\\
=&2\epsilon \ln\left(w B\right)\theta\left(w B-1\right)+\mathcal O(\epsilon^2)\;.
\end{align}
In Eqs.~(\ref{eq:Iay}) and (\ref{eq:Iay_aa}) we have defined the adimensional variable $w$ as
\begin{align}
w=\frac{m^2}{p_{j,-}^2}\frac{k_-^2}{k_T^2}\;.
\end{align}
The kinematical invariants can be written in terms of $w$ as
\begin{align}
&\pik=\frac{\pipj}{m} k_T\sqrt w\;,\\
&\pjk=\frac{m}{2} k_T \left(\frac 1{\sqrt w}+\sqrt{w}\right)\;,
\end{align}
while the explicit expression of the coefficients $\RRt_{ij}^{(n)}$ presented in Ref.~\cite{Bierenbaum:2011gg} in terms of $w$ reads
\begingroup\allowdisplaybreaks
\begin{align}
\label{eq:Ry-2}
&\RRt_{ij}^{(-2)}\,=\,-\frac 12\;,\\
\label{eq:Ry-1}
&\RRt_{ij}^{(-1)}\,=\,0\;,\\
\label{eq:Ry0}
&\RRt_{ij}^{(0)}\,=\,\frac{1}{24} \left(5 \pi ^2-6 w \ln ^2\left(\frac{w}{w+1}\right)\right)\;,\\
\label{eq:Ry1}
&\RRt_{ij}^{(1)}\,=\,\frac{1}{12} \Bigg(6 (w-1) \text{Li}_3\left(\frac{w}{w+1}\right)+6 (w-1)
   \text{Li}_2\left(\frac{1}{w+1}\right) \ln \left(\frac{w}{w+1}\right)+2 (7-3 w) \zeta_3\nonumber\\
   &\phantom{\RRt_{ij}^{(1)}\,=\,}+\ln \left(\frac{w}{w+1}\right) \left(\pi ^2 (6 w+1)-3 (w-1) \ln
   \left(\frac{w}{w+1}\right) \ln (w+1)\right)\Bigg)\;.
\end{align}
\endgroup

Our task is now to integrate Eq.~(\ref{eq:Iay_aa}) with the expansion of $\RRt$ defined in Eqs.~(\ref{eq:Ry-2})--(\ref{eq:Ry1}).
The final result reads
\begingroup\allowdisplaybreaks
\begin{align}
  \label{eq:risI1ijij}
  \nonumber
\braket{I_{ij,ij}^{(1)}({\vec b})}_{\rm av.}=&
(m^2)^{-3\lambda}\pi^{1-\epsilon} \pipj^{2\lambda}\left(\frac{b^2}4\right)^{2\epsilon-\lambda}\frac{\Gamma(-2\epsilon+\lambda)}{2\Gamma(1+\epsilon-\lambda)}
\Bigg\{
\frac{1}\lambda\left[-\frac{1}{2 \epsilon ^2}+\frac{5 \pi ^2}{24}+\frac{7 \zeta_3 \epsilon
   }{6}+\mathcal{O}\left(\epsilon ^2\right)\right]
\\\nonumber&
   +\Bigg[\frac1\epsilon\left(-\text{Li}_2\left(-\frac{1}{B}\right)-\frac{1}{2} \ln ^2(B)-\frac{\pi
   ^2}{12}\right)
+\frac{1}{6} \Bigg(-6 \text{Li}_3\left(\frac{B}{B+1}\right)
\\\nonumber&
-6\text{Li}_2\left(-\frac{1}{B}\right) \ln (B+1)
+\ln^3\left(\frac{1}{B}+1\right)+\ln ^3(B)
+3 \ln (B) \ln ^2(B+1)
\\ \nonumber &
-6 \ln ^2(B) \ln(B+1)
   -2 \pi ^2 \ln \left(\frac{1}{B}+1\right)
   -2 \pi ^2 \ln (B)
   +\pi ^2 \ln (B+1)-6
   \zeta_3\Bigg)
\\\nonumber &
   +\epsilon  \Bigg(\zeta_3 \ln (B)+ \zeta_3 \ln
   (B+1)-\frac 12 \text{Li}_2\left(-\frac{1}{B}\right){}^2+\frac 14 \pi ^2
   \text{Li}_2\left(-\frac{1}{B}\right)
\\\nonumber&
-2 \text{Li}_4\left(-\frac{1}{B}\right)-
   \text{Li}_4\left(\frac{1}{B+1}\right)- \text{Li}_4\left(\frac{B}{B+1}\right)-\frac 12
   \text{Li}_2\left(-\frac{1}{B}\right) \ln ^2(B)
\\\nonumber&
  -
   \text{Li}_3\left(-\frac{1}{B}\right) \ln (B)-
   \text{Li}_3\left(-\frac{1}{B}\right) \ln (B+1)-2
   \text{Li}_3\left(\frac{1}{B+1}\right) \ln (B+1)
\\\nonumber&
   -2
   \text{Li}_3
   \left(\frac{B}{B+1}\right) \ln (B+1)+
   S_{2,2}\left(-\frac{1}{B}\right)-\frac{1}{24} \ln ^4(B)+\frac{7}{24} \ln ^4(B+1)
   \\\nonumber&
   +\frac 13 \ln (B+1) \ln
   ^3(B)+\frac 13 \ln ^3\left(\frac{1}{B}+1\right) \ln (B+1)-\frac 34 \ln ^2(B+1) \ln ^2(B)
   \\\nonumber&
   -\frac{73}{24}
   \pi ^2 \ln ^2(B)+\frac{19}{6} \pi ^2 \ln ^2\left(\frac{1}{B}+1\right)-\frac{11}{4} \pi ^2 \ln
   ^2(B+1)
   \\\nonumber&
   -\frac{1}{24} \ln ^2\left(\frac{1}{B}+1\right) \ln ^2\left(\frac{2 B}{2 B+1}\right)-\frac{1}{24}
   \ln ^2\left(\frac{1}{B}+1\right) \ln ^2\left(\frac{1}{2 B+1}+1\right)
\\\nonumber&
   +\frac{1}{12} \ln
   ^2\left(\frac{1}{B}+1\right) \ln \left(\frac{2 B}{2 B+1}\right) \ln
   \left(\frac{1}{2 B+1}+1\right)+\frac{35}{6} \pi ^2 \ln (B+1) \ln (B)
    \\&
    -\frac 23 \pi ^2 \ln
   \left(\frac{1}{B}+1\right) \ln (B+1)-\frac{23}{240} \pi ^4\Bigg)+\mathcal{O}\left(\epsilon
   ^2\right)\Bigg]+\mathcal{O}\left(\lambda\right)\Bigg\}\;.
\end{align}
\endgroup
with $B$ defined as in Eq.~(\ref{eq:var_B}) and $S_{2,2}$ being the Nielsen generalised polylogarithm function.

We now consider the integral $I_{ij,jj}^{(1)}({\vec b})$.
The only difference with $I_{ij,ij}^{(1)}({\vec b})$ consists in the replacement $\frac \pipj{\piq\pjq} \to \frac {m^2}{\pjq^2}$, which in terms of our integration variable $w$ implies
\begin{equation}
\label{eq:subii}
 \frac{2}{k_T^2 (1+w)} \;
\longrightarrow \;
 \frac{2}{k_T^2 (1+w)} \frac{2 w}{(1+w)} \;,
\end{equation}
that is, we simply need to multiply the integrand by a factor $2w/(1+w)$.
In addition, the presence of only final-state emitters in the integrand implies that we can set $\lambda = 0$ throughout.
We can use this method to obtain from Eq.~(\ref{eq:Iay}) an expression for $I_{ij,jj}^{(1)}$ as an integral over $w$
\begin{align}
\label{eq:Iby}
I_{ij,jj}^{(1)}({\vec b})=& \pi^{1-\epsilon} \left(\frac{b^2}4\right)^{2\epsilon}\frac{\Gamma(-2\epsilon)}{\Gamma(1+\epsilon)} \int_0^{\infty}\frac{\RRt_{ij}(w)}{(1+w)^{2+\epsilon}}\Bigg\{\Bigg[{}_2F_1\left(-2\epsilon,-\epsilon;\frac 12;c_{jb}^2 w\right)
\nonumber\\&
-4\epsilon i c_{jb} \sqrt w\, \frac{\Gamma(\frac 12-2\epsilon)\Gamma(1+\epsilon)}{\Gamma(1-2\epsilon)\Gamma(\frac 12+\epsilon)}{}_2F_1\left(\frac 12-2\epsilon,\frac 12 -\epsilon;\frac 32;c_{jb}^2 w\right)\Bigg]\Bigg\}\;,
\end{align}
where we have defined
\begin{equation}
\label{eq:cjb}
c_{jb}=\frac{{\vec b}\cdot {\vec p}_{T,j}}{b\,m}\, .
\end{equation}
The azimuthally averaged equivalent of Eq.~(\ref{eq:Iby}) can be obtained from Eq.~(\ref{eq:Iay_aa})
\begin{align}
\braket{I_{ij,jj}^{(1)}({\vec b})}_{\rm av.}=& \pi^{1-\epsilon} \left(\frac{b^2}4\right)^{2\epsilon}\frac{\Gamma(-2\epsilon)}{\Gamma(1+\epsilon)}\int_0^{\infty}\,dw\,\frac{\RRt_{ij}(w)}{(1+w)^{2+\epsilon}}
\nonumber\\&\times
\left\{1+
\text{Re}\left\{\left[{}_2F_1\left(-2\epsilon, -\epsilon; 1-\epsilon;w B\right)-1\right]\left(1+i\cot(\pi\epsilon)\right)\right\}
\right\}
\;.
\end{align}
We obtain
\begingroup\allowdisplaybreaks
\begin{align}
\nonumber\braket{I_{ij,jj}^{(1)}({\vec b})}_{\rm av.}=&\pi^{1-\epsilon} \left(\frac{b^2}4\right)^{2\epsilon}\frac{\Gamma(-2\epsilon)}{\Gamma(1+\epsilon)}
\times \Bigg\{\frac{1}{\epsilon ^2}+\frac1\epsilon\left(2 \ln (B+1)-1\right)
\\\nonumber&
+\left(2
   \text{Li}_2\left(-\frac{1}{B}\right)+\ln ^2(B)+\ln ^2(B+1)-2 \ln (B+1)-\frac{\zeta_3}{2}+\frac{13 \pi ^2}{24}+\frac{3}{2}\right)
\\\nonumber&
   +\epsilon  \Bigg(-3 \text{Li}_3(-B)+2
   \text{Li}_3\left(\frac{1}{B+1}\right)-\text{Li}_3\left(\frac{B}{B+1}\right)+\text{Li}_
   4\left(-\frac{1}{B}\right)
   \\\nonumber&
   -\text{Li}_4\left(\frac{B}{B+1}\right)
   -2
   \text{Li}_4\left(\frac{1}{B+1}\right)
   +\frac{1}{2
   } \text{Li}_2\left(-\frac{1}{B}\right) \left(\ln ^2(B+1)-2 \ln (B+1)-6\right)
\\\nonumber&
   -2
   \text{Li}_3\left(\frac{1}{B+1}\right) \ln (B+1)-\text{Li}_3\left(\frac{B}{B+1}\right)
   \ln (B+1)+\frac{1}{24}\ln ^4(B)
   +\frac{3}{8} \ln ^4(B+1)
\\\nonumber&
   -\frac{2}{3} \ln ^3(B+1)
   \ln (B)+\frac{1}{3} \ln ^3(B+1)+\frac{1}{4} \ln ^2(B+1) \ln ^2(B)
   -\frac{1}{2} \ln
   (B+1) \ln ^2(B)
   \\\nonumber&
   +\frac{1}{12} \pi ^2 \ln ^2(B)-\frac{3 \ln ^2(B)}{2}+\ln ^2(B+1)
   \ln (B)-\frac{1}{12} \pi ^2 \ln ^2(B+1)
   \\\nonumber&
   -\frac{1}{2} \ln ^2(B+1)
+\frac{7}{12} \pi ^2
   \ln (B+1)+3 \ln (B+1)-\frac{7}{2} - \frac{\zeta_2}4 + \frac{\zeta_3}6 - 5 \zeta_4 \Bigg)
      \\&
      +\mathcal{O}\left(\epsilon ^2\right)\Bigg\}\;.
\end{align}
\endgroup

We finally consider $I_{ij,i(12)}^{(1)}({\vec b})$. To obtain a one-fold integral representation of this contribution we can again take advantage of the result for $I_{ij,ij}^{(1)}$, identifying $p_j = p_1 + p_2$. This also means setting $\vec b \cdot \vec p_{T,j} = 0$ due to the absence of a transverse component in $p_1 + p_2$.
Nevertheless, the identification between the results is not completely straightforward because of the additional difference in the integrand
\begin{equation}
    \left(
\frac{\pipj}{\piq \pjq}
\right)^\epsilon
\;\longrightarrow\;
\left(
\frac{\pupd}{(p_1\cdot k) (p_2\cdot k)}
\right)^\epsilon\;.
\end{equation}
However, we can notice that
\begin{equation}
\left(
\frac{\pipj}{\piq \pjq}
\right)^\epsilon
=
\left(
\frac{2}{k_T^2}
\right)^\epsilon (1+w)^{-\epsilon}
\end{equation}
while
\begin{equation}
\left(
\frac{\pupd}{(p_1\cdot k) (p_2\cdot k)}
\right)^\epsilon
=
\left(
\frac{\pipj}{\piq (\pjq - \piq)}
\right)^\epsilon
=
\left(
\frac{2}{k_T^2}
\right)^\epsilon \;.
\end{equation}
Thus we can take care of this additional difference by adding a factor $(1+w)^{\epsilon}$ to the one-fold representation of $I_{ij,ij}^{(1)}$. Therefore, we have
\begin{align}
\langle I_{ij,i(12)}^{(1)}({\vec b})\rangle_{\rm av.}=& \left(\frac{p_1\cdot p_2}{2 m^4}\right)^\lambda\pi^{1-\epsilon} \left(\frac{b^2}4\right)^{2\epsilon-\lambda}\frac{\Gamma(-2\epsilon+\lambda)}{2\Gamma(1+\epsilon-\lambda)} \int_0^{\infty} \,dw\, \frac{w^{-1+\lambda}}{(1+w)}\RRt_{12}(w)\;.
\end{align}
The expression for $\RRt_{12}(w)$ can be obtained taking the massless limit of Eqs.~(\ref{eq:Ry-2})--(\ref{eq:Ry1}), which leads to the simplified expressions
\begin{align}
&\RRt_{12}^{(-2)} = -\frac{1}{2} \\
&\RRt_{12}^{(-1)} = 0 \\
&\RRt_{12}^{(0)} = \frac{5}{4}\zeta_2 \\
&\RRt_{12}^{(1)} = \frac{7}{6}\zeta_3  \;,
\end{align}
and therefore straightforwardly
\begin{equation}
\int_0^{\infty}\frac{dw\,w^{-1+\lambda}}{(1+w)}\RRt_{12}(w)
=
\frac{1}{\lambda}\left[
-\frac{1}{2 \epsilon^2} + \frac{5}{4}\zeta_2 + \epsilon \frac{7}{6}\zeta_3 + {\cal O}(\epsilon^2)
\right] + {\cal O}(\lambda)\;.
\end{equation}
By comparing with Eq.~(\ref{eq:risI1ijij}) we see that the $\lambda\to 0$ singular terms cancel out as expected.
By using Eq.~(\ref{eq: Iij1L_2}) we can write the final result for $I_{ij}^{(1)}({\vec b})$ as
\begingroup
\allowdisplaybreaks
\begin{align}
\nonumber\braket{I_{ij}^{(1)}({\vec b})}_{\rm av.}=&\pi ^{1-\epsilon }\left(\frac{b^2}{4}\right)^{2 \epsilon }
\frac{\Gamma (-2 \epsilon )}{2\Gamma (1+\epsilon)}
\Bigg\{
\frac1{\epsilon ^2}\left(1-2\ln \left(\frac{2 \pipj}{\sqrt{s} m}\right)\right)
\\\nonumber&
+\frac{1}{\epsilon }\left(-1-\frac{\pi ^2}{6}+2 \ln (B+1)-\ln ^2(B)
-2 \text{Li}_2\left(-\frac{1}{B}\right)\right)
\\\nonumber&
+\Bigg(\frac{5}{6} \pi ^2 \ln \left(\frac{2 \pipj}{\sqrt{s} m}\right)-\frac{13 \pi ^2}{12}-6 \text{Li}_2(-B)+2 \text{Li}_3\left(-\frac{1}{B}\right)+2 \text{Li}_3\left(\frac{1}{B+1}\right)
\\\nonumber&
-\frac{1}{3}
   \ln ^3(B)-\frac{1}{3} \ln ^3(B+1)-\ln (B+1) \ln ^2(B)-2 \ln ^2(B)+\ln ^2(B+1) \ln (B)
   \\\nonumber&
   +\ln ^2(B+1)-\frac{1}{3} \pi ^2 \ln (B)-2 \ln (B+1)-3 \zeta
   _3
   -2 \text{Li}_2\left(-\frac{1}{B}\right) (\ln (B+1)+2)
   \Bigg)
   \\\nonumber&
   +\epsilon \Bigg(
   \frac{14}{3} \zeta_3
   \ln \left(\frac{2 \pipj}{\sqrt{s} m}\right)+
   2 \zeta_3 \ln (B)-\text{Li}_2^2\left(-\frac{1}{B}\right)+6 \text{Li}_2(-B)-2 \text{Li}_3\left(-\frac{1}{B}\right)
   \\\nonumber&   
   -6 \text{Li}_4\left(-\frac{1}{B}\right)    +\text{Li}_2\left(-\frac{1}{B}\right) 
   \left(-\ln ^2(B)-\ln ^2(B+1)+2 \ln (B+1)+\frac{\pi^2}2+6\right)
      \\\nonumber&
      +2
   \text{Li}_4\left(\frac{1}{B+1}\right)+2 S_{2,2}\left(-\frac{1}{B}\right)
   -2
   \text{Li}_3\left(-\frac{1}{B}\right) \ln (B)
   +2 \text{Li}_3\left(\frac{1}{B+1}\right) \ln (B+1)
   \\\nonumber&
   -\frac{1}{4} \ln
   ^4(B)-\frac{1}{4} \ln ^4(B+1)
      +\frac{1}{3}\ln ^3(B)
   +\frac{2}{3} \ln ^3(B+1) \ln (B)+\frac{1}{12} \pi ^2
   \ln ^2(B)
   \\\nonumber&
   -\frac{1}{2} \ln ^2(B+1) \ln ^2(B)
   +\ln (B+1) \ln ^2(B)   +3 \ln ^2(B)-2 \ln ^2(B+1)
   \\&
+\frac{1}{3} \pi ^2 \ln (B)
   -\frac{1}{2} \pi ^2 \ln (B+1)-\frac{13 \zeta_3}{3}-\frac{29 \pi ^4}{360}
   +\frac{\pi ^2}{12}+4
   \Bigg)
   + \mathcal{O}\left(\epsilon ^2\right)
\Bigg\}
\;.
\end{align}
\endgroup

\subsubsection{Massive-massive contribution: \texorpdfstring{$I_{34}^{(1)}$}{I341L}}
\label{sec: I341l}
Let us now consider the purely massive contribution, i.e. $I_{34}^{(1)}({\vec b})$ in Eq.~(\ref{eq: I341L}), which can also be written as
\begin{align}
\label{eq:I341L}
I_{34}^{(1)}=& \int d^Dk\,\delta_+(k^2) \left( \frac {2 (p_3\cdot p_4)}{(p_3\cdot k)(p_4\cdot k)} - \frac {m^2}{(p_3\cdot k)^2} - \frac {m^2}{(p_4\cdot k)^2}\right) \,\left(\frac{(p_3\cdot p_4)}{2(p_3\cdot k)(p_4\cdot k)}\right)^\epsilon\, \RRt_{34} e^{i \vec b\cdot\vec k_T}  \;,
\end{align}
where the functions $\RRt_{34}$ have been presented for the first time in Ref.~\cite{Bierenbaum:2011gg}, while in Ref.~\cite{Czakon:2018iev} a simplified expression has been proposed.
The coefficients $\RRt_{34}^{(n)}$ read
\begingroup
\allowdisplaybreaks
\begin{align}
  \label{eq:r34-2}
\RRt_{34}^{(-2)} =&1\;,\\[0.2cm]
\RRt_{34}^{(-1)} =& \, \ln(v_+) - \frac{v_-}{v} \Big(\lni{} + \lnj{}\Big) \; ,
\\[0.2cm]
\label{eq:r340}
\RRt_{34}^{(0)} =& \,\frac{1}{2} \ln^2(v_+)+ \frac{1}{v} \Bigg[ \frac{1}{(d_3+d_4)} \Big( (\alpha_3 v_+ - \alpha_4
v_-) \lni2 + \big( \alpha_4 v_+ - \alpha_3 v_- \big) \lnj2 \Big) \nonumber\\&
+ \Big( \lni{} + \lnj{} \Big) \big( v_+ \ln(v_+) - \ln(v) \big) - \lix2
\Bigg] + \zeta_2 \Big( \frac{7}{v} - \frac{19}{2}
\Big)\; ,\\[0.2cm]
\label{eq:r341}
\RRt_{34}^{(1)} =& \, \frac{1}{d_3+d_4} \Bigg\{ \big (1 - (d_3+d_4) \big) \Bigg[
\lnip{} \lni2 + \lnjp{} \lnj2\nonumber \\&
+  2 \Big( \lni{} \lii2 + \lnj{} \lij2 \Big) - \lix2 \Big( \lni{} +
\lnj{} \Big) \nonumber\\&
+  2 \Big( \lix3 - \lii3 - \lij3 + \zeta_3 \Big) \Bigg] - 7 \zeta_2
\Big( \lni{} + \lnj{} \Big) \nonumber\\&
+ \frac{1}{v} \Bigg[ \Big( \big(\alpha_4 v_+ - \alpha_3 v_- \big) \lni2 + \big(
\alpha_3 v_+ - \alpha_4 v_- \big) \lnj2 \Big) \ln(v_+) 
\nonumber\\&
+ \big( \alpha_3 - \alpha_4 \big)
\Big( \lni2 - \lnj2 \Big) \ln(v)
\nonumber\\&
- \Big( d_3 \lni{} + d_4 \lnj{} \Big)
\big( \lix2 - 7 \zeta_2 \big) \Bigg] \Bigg\} \nonumber\\&
+ \frac{1}{v} \Bigg\{ \Bigg[ \ln(v_+) \Big( \frac{3 + v}{4} \ln(v_+) - \ln(v)
\Big) - \frac{9 v_-}{2} \zeta_2 \Bigg] \Big( \lni{} + \lnj{} \Big)  \nonumber\\&
-\frac{v_-}{6} \Big( \lni3 + \lnj3 \Big) + 2 \lixp3 + \lix3
\nonumber\\&
- \Bigg[ \lix2 +
\zeta_2 \Big( 5 + \frac{19}{2} v \Big) \Bigg] \ln(v_+)
+ 12 \zeta_2 \ln(v) \Bigg\} + \frac{1}{6} \ln^3(v_+) - \Big( \frac{7}{3} +
\frac{1}{v} \Big) \zeta_3 \; .
\end{align}
\endgroup
In Eqs.~(\ref{eq:r34-2})--(\ref{eq:r341}) we used the same notation of Refs.~\cite{Bierenbaum:2011gg,Czakon:2018iev}, introducing the variables
\begin{align}
&\alpha_3 = \frac{m^2(p_4\cdot k)}{(p_3\cdot k) (p_3\cdot p_4) }\;,\\
&\alpha_4 = \frac{m^2(p_3\cdot k)}{ (p_4\cdot k) (p_3\cdot p_4) }\;,\\
&v_\pm=\frac{1\pm v}{2}\;,\\
& d_3=1-2\alpha_3\;,\\
& d_4=1-2\alpha_4\;,
\end{align}
with $v$ defined as in Eq.~(\ref{eq:var_v}).

We first discuss the contributions of $\RRt^{(-2)}_{34}$ and $\RRt^{(-1)}_{34}$. Both these coefficients are independent of the gluon momentum $k$ (note that $\alpha_3\alpha_4=m^4/(p_3\cdot p_4)^2$),
and, therefore, the corresponding integrals can be evaluated with the same method.
We start from the generalised Feynman parametrisation
\begin{equation}\label{eq:genfeyn1L}
\frac 1{A^m B^n}=\frac{\Gamma(m+n)}{\Gamma(m)\Gamma(n)}\,\int_0^1dx\, \frac{x^{m-1}(1-x)^{n-1}}{(x A+(1-x) B)^{m+n}}\;,
\end{equation}
to write the denominators in terms of a single scalar product.

For the term proportional to $(p_3\cdot p_4)/(p_3\cdot k \;p_4\cdot k)$ in $I_{34}^{(1)}$, dropping overall constant terms, the relevant integral is
\begin{align}
  \label{eq:e34}
  \int d^Dk\,\delta_+(k^2)\frac{e^{i \vec b\cdot \vec k_T}}{\ptk^{1+\epsilon}\pqk^{1+\epsilon}}=\frac{\Gamma(2+2\epsilon)}{\Gamma^2(1+\epsilon)}\int_0^1 dx\! &\int d^Dk\,
  \frac{\delta_+(k^2)(1-x)^\epsilon x^\epsilon\,e^{i \vec b\cdot \vec k_T}}{((1-x)\pqk+x\ptk)^{2+2\epsilon}}\;,
\end{align}
while when considering the term proportional to $m^2/\pjq^2$, with $j=3,4$ we need to evaluate
\begin{equation}
  \label{eq:ejj}
  \int \,d^Dk\,\delta_+(k^2)\frac{e^{i \vec b\cdot \vec k_T} }{\pjk^{2+\epsilon}\pik^{\epsilon}}=\frac{\Gamma(2+2\epsilon)}{\Gamma(2+\epsilon)\Gamma(\epsilon)}\int_0^1\,dx\!\int d^Dk\,\frac{\delta_+(k^2)(1-x)^{-1+\epsilon} x^{1+\epsilon}\,e^{i \vec b\cdot \vec k_T}}{((1-x)\pqk+x\ptk)^{2+2\epsilon}}\;.
\end{equation}
We see that both Eq.~(\ref{eq:e34}) and Eq.~(\ref{eq:ejj}) depend on the same integral
\begin{equation}
  I_k^{(1)}(x)=\int\,d^Dk\,\delta_+(k^2)\frac{e^{i \vec b\cdot \vec k_T}}{(p(x)\cdot k)^{2+2\epsilon}}\;,
\end{equation}
with
\begin{equation}
p^\mu(x)=x\, p_3^\mu+(1-x)\,p_4^\mu\;.
\end{equation}
This integral can be evaluated with the techniques used in Sec.~\ref{sub:nlo} and we find
\begin{equation}
\label{eq:Ik1L}
\langle I_k^{(1)}(x)\rangle_{\rm av.}=- \frac{4^{-\epsilon} b^{4\epsilon}\pi^{2-\epsilon}}{\sin(\epsilon \pi)}\frac{\Gamma(-2\epsilon)}{\Gamma(-\epsilon)\Gamma(2+2\epsilon)} (p^2(x))^{-1-\epsilon}{}_2F_1\left(-2\epsilon,1+\epsilon,1-\epsilon;-\frac{p_T^2(x)}{p^2(x)}\right)\;.
\end{equation}
We are now left with the integration over the Feynman parameter $x$. It is convenient to expand in $\epsilon$ the hypergeometric function
\begin{align}
  {}_2F_1\left(-2\epsilon,1+\epsilon,1-\epsilon;-X\right)=&1+2\ln(1+X)\,\epsilon-4\liD(-X)\,\epsilon^2
  \nonumber\\&
  +\frac 43\Bigg(\ln^3(1+X)+3\ln(1+X)\liD(-X)-9\liT(-x)
  \nonumber\\&
  -6\liT\left(\frac X{1+X}\right)\Bigg)\,\epsilon^3+{\cal O}(\epsilon^4)\;.
\end{align}
By substituting Eq.~(\ref{eq:Ik1L}) in Eqs.~(\ref{eq:e34}), (\ref{eq:ejj}) we obtain a sum of integrals
that in most cases can be computed in terms of multiple polylogarithms. The remaining finite integrals are computed numerically.

We now focus on the contribution of $\RRt_{34}^{(0)}$.
We can split $\RRt_{34}^{(0)}$ in a part independent on $k$, $\RRt_{34;\text{const}}^{(0)}$, and one with an explicit $k$ dependence, $\RRt_{34; k}^{(0)}$
\begin{equation}
  \RRt_{34}^{(0)}=\RRt_{34;\text{const}}^{(0)}+\RRt_{34; k}^{(0)}\, .
\end{equation}
We define
\begin{align}
\RRt_{34;\text{const}}^{(0)}=&\frac 1v \Bigg[\Big( \lni{} + \lnj{} \Big) \big( v_+ \ln(v_+) - \ln(v) \big) - \lix2
\Bigg] + \frac{1}{2} \ln^2(v_+)
\nonumber\\&
+ \zeta_2 \Big( \frac{7}{v} - \frac{19}{2}
\Big) \;.
\end{align}
and
\begin{align}
    \RRt_{34; k}^{(0)}=\frac{1}{(d_3+d_4)v} \Bigg[ (\alpha_3 v_+ - \alpha_4
v_-) \lni2 + \big( \alpha_4 v_+ - \alpha_3 v_- \big) \lnj2 \Bigg]\equiv\frac{1}{d_3+d_4}\rrt_{34}^{(0)}\;.
\end{align}
The contribution of $\RRt_{34;\text{const}}^{(0)}$ can be evaluated as those of $\RRt_{34}^{(-2)}$ and $\RRt_{34}^{(-1)}$.
The singular part of the contribution of $\RRt_{34; k}^{(0)}$ can be computed by using the following identity
\begin{equation}
  \label{eq:theta_trick}
  \int\,d^Dk\,(k^2)^{-1-\epsilon} f(k)\,e^{i\vec b\cdot\vec k_T}=\int\,d^Dk\,(k^2)^{-1-\epsilon} f(k)\theta(\mu-k_T)
  +\mathcal{O}(\epsilon^0)\;,
\end{equation}
$\mu$ being an arbitrary mass scale and $f(k)$ an arbitrary function of the momentum $k$.

We consider the integral of the $k$-dependent part of $\RRt_{34}^{(0)}$
\begin{align}
    \label{eq:before_trick_applied}
    I_{34;k}^{(1,0)}({\vec b})=\int d^Dk & \,\delta_+(k^2)\frac{1}{d_3+d_4} \left(\frac {2(p_3\cdot p_4)}{(p_3\cdot k)(p_4\cdot k)} - \frac {m^2}{(p_3\cdot k)^2} - \frac {m^2}{(p_4\cdot k)^2}\right)\\\nonumber&
    \times\left(\frac{(p_3\cdot p_4)}{2(p_3\cdot k)(p_4\cdot k)}\right)^\epsilon \, \rrt_{34}^{(0)} e^{i \vec b\cdot\vec k_T} \;,
\end{align}
up to order $1/\epsilon$. The reason to pull out the factor $1/(d_3+d_4)$ is because it allows us to use the identity
\begin{equation}
  \label{eq:eee}
  \frac {2(p_3\cdot p_4)}{(p_3\cdot k)(p_4\cdot k)} - \frac {m^2}{(p_3\cdot k)^2} - \frac {m^2}{(p_4\cdot k)^2}=\frac {(p_3\cdot p_4)}{(p_3\cdot k)(p_4\cdot k)}(d_3+d_4)\;,
\end{equation}
which makes the integrand considerably simpler.
If now we extract the pole structure of the integral by using Eq.~(\ref{eq:theta_trick}), we get
\begin{equation}
  I_{34;k}^{(1,0)}({\vec b})=\int\,d^Dk\,\delta_+(k^2)\left(\frac{p_3\cdot p_4}{\ptk \pqk}\right)^{1+\epsilon} \rrt_{34}^{(0)}
  \theta(\mu^2-k_0^2) + {\cal O}(\epsilon^0)\;.
\end{equation}
There is no singularity associated to the angular variables. We can thus safely set $\epsilon=0$ in the angular integral, obtaining
\begin{align}
  I_{34;k}^{(1,0)}({\vec b})=2\pi&\int_0^\infty\,\frac{dk_0}{k_0^{1+4\epsilon}} \theta(\mu^2-k_0^2)
  \int_0^1\,dt\int_{-1}^1\,d\cos\theta\,t^{2-2\epsilon} \delta(1-t^2)\frac{1}{1-v\cos\theta}\rrt_{34}^{(0)} + {\cal O}(\epsilon^0) \;,
\end{align}
with $t=|\vec k|/k_0$. Now we can perform the integral over $t$ by using the delta function, while the (otherwise divergent) integration over $k_0$ is regulated by the cutoff we inserted. We find for the pole
\begin{equation}
  \label{eq:trick_applied}
  I_{34;k}^{(1,0)}({\vec b})\big|_{\text{pole}}=-\frac\pi{4\epsilon}\int_{-1}^1\,d\cos\theta\,\frac{1}{1-v\cos\theta}\,\rrt_{34}^{(0)}\;.
\end{equation}
The integration of the pole part of \mbox{Eq.~(\ref{eq:r340})} is thus finally reduced to a one-fold integral that can be computed with standard methods.
In order to write $\rrt_{34}^{(0)}$ in terms of $v$, $\cos\theta$ the following relations are useful
\begin{align}
  &\alpha_3=\frac{1-v\cos\theta}{2}\;,
  \\&
  \alpha_4=\frac{1-v^2}2\,\frac{1}{1-v\cos\theta}\;.
\end{align}
The result for the pole part of this contribution reads
\begin{align}
    \langle I_{34;k}^{(1,0)}({\vec b})\big|_{\text{pole}} \rangle_{\rm av.} = \pi ^{1-\epsilon }\left(\frac{b^2}{4}\right)^{2\epsilon}\frac{\Gamma (-2 \epsilon )}{\Gamma (\epsilon +1)}\left(2-\frac{\left(1-\beta ^2\right)^2 }{2 \beta ^2}\ln
   ^2\left(\frac{1-\beta }{\beta +1}\right)\right)\;.
\end{align}
The finite part in $\epsilon$ of the contribution of $\RRt_{34}^{(0)}$ can be integrated numerically.

The last contribution to be computed is that from $\RRt_{34}^{(1)}$. Since it comes with an overall $\epsilon$ factor
we can directly apply Eq.~(\ref{eq:theta_trick}) to evaluate it. The analytic result is too lengthy to be reported.

We conclude this subsection discussing the contribution of the one-loop heavy-quark vacuum polarization. Such term can be inserted in the radiated soft-gluon line, thus leading to an additional virtual contribution to the one-loop soft-gluon current. Then
such contribution has to be consistently taken into account through
the renormalization procedure, which amounts to the wave function 
renormalization of the soft-gluon line and the $\msbar$
renormalization of $\as$ with $n_f+1$ quark flavours (the $n_f$ massless quarks and the heavy quark $Q$). Finally, we can apply
the decoupling relation of the heavy quark \cite{Baernreuther:2013caa}
and introduce the running coupling $\as^{(n_f)}(\mu_R^2)$
that we use throughout this paper (see the comment at the beginning of Sect.~\ref{sec:soft}). To the purpose of computing the soft contributions at small $q_T$, the final result of this entire procedure is equivalent to avoiding the introduction of the heavy-quark vacuum polarization and to directly renormalizing the QCD coupling with $n_f$ light-quark flavours as in 
Eq.~(\ref{eq:renorm}).

\subsection{Light-quark pair production}
\label{sub:qq}

We start the analysis of the double real contribution by focusing on the process in which a massless soft quark-antiquark pair is radiated
\begin{equation}
c(p_1) {\bar c}(p_2)\quad\rightarrow\quad Q(p_3) \bar Q (p_4)\,q(k_1)\bar q(k_2)\;.
\end{equation}
The corresponding factorisation formula for the squared matrix element is \cite{Catani:1999ss}
\begin{align}
\label{eq:qqfact}
&\left|{\cal M}^{(0)}_{c{\bar c}\to Q{\bar Q}q{\bar q}}\right|^2\sim (g_0\mu_0^\epsilon)^4\langle {\cal M}_{c{\bar c}\to Q{\bar Q}}^{(0)}|{\boldsymbol I}^{(0)}_{q{\bar q}}(k_1,k_2) |{\cal M}_{c{\bar c}\to Q{\bar Q}}^{(0)}\rangle
\end{align}
where the singular contributions are controlled by the soft factor
\begin{equation}
\label{eq:taskqq_2}
{\boldsymbol I}_{q\bar q}^{(0)}(k_1,k_2)=\left[\mathbf{J}_{g,\mu}^{(0)}(k_1+k_2)\right]^\dagger \Pi^{\mu\nu}(k_1,k_2)\,\mathbf{J}^{(0)}_{g,\nu}(k_1+k_2)+...\;.
\end{equation}
In Eq.~(\ref{eq:taskqq_2}) $\mathbf{J}^{(0)}_{g,\mu}$ is the tree-level soft current in Eq.~(\ref{eq:nlo current})
and we have defined the tensor $\Pi^{\mu\nu}$ as:
\begin{equation}
\Pi^{\mu\nu}(k_1,k_2)=\frac{T_R}{(k_1\cdot k_2)^2}\left(-g^{\mu\nu} k_1\cdot k_2 +k_1^\mu k_2^\nu+k_1^\nu k_2^\mu\right)\;.
\end{equation}
The dots in Eq.~(\ref{eq:taskqq_2}) stand for gauge dependent contributions that are proportional to the total charge of the hard partons, thereby vanishing when evaluated on the
$c{\bar c}\to Q{\bar Q}$ matrix element.
Our task is now to integrate Eq.~(\ref{eq:taskqq_2}) over the phase space of the $q{\bar q}$ pair after subtracting the initial-state contribution, i.e., to evaluate
the integral $\mathbf{I}^{(0)}_{q{\bar q}}({\vec b})$ in Eq.~(\ref{eq:Ibomqq}).

To perform this calculation, we first integrate over the light-quark momenta $k_1$ and $k_2$ while keeping their total momentum $k=k_1+k_2$ fixed.
This procedure will leave us with expressions similar to the ones for the NLO-like contribution already described in Sect.~\ref{sub:nlo} and will be useful in order to organise the final integration over $k$ in a similar way.

To proceed in this direction, we rewrite the integration of the soft factor in Eq.~(\ref{eq:Ibomqq}) in the following way:
\begin{align}
\label{eq: Iqq_nosub}
\int \frac{d^Dk_1}{(2\pi)^{D-1}}\frac{d^Dk_2}{(2\pi)^{D-1}}&\delta_+(k_1^2)\delta_+(k_2^2){\boldsymbol I}^{(0)}_{q \bar q}\left( k_1, k_2\right)e^{i \vec{b}\cdot\left(\vec k_{T1}+\vec k_{T2}\right)}
=\nonumber\\
&=\int \frac{d^D k}{(2\pi)^{D-1}}\mathbf{J}^{(0)}_{g,\mu}(k) \mathbf{J}^{(0)}_{g,\nu}(k)\, F^{\mu\nu}(k)\, e^{i \vec b\cdot \vec k_T}\;,
\end{align}
obtained by inserting the  identity
\begin{equation}
1=\int d^D k\,\delta^{(D)} (k-k_1-k_2)\;,
\end{equation}
and by isolating the integral on the soft-quark momenta in the tensor $F^{\mu\nu}(k) $, defined as:
\begin{equation}
\label{eq:F1}
F^{\mu\nu}(k)=\frac 1{(2\pi)^{D-1}}\int d^Dk_1 \int d^Dk_2\,\Pi^{\mu\nu}(k_1,k_2) \delta_+(k_1)\delta_+(k_2) \delta^{(D)}(k-k_1-k_2)\;.
\end{equation}

We now continue with the computation of the tensor $F^{\mu\nu}$. Since $F^{\mu\nu}$ is a symmetric tensor fulfilling $k_\mu F^{\mu\nu}=k_\nu F^{\mu\nu}=0$ it must take the form
\begin{equation}\label{eq:F2}
F^{\mu\nu}(k)=C\left(-g^{\mu\nu}+\frac{k^\mu k^\nu}{k^2}\right)\;.
\end{equation}
The normalisation factor $C$ can be fixed by evaluating the quantity $g_{\mu\nu}F^{\mu\nu}$  using Eq.~(\ref{eq:F1}) and Eq.~(\ref{eq:F2}) and comparing the results.
From Eq.~(\ref{eq:F2}) we immediately obtain
\begin{equation}\label{eq:Fmunu2}
g_{\mu\nu}F^{\mu\nu}=-C(3-2\epsilon)\;,
\end{equation}
while from Eq.~(\ref{eq:F1})
\begin{equation}\label{eq:Fmunu1}
g_{\mu\nu}F^{\mu\nu}=-T_R \,\frac{2-2\epsilon}{(k^2)^{1+\epsilon} \Gamma(\frac{3}{2}-\epsilon)16^{1-\epsilon} \pi^{\frac{3}{2}-\epsilon}}\;.
\end{equation}
We can therefore write
\begin{equation}
C=\frac{F(\epsilon)}{(k^2)^{1+\epsilon}}\;,
\end{equation}
with
\begin{equation}
F(\epsilon)=\frac{T_R(1-\epsilon)}{\Gamma\left(\frac 52-\epsilon\right)16^{1-\epsilon}\pi^{\frac 32 -\epsilon}}\;.
\end{equation}
With the explicit expression for $F^{\mu\nu}$ just obtained, the right-hand side of Eq.~(\ref{eq: Iqq_nosub}) reads:
\begin{equation}
\label{eq:qintegrated}
\int \frac{d^D k}{(2\pi)^{D-1}}\mathbf{J}^{(0)}_{g,\mu}(k)\mathbf{J}^{(0)}_{g,\nu}(k)F^{\mu\nu}(k) \, e^{i \vec b\cdot \vec k_T}
=-\int \frac{d^Dk}{(2\pi)^{D-1}}\frac {F(\epsilon)}{(k^2)^{1+\epsilon}} \sum_{i,j=1}^4 \textbf{T}_i\cdot\textbf{T}_j
\frac{p_i\cdot p_j}{(p_i\cdot k)(p_j\cdot k)}\, e^{i \vec b\cdot \vec k_T}
\;,
\end{equation}
where the term in $F^{\mu\nu}$ proportional to $k^\mu k^\nu$ gives no contribution because of colour conservation.

We observe that Eq.~(\ref{eq:qintegrated}) has a similar structure to Eq.~(\ref{eq:I0g}), the corresponding NLO integral for single soft-gluon emission at tree-level, after the substitution
\begin{equation}\label{eq:substitution}
\delta_+(k^2)\quad\rightarrow\quad \frac{F(\epsilon)}{(k^2)^{1+\epsilon}}\;,
\end{equation}
which removes the on-shell constraint for the gluon.
We can thus apply for the computation a similar strategy as the one already employed in Sect.~\ref{sub:nlo}, when dealing with the NLO-like contribution.

Before performing the final integration over $k$ of the expression in Eq.~(\ref{eq:qintegrated}), we need to subtract the initial-state contribution from the soft current.
We therefore write the integral $\mathbf{I}_{q\bar q}^{(0)}({\vec b})$ in Eq.~(\ref{eq:Ibomqq}) as 
\begin{align}
\label{eq:integrals1}
\mathbf{I}^{(0)}_{q\bar q}({\vec b})=&-F(\epsilon)\int \frac{d^D k}{(2\pi)^{D-1}}\frac{1}{(k^2)^{1+\epsilon}} \Bigg\{\sum_{j=3,4}\left[\frac{m^2}{(p_j\cdot k)^2}\mathbf{T}_j^2+2\sum_{i=1,2}\left(\frac{p_i\cdot p_j}{p_j\cdot k}-\frac{p_1\cdot p_2}{(p_1+p_2)k}\right)\frac{\mathbf{T}_i\cdot\mathbf{T}_j}{p_i\cdot k}\right]\nonumber\\
&+\frac {2p_3\cdot p_4}{(p_3\cdot k) (p_4\cdot k)}\mathbf{T}_3\cdot\mathbf{T}_4 \Bigg\}\,e^{i \vec b\cdot \vec k_T}\;.
\end{align}
We can split Eq.~(\ref{eq:integrals1}) in different integrals according to the different colour factors
\begin{align}
\label{eq:ijjqspace}
&I^{q\bar q}_{jj}({\vec b}) =
\int \frac {d^D k}{(k^2)^{1+\epsilon}} \, \frac{m^2}{(p_j \cdot k)^2}\,e^{i \vec b\cdot \vec k_T} \;, \\
\label{eq:iijqspace}
&I^{q\bar q}_{ij}(\vec b) = \int\frac {d^D k}{(k^2)^{1+\epsilon}}\,
\frac{1}{p_i \cdot k}
\left(
\frac{p_i \cdot p_j}{p_j \cdot k} - \frac{p_1 \cdot p_2}{(p_1+p_2) \cdot k}
\right)\,e^{i \vec b\cdot \vec k_T}\;, \\
\label{eq:i34qspace}
&I^{q\bar q}_{34}(\vec b) = \int \frac{d^D k}{(k^2)^{1+\epsilon}}  \,
\frac{p_3\cdot p_4}{(p_3 \cdot k) (p_4 \cdot k)}\, e^{i \vec b\cdot \vec k_T}
\;.
\end{align}
In terms of these integrals, Eq.~(\ref{eq:integrals1}) reads
\begin{equation}
    \mathbf{I}^{(0)}_{q\bar q}({\vec b})=-\frac{F(\epsilon)}{(2\pi)^{D-1}}\left\{\sum_{j=3,4} \left[I^{q\bar q}_{jj}({\vec b})\,\mathbf{T}_j^2+2\sum_{i=1,2} I^{q\bar q}_{ij}({\vec b})\,\mathbf{T}_i\cdot\mathbf{T}_j\right]+2I^{q\bar q}_{34}({\vec b})\,\mathbf{T}_3\cdot \mathbf{T}_4\right\}\;.
\end{equation}
The integrals in Eqs.~(\ref{eq:ijjqspace})--(\ref{eq:i34qspace}) can be evaluated with a similar strategy as to the one used for the integrals in Eqs.~(\ref{eq:ijjnlo})--(\ref{eq:i34nlo}). The azimuthally averaged results are
\begin{align}
\label{eq:iqqjj_result}
\nonumber \langle I^{q\bar q}_{jj}(\vec b) \rangle_{\rm av.} =&\pi ^{1-\epsilon } \Gamma (1-\epsilon ) \Gamma (-2 \epsilon )\left(\frac{b^2}{4}\right)^{2\epsilon} \left[-\frac{1}{\epsilon}\, {}_2 F_1\left(1,-2\epsilon; 1-\epsilon; -B\right)\right]
\\
=& \pi ^{1-\epsilon } \Gamma (1-\epsilon ) \Gamma (-2 \epsilon )\left(\frac{b^2}{4}\right)^{2\epsilon} \Bigg\{-\frac{1}{\epsilon} - 2\ln\left(1+B\right)
 \nonumber
 \\
&+ \epsilon\left[2 \, \text{Li}_2 \left(-B\right) - \ln^2\left(1+B\right)\right] + {\cal O}(\epsilon^2) \Bigg\}\;,\\
\label{eq:iqqij_result}
\nonumber\langle I^{q\bar q}_{ij}(\vec b)\rangle_{\rm av.} =&\frac{1}{2} \pi ^{1-\epsilon } \Gamma (1-\epsilon ) \Gamma (-2 \epsilon )
\left(\frac{b^2}{4}\right)^{2\epsilon}\frac{\Gamma(\tfrac{\lambda}{2}-2\epsilon)\Gamma(\tfrac{\lambda}{2})}{\Gamma(1-2\epsilon)}
\\
&\times2\left[\left(\frac{p_i \cdot p_j}{m^2}\right)^\lambda{}_2 F_1\left(\frac{\lambda}{2} , \frac{\lambda}{2}-2\epsilon ; 1\epsilon; -B\right)-\left(\frac{p_1 \cdot p_2}{\sqrt{s}}\right)^\lambda{}_2 F_1\left(\frac{\lambda}{2} , \frac{\lambda}{2}-2\epsilon ; 1-\epsilon; 0 \right)\right]\nonumber
\\
=& \frac{1}{2} \pi ^{1-\epsilon } \Gamma (1-\epsilon ) \Gamma (-2 \epsilon )\left(\frac{b^2}{4}\right)^{2\epsilon}\Bigg\{-\frac{2}{\epsilon} \ln \left( \frac{2\,p_i\cdot p_j}{m\,\sqrt{s}} \right)+2\text{Li}_2 \left(-B\right)\nonumber   \\
&+   \frac\epsilon3\left[\ln^3\left(1+B\right)+6\ln\left(1+B\right)\text{Li}_2\left(-B\right)-6\,\text{Li}_3\left(\frac {B}{B+1}\right)\right]+ {\cal O}(\epsilon^2)\Bigg\}\;,\\
\label{eq:iqq34_result}
\langle I^{q\bar q}_{34}(\vec b)\rangle_{\rm av.}=&\pi ^{1-\epsilon } \Gamma (1-\epsilon ) \Gamma (-2 \epsilon )\left(\frac{b^2}{4}\right)^{2\epsilon} \frac{1+\beta^2}{2\beta}
\left\{-\frac{1}{\epsilon} L_0 - 2 L_1+ \epsilon ( 2 P_2 - L_2 )+ {\cal O}(\epsilon^2) \right\}\;.
\end{align}
The functions $L_n$ and $P_n$ have been defined in Eq.~(\ref{eq:ln}) and Eq.~(\ref{eq:pn}), respectively, while their explicit expressions are reported in Eqs.~(\ref{eq:L0_result})--(\ref{eq:P2_result}).
In the present case we also need the function $L_2(\beta,\theta)$, which reads
\begin{align}
\label{eq:L2_result}
    \nonumber L_2(&\beta,\theta)=
    2 (G(1,-1,-\sec (\theta ),\beta )+G(1,-1,\sec (\theta ),\beta )+G(1,1,-\sec (\theta ),\beta )
   \\ & +G(1,1,\sec(\theta ),\beta )
   +G(1,-\sec (\theta ),-1,\beta )+G(1,-\sec (\theta ),1,\beta ) -G(1,1,1,\beta )
   \nonumber\\ &
   -G(1,-\sec (\theta ),-\sec
   (\theta ),\beta )
   +G(1,\sec (\theta ),-1,\beta )+G(1,\sec
   (\theta ),1,\beta )-G(1,1,-1,\beta )
   \nonumber\\ &
   -G(1,\sec (\theta ),-\sec (\theta ),\beta )
   -G(1,\sec (\theta ),\sec (\theta ),\beta)
   -G(1,-\sec (\theta ),\sec (\theta ),\beta )
   \nonumber\\ &
   -G(1,-1,-1,\beta )-G(1,-1,1,\beta ))
    \;.
\end{align}
Note that in Eq.~(\ref{eq:iqqij_result}), the expression for $\langle I^{q\bar q}_{ij}(\vec b)\rangle_{\rm av.}$ before the $\epsilon$-expansion
depends on the regularisation parameter $\lambda$.
As in Sec.~\ref{sub:nlo} the integration in Eq.~(\ref{eq:iijqspace}) needs to be carried out separately for the two terms, by using
the regulator factor in Eq.~(\ref{eq: lambda regulator}). We can then perform the limit $\lambda\to0$ in the expanded result.

\subsection{Double gluon emission}
\label{sub:gg}

\def\f{{\mathsf f}}
\def\g{{\mathsf g}}

We finally consider the contribution due to the emission of a soft-gluon pair, i.e. we consider the process
	\begin{equation}
	c(p_1) {\bar c}(p_2)\quad\rightarrow\quad Q(p_3) \bar Q (p_4) g(k_1) g(k_2)\;.
	\end{equation}
In the limit in which the two gluons become soft the singular behaviour is controlled by the double-soft current $\mathbf{J}_{gg}^{(0)\mu\nu}(k_1,k_2)$~\cite{Catani:1999ss, Czakon:2011ve}. The general expression of the squared soft current reads
\begin{equation}
\label{eq:gg current squared}
	 \left[J_{gg,\mu\nu}^{(0)a_1a_2}(k_1,k_2)\right]^\dagger d^{\sigma\mu}(k_1)d^{\rho\nu}(k_2)J_{gg,\sigma\rho}^{(0)a_1 a_2}(k_1,k_2) = \frac{1}{2}
	\left\{ {\mathbf{J}_g^{(0)}}^2(k_1) , {\mathbf{J}_g^{(0)}}^2(k_2) \right\}
	+\mathbf{W}_{gg}^{(0)}(k_1,k_2)+...\; ,
\end{equation}
where the purely non-abelian two-parton correlations are controlled by the function $\mathbf{W}_{gg}^{(0)}(k_1,k_2)$,
which is defined as
\begin{equation}
\label{eq:ggW}
\mathbf{W}^{(0)}_{gg}(k_1,k_2)=-C_A \sum_{i,j=1}^n \mathbf{T}_i \cdot \mathbf{T}_j \; {\cal
		S}_{ij}(k_1,k_2)\, .
\end{equation}
The dots in Eq.~(\ref{eq:gg current squared}) stand for gauge-dependent terms proportional to the total colour charge of the hard partons and, thus, give a vanishing contribution when evaluated on the $c{\bar c}\to Q{\bar Q}$ matrix element.
The soft factor can be separated into massless and massive contributions
	\begin{equation}
	{\cal S}_{ij}(k_1,k_2) = {\cal S}^{m=0}_{ij}(k_1,k_2) + \left( m_i^2
	\; {\cal S}^{m \neq 0}_{ij}(k_1,k_2) + m_j^2 \; {\cal S}^{m \neq
		0}_{ji}(k_1,k_2) \right) \;,
\end{equation}
where $m_i(m_j)=0$ for $i(j)=1,2$ and $m_i(m_j)=m$ for $i(j)=3,4$.
The massless contribution reads \cite{Catani:1999ss}
	\begin{align}
	\label{eq:sij0}
 \nonumber
	{\cal S}^{m=0}_{ij}(k_1,k_2) =& \frac{(1-\epsilon)}{(k_1 \cdot k_2)^2}
	\frac{p_i \cdot k_1 \; p_j \cdot k_2 + p_i \cdot k_2 \; p_j \cdot
		k_1}{p_i \cdot (k_1+k_2) \; p_j \cdot (k_1+k_2)}
	\\
	& - \frac{(p_i \cdot p_j)^2}{2 \; p_i \cdot k_1 \; p_j \cdot k_2 \; p_i
		\cdot k_2 \; p_j \cdot k_1} \left[ 2 - \frac{p_i \cdot k_1 \; p_j
		\cdot k_2 + p_i \cdot k_2 \; p_j \cdot k_1}{p_i \cdot (k_1+k_2) \;
		p_j \cdot (k_1+k_2)} \right] \nonumber \\
	& + \frac{p_i \cdot p_j}{2 \; k_1 \cdot k_2} \left[ \frac{2}{p_i
		\cdot k_1 \; p_j \cdot k_2} + \frac{2}{p_j \cdot k_1 \; p_i \cdot
		k_2} - \frac{1}{p_i \cdot (k_1+k_2) \; p_j \cdot (k_1+k_2)}
	\right. \nonumber \\
	& \times \left. \left( 4 + \frac{(p_i \cdot k_1 \; p_j \cdot k_2 +
		p_i \cdot k_2 \; p_j \cdot k_1)^2}{\; p_i \cdot k_1 \; p_j \cdot
		k_2 \; p_i \cdot k_2 \; p_j \cdot k_1} \right) \right] \; ,
	\end{align}
	$p_i$, $p_j$ being the momenta of the emitters. 
The massive contribution is \cite{Czakon:2011ve}
	\begin{align}
	\label{eq:sijm}\nonumber
	{\cal S}^{m\neq0}_{ij}(k_1,k_2) =&  - \frac{1}{4  \; k_1 \cdot k_2 \; p_i \cdot
		k_1 \; p_i \cdot k_2} + \frac{p_i \cdot p_j \; p_j
		\cdot (k_1+k_2)}{2 \; p_i \cdot k_1 \; p_j \cdot k_2 \; p_i \cdot k_2 \;
		p_j \cdot k_1 \; p_i \cdot (k_1+k_2)} \\
	& - \frac{1}{2 \; k_1 \cdot k_2 \; p_i \cdot (k_1+k_2) \; p_j \cdot
		(k_1+k_2) } \left( \frac{(p_j \cdot k_1)^2}{p_i \cdot k_1 \; p_j
		\cdot k_2} + \frac{(p_j \cdot k_2)^2}{p_i \cdot k_2 \; p_j \cdot
		k_1} \right) \; . 
	\end{align}

In the right-hand side of Eq.~(\ref{eq:gg current squared}), 
$\mathbf{W}_{gg}^{(0)}$ is the irreducible correlation component of 
double-soft radiation, while the anticommutator term corresponds 
to the independent-emission component. We have to evaluate 
the $b$-space contribution of the squared current in 
Eq.~(\ref{eq:gg current squared}). Going to $b$-space, the 
phase space for double-parton emission factorizes
in terms of single-parton factors 
(see Eq.~(\ref{eq:phasespace})). Therefore the $b$-space integral of the
independent-emission component of Eq.~(\ref{eq:gg current squared})
is fully factorized and it leads to the straightforward exponentiation
of the tree-level single-emission contribution $\mathbf{F}_{{\rm ex},1}$
in Eq~(\ref{eq:Fex1I}). Consequently the $b$-space contribution 
$\mathbf{I}_{gg}^{(0)}(\vec{b})$ of double soft-gluon emission to 
$\mathbf{F}_{{\rm ex},2}$ in Eq.~(\ref{eq:Fex2I}) is entirely due 
to the correlation component $\mathbf{W}_{gg}^{(0)}$
of Eq.~(\ref{eq:gg current squared}). More precisely,
we have to perform the integral in Eq.~(\ref{eq:Ibomgg})
where $\mathbf{W}^{(0)}_{gg}(k_1,k_2)\big|_{\rm sub}$ is defined
from $\mathbf{W}^{(0)}_{gg}\left(k_1, k_2\right)$ in Eq.~(\ref{eq:ggW}) after the proper subtraction of the contribution from initial-state radiation.

Part of the contribution to $\mathbf{I}_{gg}^{(0)}(\vec{b})$
is similar to $\mathbf{I}_{q{\bar q}}^{(0)}(\vec{b})$.
The soft term in Eq.~(\ref{eq:taskqq_2}) involves the factor
\begin{align}
\label{eq:jpj}
\frac{p_i^\mu p_j^\nu\, \Pi_{\mu\nu}\,(k_1,k_2)}{p_i\cdot(k_1+k_2)\,p_j\cdot(k_1+k_2)}  
=\frac{T_R}{\quqd^2}\frac{-\pipj \quqd +\piqu \pjqd + \piqd\pjqu}{p_i\cdot(k_1+k_2)p_j\cdot(k_1+k_2)}\;.
\end{align}
In Eq.~(\ref{eq:sij0}) we have some terms with a similar structure as the ones in Eq.~(\ref{eq:jpj}).
Those are
\begin{equation}
\label{eq:jpj2}
{\cal S}^{m=0}_{ij}(k_1,k_2)\Big|_{12}=\frac{4}{\quqd}\frac{\pipj}{\piq\pjq}
-\frac{(1-\epsilon)}{(k_1 \cdot k_2)^2}
\frac{(p_i \cdot k_1) \; (p_j \cdot k_2) + (p_i \cdot k_2) \; (p_j \cdot
	k_1)}{p_i \cdot (k_1+k_2) \; p_j \cdot (k_1+k_2)}\;.
\end{equation}
Indeed by defining
\begin{equation}
\widetilde \Pi^{\mu\nu}(k_1,k_2)=-\frac {1}{\quqd^2}\left(-4 g^{\mu\nu}\quqd+(1-\epsilon)k_1^\mu k_2^\nu+(1-\epsilon)k_2^\mu k_1^\nu\right)
\end{equation}
we can rewrite Eq.~(\ref{eq:jpj2}) as
\begin{equation}
\label{eq:pimunu2}
{\cal S}^{m=0}_{ij}(k_1,k_2)\Big|_{12}=\frac{p_i^\mu}{p_i\cdot(k_1+k_2)}\, \widetilde\Pi_{\mu\nu}(k_1,k_2)\,\frac{p_j^\nu}{p_j\cdot(k_1+k_2)}\;.
\end{equation}
Now the integration of Eq.~(\ref{eq:pimunu2}) can be performed exactly in the same way as the integration of Eq.~(\ref{eq:taskqq_2}) in the case of the emission of a soft $q\bar q$ pair in Sect.~\ref{sub:qq}, leading to the same results with an overall multiplicative factor. 

By following the strategy of integrating over $k_1$ and $k_2$ at a fixed value of $k=k_1+k_2$, similarly to what was done in Eq.~(\ref{eq:F1}), we isolate the following integral
\begin{align}
	\label{F1}
\widetilde F^{\mu\nu}(k)=&\frac 1{(2\pi)^{D-1}}\int d^Dk_1 \int d^Dk_2\,\widetilde\Pi^{\mu\nu}(k_1,k_2) \delta_+(k^2_1)\delta_+(k^2_2) \delta^{(D)}(k-k_1-k_2)\;.
\end{align}
The structure of $\widetilde F^{\mu\nu}(k)$ must be of the form
\begin{equation}
\label{F2}
\widetilde F^{\mu\nu}(k)=a g^{\mu\nu} + b\frac{k^\mu k^\nu}{k^2}\;.
\end{equation}
The coefficients $a$ and $b$ can be obtained by contracting Eq.~(\ref{F1}) with $g_{\mu\nu}$ and $k_\mu k_\nu$.
We find
\begin{equation}
\label{F3}
\widetilde F^{\mu\nu}=
\frac{1}{(k^2)^{1+\epsilon}} 
\frac{1}{\Gamma\left(\frac 52-\epsilon\right)16^{1-\epsilon}\pi^{\frac 32 -\epsilon}}
  \left(\frac{11-7\epsilon}{2}g^{\mu\nu}+(1-\epsilon)\frac{k^\mu k^\nu}{k^2}\right)\;.
\end{equation}
Because of current conservation, the second term will give no contribution to the integrals. The first term, on the other hand, is exactly the same we obtained in the computation for the soft-quark pair emission, but with an overall multiplicative factor: $-(11-7\epsilon)/(1-\epsilon)$.

Hence, to compute the integral of the contribution in Eq.~(\ref{eq:jpj2}), we can take the result for the $q\bar q$ pair production and perform the formal substitution
\begin{equation}
n_f T_R\longrightarrow - C_A\, \frac{11-7\epsilon}{4(1-\epsilon)}\;,
\end{equation}
where we also included the Bose factor $1/2$ of 
Eq.~(\ref{eq:Ibomgg}),
which is due to the production of two identical particles.

We can now define a new soft factor, in which we subtract the contribution that can be computed as described above
\begin{equation}
\label{eq:S_tilde}  
  \widetilde{\cal S}_{ij}(k_1,k_2) = \widetilde{\cal S}^{m=0}_{ij}(k_1,k_2) + \left(m_i^2\,
  \widetilde{\cal S}^{m \neq 0}_{ij}(k_1,k_2) + m_j^2\,\widetilde{\cal S}^{m \neq
    0}_{ji}(k_1,k_2) \right)\,,
\end{equation}
where
\begin{align}
  \widetilde{\cal S}^{m=0}_{ij}(k_1,k_2)&={\cal S}^{m=0}_{ij}(k_1,k_2)-{\cal S}^{m=0}_{ij}(k_1,k_2)\Big|_{12}\,,\\
   \widetilde{\cal S}^{m\neq 0}_{ij}(k_1,k_2)&={\cal S}^{m\neq 0}_{ij}(k_1,k_2)\, .
\end{align}
We have
\begin{align}
\label{eq:sm0}\nonumber
	\tilde{\cal S}^{m=0}_{ij}(k_1,k_2)=&-\frac{\pipj^2}{2\piq \pjk}\left(\frac 2{\piqu\pjqu}+\frac 1{\piqu\pjqd}\right)
	\\\nonumber&
	-\frac{\pipj}{2 k^2 \piq\pjk}\frac{(\piqu\pjqd-\piqd\pjqu)^2}{\piqu\pjqd\piqd\pjqu}
 	\\&
	+\frac {\pipj}{k^2}\frac{2}{\piqu\pjqd}+(1\leftrightarrow 2)\;,\\
	\label{eq:sm}
		\tilde{\cal S}^{m\neq 0}_{ij}(k_1,k_2)=&\frac{\pipj}{2\piq^2}\left(\frac 1{\piqu\pjqu}+\frac 1{\piqu\pjqd}\right)\nonumber\\
		&-\frac 1{k^2 \piq}\frac1{\piqu}\left(\frac{\pjqu^2}{\pjk\pjqd}-\frac{\piqu^2}{\piq\piqd}\right)+(1\leftrightarrow 2)\;,
\end{align}
where we have introduced $k=k_1+k_2$.
We now need to subtract the contribution from initial-state radiation.
We can use the same technique already used in the previous Sections.
The sum over the colour configurations can be organised as
\begin{align}
\sum_{i,j=1}^4  \tilde{\cal S}_{ij}(k_1,k_2) \mathbf{T}_i \cdot \mathbf{T}_j =
\left[\sum_{i,j=1}^4  \tilde{\cal S}_{ij} \mathbf{T}_i \cdot \mathbf{T}_j-\left(-\tilde{\cal S}_{12}(\mathbf{T}_1^2+\mathbf{T}_2^2)\right)  \right]+\left(-\tilde{\cal S}_{12}(\mathbf{T}_1^2+\mathbf{T}_2^2)\right)\;.
\end{align}
The second term on the r.h.s. is the same we would have for a colourless final state. The first term is the new contribution to the subtracted current we have to compute and, by using colour conservation, we can rewrite it as
\begin{align}
\label{eq:colourless_sub}
\sum_{j=3,4}\left[\tilde{\cal S}_{jj}\mathbf{T}_j^2+\sum_{i=1,2}\left(2\,\tilde{\cal S}_{ij}-\tilde{\cal S}_{12}\right)\mathbf{T}_i\cdot\mathbf{T}_j\right]+2\,\tilde{\cal S}_{34}\mathbf{T}_3\cdot\mathbf{T}_4\;.
\end{align}
Hence we need now to compute
\begin{equation}
\int\;d^Dk_1\;d^Dk_2\;\widetilde{\cal S}_{ij}(k_1,k_2) \delta_+(k_1^2)\,\delta_+(k_2^2)\;,
\end{equation}
for all the contributions involved in Eq.~(\ref{eq:colourless_sub}). This means we have to consider the following combinations of emitters $i$ and $j$:
\begin{itemize}
	\item $i$ and $j$ being the two initial-state massless emitters;
	\item $i$ being an initial-state massless emitter, $j$ a final-state massive emitter;
	\item $i=j$ being the same final-state massive emitter;
	\item $i$ and $j$ being the two final-state massive emitter.
\end{itemize}

It is convenient to integrate over the soft-gluon momenta $k_1$ and $k_2$ at fixed $k^\mu=k_1^\mu+k_2^\mu$: after that, we are left with only the integration over $k$.
With this goal in mind, we define the shorthand notation
\begin{equation}
\int_{(12)} f(k_1,k_2)\equiv\frac{\Gamma(\frac 12 -\epsilon)}{4^\epsilon \pi^{\frac 12-\epsilon}}\int d^D k_1\,d^D k_2\,f(k_1,k_2)\,\delta_+(k_1^2)\,\delta_+(k_2^2)\,\delta^{(D)}(k-k_1-k_2)\;,
\end{equation}
and we apply it to the functions $\tilde{\mathcal S}_{ij}^{m=0}$ and $\tilde{\mathcal S}_{ij}^{m\neq 0}$.
To perform this computation, we can first integrate one of the soft-gluon momenta (e.g. $k_1$) using the delta function $\delta^{(D)}(k-k_1-k_2)$.
Afterwards, we can go in the rest frame of $k$ and integrate over the energy component and the modulus of $\vec{k}_2$ by using the two remaining delta functions: this way only angular integrals are left.

By following these steps, we obtain
\begin{align}
\label{eq:sumangulararm=0}
\int_{(12)} \tilde{\cal S}^{m=0}_{ij}=&\frac{(k^2)^{-1-\epsilon}\pipj}{\pik\pjk}\left[\left(1+\vec n_i\cdot\vec n_j\right){\cal A}_{1,1}^+-2\left(1-\vec n_i\cdot \vec n_j\right)\mathcal A_{1,1}^-+\mathcal A_{1,0}+\mathcal A_{0,1}\right]\nonumber\\
\equiv &\frac{(k^2)^{-1-\epsilon}\pipj}{\pik\pjk}\,\f_{ij}^{gg}(\vec n_i\cdot \vec n_j,\vec n_i^2,\vec n_j^2)\,,\\
\label{eq:sumangulararm!=0}
\int_{(12)} \tilde{\cal S}^{m\neq0}_{ij}=&\frac{(k^2)^{-1-\epsilon}}{\pjk^2}\left[\left(1-\vec n_i\cdot \vec n_j\right)\mathcal A_{1,1}^--\left(1+\vec n_i\cdot\vec n_j\right)\mathcal A_{1,1}^+- \frac 12 \mathcal A^+_{1,-1}+3 \mathcal A_{1,0}-\frac 12 \mathcal A_{0,0}\right]\nonumber\\
\equiv &\frac{(k^2)^{-1-\epsilon}}{\pjk^2}\,\g_{ij}^{gg}(\vec n_i\cdot \vec n_j, \vec n_i^2,\vec n_j^2)\;,
\end{align}
where we defined $\vec n_i$ and $\vec n_j$ as vectors in the centre-of-mass frame of $k$ via $p_i = E_i (1, \vec n_i)$ and $p_j = E_j (1, \vec n_j)$. In terms of invariants, we have
\begin{align}
&\nini = 1-\frac{k^2 m_i^2}{\pik^2}\;,\\
&\njnj = 1-\frac{k^2 m_j^2}{\pjk^2}\;,\\
\label{eq:ninj}
&\ninj = 1-\frac{k^2 \pipj}{\pik \pjk}\;.
\end{align}
The functions $\f_{ij}^{gg}$, $\g_{ij}^{gg}$ are defined as the sum of the angular integrals (with appropriate multiplicative factors) for the massless and massive case respectively. The angular integrals $\mathcal A_{k,l}^\pm$ are defined as
 \begin{equation}\label{eq:angular}
 \mathcal A_{k,l}^\pm=\int_0^\pi d\theta \int_0^\pi d\phi \, \frac{\sin^{D-3}\theta \sin^{D-4}\phi}
{(1-a_i \cos\theta)^k(1\pm a_j \cos\chi \cos\theta\pm a_j \sin\chi \sin\theta \cos\phi)^l}\;,
 \end{equation}
with:
\begin{equation}
a_i=\sqrt{\nini}
\;
\hspace*{1cm}
\cos\chi = \frac{\ninj}{\sqrt{\nini \njnj}}
\;.
\end{equation}
The expression of the angular integral in Eq.~(\ref{eq:angular}) in many cases of interest can be found in Ref.~\cite{Somogyi:2011ir}.
Observe that $\mathcal A_{1,0}$ and $\mathcal A_{0,1}$ only depend on $a_i$ and $a_j$ respectively, and are independent of the label $\pm$ in Eq.~(\ref{eq:angular}).

We now need to perform the integration over $k$ (and, when needed, the explicit evaluation of the angular integrals) of the expressions in Eqs.~(\ref{eq:sumangulararm=0}) and (\ref{eq:sumangulararm!=0}) for all the possible emitters.

\subsubsection{Massless-massless contribution: \texorpdfstring{$\tilde{\cal S}_{12}$}{S12}}
\label{sub:massless-massless}
We start by addressing the problem of the integration of Eq.~(\ref{eq:sumangulararm=0}) in the case in which both the emitters are massless. The first step is to write explicitly the function $\f_{ij}^{gg}$ for this configuration. By using the results of Ref.~\cite{Somogyi:2011ir}
we find
\begin{align}
\label{eq:resultsm0}
\int_{(12)}\tilde{\cal S}^{m=0}_{12}(k_1,k_2)=&
\frac \pi 2
\frac{\pupd\,(k^2)^{-1-\epsilon}}{\puk\pdk}
\Bigg\{ -\frac 8\epsilon \left[1-\Gamma(1+\epsilon)\Gamma(1-\epsilon)\left(\frac{1+\vec n_1 \cdot \vec n_2}{1-\vec n_1\cdot\vec n_2}\right)^\epsilon\right]\nonumber\\
& -4 \left(\frac{1-\vec n_1\cdot \vec n_2}2\right)\int_0^1\frac{d t}{1-\left(\frac{1-\vec n_1\cdot\vec n_2}2\right)t}\left[(1-t)^{-\epsilon}-2(1-t)^\epsilon\right]\Bigg\}\;,
\end{align}
where the integration over $t$ in the last line is the integral representation of an hypergeometric function.

Our task is now to compute
\begin{align}
\int d^D k \,e^{i \vec b\cdot \vec k_T}\int_{(12)} \tilde{\cal S}^{m=0}_{12}(k_1,k_2)\;.
\end{align}
We observe that, while the first term on the right-hand side of Eq.~(\ref{eq:resultsm0}) is singular in the limit $k^2\to 0$, the second term is regular since $\vec n_1\cdot\vec n_2 \to 1$ as $k^2\to 0$ (see Eq.~(\ref{eq:ninj})) and thus it can be safely expanded in $\epsilon$.

To proceed further, we need to regularise the additional collinear singularity due to the presence of massless emitters, and we do it by partial fractioning
\begin{align}
\frac 1{\puk\pdk}=\frac 1\pupdk\left(\frac 1{\puk}+\frac 1\pdk\right)\;,
\end{align}
and by multiplying each singular contribution by the regulator already introduced in Eq.~(\ref{eq: lambda regulator}).
The next step is, after switching to light-cone coordinates, to add the integration over the delta function of $k^2$
\begin{equation}
  \label{eq:1delta}
\int d K^2 \;\delta(k^2-K^2)\;.
\end{equation}
which is used to integrate over $k_+$.
Then we can introduce the dimensionless variables
\begin{equation}
  \label{eq:adimj}
x=\frac {K^2}{k_T^2}~~~~~~~~
y=\frac{k_-^2}{k_T^2}\,,
\end{equation}
obtaining expressions where the integrals over $k_T$ and the one over $x$ and $y$ are factorised. The calculation can be completed with standard techniques: we find
\begin{align}
\label{eq:s12}\nonumber
\langle\int d^D k\;e^{i \vec b\cdot \vec k_T}&\int_{(12)}\tilde{\cal S}^{m=0}_{ij} \left(\frac{p_i \cdot k}{m^2}\right)^{2\lambda} \rangle_{\text{av.}}=
\;\left(\frac{b^2}4\right)^{2\epsilon-\lambda}
\left(\frac{\sqrt{s}}{2 m^2}\right)^{2\lambda}\pi^{2-\epsilon}
\frac{\Gamma(-2\epsilon+\lambda)}{2\Gamma(1+\epsilon-\lambda)}
\\&
\times\Bigg\{\frac1\lambda\left[\frac 4{\epsilon^2}-8\zeta_2-28\zeta_3\epsilon+\mathcal{O}(\epsilon^2)\right]+\left(31\zeta_4\epsilon+\mathcal O(\epsilon^2)\right)+\mathcal O (\lambda)\Bigg\}\;.
\end{align}

\subsubsection{Massless-massive contribution: \texorpdfstring{$\tilde{\cal S}_{ij}$}{Sij}}
\label{sub:mass-massless}
We now consider the massless-massive contribution.
In this case we have to integrate both Eq.~(\ref{eq:sumangulararm=0}) and Eq.~(\ref{eq:sumangulararm!=0}) for $i=1,2$ and $j=3,4$.

\paragraph{Mass-independent part}
We start our analysis with the massless-like part of the soft factor.
After writing explicitly the function $\f_{ij}^{gg}$ for this configuration by using the results of Ref.~\cite{Somogyi:2011ir}, we split it into a regular and a singular part as done for the massless-massless contribution in Sec.~\ref{sub:massless-massless}, immediately expanding in $\epsilon$ the regular part.
Because of the $\epsilon$-pole coming from phase-space integration, the expansion of the integrand needs to be performed up to order $\epsilon$.

We now describe the integration over $k$ of the angular function $\f_{ij}^{gg}$.
The collinear singularity due to the presence of the massless momentum $p_i$ is regularised as before with the regulator factor in Eq.~(\ref{eq: lambda regulator}).
We then introduce a delta function $\delta(k^2-K^2)$ as in Eq.~(\ref{eq:1delta})
and we use it to perform the integral over $k_+$.

Unlike the massless-massless contribution to the double gluon soft current but similarly to the single-gluon computation, the emitter has a non-zero transverse momentum and hence we have a dependence on $\vec k_T$ in $\pjk$.
As it is by now customary, we remove it with the shift
\begin{equation}
\vec k_T\to \vec k_T+\frac {k_-}{p_{j,-}}\vec p_{T,j}\;.
\end{equation}
This way the only dependence left on the angular part of $\vec k_T$ is in the exponential and the angular integral can now be easily performed.

The integrals that are left are now the one over $K^2$, over $k_T$ and over $k_-$. We introduce the dimensionless variables
\begin{equation}
\label{eq:uv_def}
 u=\frac {K^2}{k_T^2}~~~~~~~
 w=\frac {k_-^2}{k_T^2}\frac{m^2}{p_{j,-}^2}\;,
 \end{equation}
 in terms of which $\f_{ij}^{gg}(\vec n_j\cdot \vec n_i,1,\vec n_j^2)$ is now independent of $k_T^2$
 \begin{equation} \f_{ij}^{gg}(\vec n_i\cdot \vec n_j,1, \vec n_j^2)\equiv f_{ij}^{gg}(u,w)\;.\end{equation}
Because of this, the integral over $k_T$ factorises in the form
 \begin{align}
 \int_0^\infty d k_T\; k_T^{-1-3\epsilon+2\lambda}J_{-\epsilon}(b k_T) e^{i\,\,\vec b\cdot \vec p_{T,j}\,\sqrt w\,k_T/m}\;,
 \end{align}
 and can be computed separately obtaining an hypergeometric function\footnote{See formula 6.621 in Ref.~\cite{gradshteyn}.}.
 As for the integral over the variables $u$ and $w$ we obtain, up to overall factors
 \begin{align}
 \label{eq:partial_uv}
 &
 \int d^Dk\,e^{i\vec b\cdot\vec k_T}\int_{(12)}\tilde{\cal S}_{ij}^{m=0}\left(\frac{p_i\cdot k}{m^2}\right)^{2\lambda}\propto 
  \nonumber\\&
 \qquad\int_0^\infty d u\;d w\;\frac{u^{-1-\epsilon}w^{-1+2\epsilon}}{1+u+w}
 {}_2F_1\left(-2\epsilon+\lambda,\frac 12 -2\epsilon+\lambda;1-\epsilon;\frac 1{w}\,\frac{b^2m^2}{(\vec b\cdot\vec p_{T,j})^2}\right)f_{ij}^{gg}(u,w)\;.
 \end{align}
 Notice that the collinear divergence, regulated by the parameter $\lambda$, is now described by the limit $w\to0$.

It is now useful to perform some manipulation on Eq.~(\ref{eq:partial_uv}) in order to move the dependence on the regulator $\lambda$ outside of the hypergeometric function.
We use the following relation
\begin{align}
{}_2F_1(a,b,c;z)=&
\frac{\Gamma(b - a) \Gamma(c)}{\Gamma(b) \Gamma(c - a)} (-z)^{-a} {}_2F_1\left(a, a - c + 1, a - b + 1; \frac1z\right)
\nonumber\\&
+ \frac{\Gamma(a - b) \Gamma(c)}{\Gamma(a) \Gamma(c - b)} (-z)^{-b}{}_2F_1\left(b, b - c + 1, b - a + 1, \frac1z\right)\;.
\end{align}
By applying it to Eq.~(\ref{eq:partial_uv}) and by exploiting the $w\to 0$ limits of the new hypergeometric functions,
the limit $\lambda\to 0$ can be easily carried out and we obtain
\begin{align}
 \label{eq:intuvaz}
\nonumber\int d^D k \,e^{i \vec b\cdot \vec k_T}\int_{(12)} &\tilde{\cal S}^{m=0}_{ij}\left(\frac{p_i\cdot k}{m^2}\right)^{2\lambda}=
\frac{(p_i\cdot p_j)^{2\lambda}}{(m^2)^{3\lambda}}\left(\frac{b^2}{4}\right)^{2\epsilon-\lambda}\pi^{1-\epsilon}\frac{\Gamma(-2\epsilon+\lambda)}{2\Gamma(1+\epsilon-\lambda)}
\\\nonumber &\times
\int_0^\infty d u \;d w \;f_{ij}^{gg}(u,w)
 \;\frac{u^{-1-\epsilon}w^{-1}}{1+u+w}\Bigg\{w^\lambda+\Bigg[{}_2F_1\left(-2\epsilon,-\epsilon;\frac 12;c_{jb}^2 w\right)-1\\
&-4\epsilon i c_{jb} \sqrt w \frac{\Gamma(\frac 12-2\epsilon)\Gamma(1+\epsilon)}{\Gamma(1-2\epsilon)\Gamma(\frac 12+\epsilon)}{}_2F_1\left(\frac 12-2\epsilon,\frac 12 -\epsilon;\frac 32; c_{jb}^2 w\right)\Bigg]\Bigg\}\;,
\end{align}
where $c_{jb}$ is defined in Eq.~(\ref{eq:cjb}).
By comparing Eq.~(\ref{eq:intuvaz}) with the expression obtained in the one loop case in Eq.~(\ref{eq:Iay}), we can observe that they share the same structure, the only differences being an overall multiplicative factor, an additional integration over $u$ and the formal substitution
\begin{align}
\frac{u^{-1-\epsilon}w^{-1}}{1+u+w}\to\frac1{w\,(1+w)^{1+\epsilon}}\;.
\end{align}
By using this relation, the azimuthal average can be directly deduced from Eq.~(\ref{eq:Iay_aa})
\begingroup
\allowdisplaybreaks
\begin{align}
\label{eq:sij2azavfolded}\nonumber
\langle\int d^D k \,e^{i \vec b\cdot \vec k_T}\int_{(12)} &\tilde{\cal S}^{m=0}_{ij} \left(\frac{p_i\cdot k}{m^2}\right)^{2\lambda}\rangle_{\text{av.}}=
\frac{\pipj^{2\lambda}}{(m^2)^{3\lambda}}
\left(\frac{b^2}{4}\right)^{2\epsilon-\lambda}
\pi^{1-\epsilon}\frac{\Gamma(-2\epsilon+\lambda)}{2\Gamma(1+\epsilon-\lambda)}
\\\nonumber&
\times
\int_0^\infty d u \;d w \;f_{ij}^{gg}(u,w)
\frac{u^{-1-\epsilon}w^{-1}}{1+u+w}\;\Bigg\{w^\lambda
+\text{Re}\Big\{\left({}_2F_1\left(-2\epsilon, -\epsilon; 1-\epsilon;w B\right)-1\right)
\\&
\times\left(1+i\cot(\pi\epsilon)\right)\Big\}
\Bigg\}\;.
\end{align}
\endgroup
The two-folded integral in Eq.~(\ref{eq:sij2azavfolded}) does not present significant complications and can be computed with standard techniques.
We obtain
\begingroup
\allowdisplaybreaks
\begin{align}
\label{eq:sijm0}
\nonumber
\langle\int d^D k \,e^{i \vec b\cdot \vec k_T}&\int_{(12)} \tilde{\cal S}^{m=0}_{ij}\left(\frac {p_i\cdot k}{m^2}\right)^{2\lambda}\rangle_{\text{av.}}
=
-\frac{(p_i\cdot p_j)^{2\lambda}}{(m^2)^{3\lambda}}
\left(\frac{b^2}{4}\right)^{2\epsilon-\lambda}\pi ^{2-\epsilon }
\frac{\Gamma (\lambda -2\epsilon )}{2\epsilon  \Gamma (\epsilon -\lambda +1)}
\\\nonumber&
\times\Bigg\{\frac 1\lambda \left(-\frac{2}{\epsilon }+4 \zeta _2 \epsilon
+14 \zeta _3 \epsilon ^2+\mathcal{O}(\epsilon^3)\right)+\Bigg(-\Bigg[2 \ln^2\left(B\right)+4 \text{Li}_2\left(-\frac 1B\right)+2 \zeta _2\Bigg]
\\\nonumber&
+\frac{2}{3}\epsilon\Bigg[\ln ^3\left(\frac 1B+1\right)+\ln ^3\left(B\right)
+3 \ln \left(B\right) \ln^2\left(B+1\right)
-6 \ln ^2\left(B\right) \ln \left(B+1\right)
\\\nonumber&
-6 \ln\left(B+1\right) \text{Li}_2\left(-\frac 1B\right)-6 \text{Li}_3\left(\frac B{B+1}\right)
-6 \left(2 \ln\left(\frac 1B+1\right)+2 \ln \left(B\right)
\right.\\\nonumber&\left.
-\ln \left(B+1\right)\right) \zeta _2\Bigg]
+\epsilon^2\Bigg[\frac{7}{12} \ln ^4\left(\frac 1B+1\right)+14 \zeta _2 \ln ^2\left(\frac 1B+1\right)-\frac{2}{3} \ln\left(B+1\right)
\\\nonumber&
 \times\ln ^3\left(\frac 1B+1\right)+\text{Li}_2\left(\frac B{B+1}\right) \ln^2\left(\frac 1B+1\right)+8 \ln \left(B+1\right) \zeta _2 \ln\left(\frac 1B+1\right)
\\\nonumber&
 -\frac{31}{12} \ln^4\left(B\right)+\frac{11}{12} \ln^4\left(B+1\right)
 -\frac{7}{3} \ln \left(B\right) \ln ^3\left(B+1\right)-2
 \text{Li}_2\left(-\frac 1B\right){}^2
\\\nonumber&
+\text{Li}_2\left(\frac B{B+1}\right){}^2
+\frac{10}{3} \ln^3\left(B\right) \ln \left(B+1\right)-4 \ln \left(B+1\right)
\text{Li}_3\left(-\frac 1B\right)
\\\nonumber&
+4 \ln \left(B+1\right) \text{Li}_3\left(\frac B{B+1}\right)-12 \text{Li}_4\left(-\frac 1B\right)
+6 \ln\left(B+1\right) \text{Li}_3\left(\frac 1{B+1}\right)
\\\nonumber&
+8\text{Li}_4\left(\frac B{B+1}\right)+10 \text{Li}_4\left(\frac 1{B+1}\right)-27 \ln^2\left(B\right) \zeta _2
+3\text{Li}_2\left(\frac 1{B+1}\right) 
\\\nonumber&
\times\left(\ln ^2\left(B+1\right)-6 \zeta _2\right)
-33 \ln ^2\left(B+1\right) \zeta _2
   +60 \ln \left(B\right) \ln
   \left(B+1\right) \zeta _2
    \\\nonumber&
    -4 \text{Li}_2\left(\frac B{B+1}\right) \zeta _2
   -\text{Li}_2\left(-\frac 1B\right)
   \left(\ln ^2\left(\frac 1B+1\right)+3 \ln ^2\left(B\right)-2 \ln ^2\left(B+1\right)\right.
   \\&
  \left.+2\text{Li}_2\left(\frac B{B+1}\right)+10 \zeta _2\right)+4 \ln \left(B+1\right) \zeta _3+\frac{\zeta _4}{2}\Bigg]+\mathcal O\left(\epsilon^2 \right)\Bigg)+\mathcal O\left(\lambda \right)\Bigg\}\;.
   \end{align}
   \endgroup
Combining the results in Eq.~(\ref{eq:sijm0}) and Eq.~(\ref{eq:s12}) to construct the second term in the square bracket of Eq.~(\ref{eq:colourless_sub}) we see that the $\lambda\to 0$ singularities cancel out, as expected.

\paragraph{Mass-dependent part}
We now address the problem of the integration of the mass-dependent part, thus considering Eq.~(\ref{eq:sumangulararm!=0}).
The structure of this integral is similar to the one we evaluated for the mass-independent part, but with some differences that simplify the computation.
In particular, the overall factor multiplying the angular functions changes according to the substitution
\begin{equation}
\label{eq:mass_sub_1}
\frac{(k^2)^{-1-\epsilon} \pipj}{\pik \pjk}\to\frac{(k^2)^{-1-\epsilon}}{\pjk^2}\;,
\end{equation}
which in terms of the dimensionless variables introduced in Eq.~(\ref{eq:uv_def}), corresponds to:
\begin{equation}\label{eq:mass_sub_2}
\frac{u}{1+u+w}\to \frac {2 u w}{(1+u+w)^2}\;.
\end{equation}
This replacement removes the dependence on the massless momentum from the denominator of the integrand: as a consequence, the integration of this term will not give rise to additional collinear singularities and thus there is no need for the regulator we introduced in Eq.~(\ref{eq: lambda regulator}), that we can safely drop.

Eq.~(\ref{eq:sumangulararm!=0}) can thus be written in terms of two-folded integrals simply by applying the substitution (\ref{eq:mass_sub_2}) in Eq.~(\ref{eq:sij2azavfolded}), setting $\lambda=0$ and considering the function $\g^{gg}_{ij}$ rather than $\f_{ij}^{gg}$.
By doing so, we obtain
\begin{align}
\label{eq:sij2azavfoldedM}\nonumber
\langle\int d^D k \,e^{i \vec b\cdot \vec k_T}&\int_{(12)} \tilde{\cal S}^{m\neq 0}_{ij}\rangle_{\text{av.}} =
\left(\frac{b^2}{4}\right)^{2\epsilon}
\pi^{1-\epsilon}\frac{\Gamma(-2\epsilon)}{\Gamma(1+\epsilon)}
\int_0^\infty d u \;d w \;\frac{u^{-1-\epsilon}}{(1+u+w)^2}
\\
\times &
\Big\{1+
\text{Re}\left\{\left[{}_2F_1\left(-2\epsilon, -\epsilon; 1-\epsilon;w B\right)-1\right]\left(1+i\cot(\pi\epsilon)\right)\right\}
\Big\}g_{ij}^{gg}(u,w)\;,
\end{align}
where we have defined $\g_{ij}^{gg}(\vec n_i\cdot \vec n_j,1,\vec n_j^2)\equiv g_{ij}^{gg}(u,w)$.
The computation of this integral does not present any particular additional challenge that cannot be solved with standard techniques.
The expression of the angular function $g_{ij}^{gg}$ can be obtained from Ref.~\cite{Somogyi:2011ir} and it is again convenient to identify and immediately expand its contribution which is regular in $\epsilon$.
To simplify the expression of the integrand, we also isolated a part that, after integration, would have been independent of any kinematical variables: we evaluated numerically this contribution to obtain its constant result $c_0$, finding $c_0=-37.73041235261383$.
The final result reads
\begingroup
\allowdisplaybreaks
\begin{align}
\nonumber
    \langle\int d^D k \,e^{i \vec b\cdot \vec k_T}&\int_{(12)} \tilde{\cal S}^{m\neq 0}_{ij}\rangle_{\text{av.}}
=
-\left(\frac{b^2}{4}\right)^{2\epsilon} \pi ^{2-\epsilon }\frac{\Gamma (-2 \epsilon )}{\epsilon \, \Gamma
   (1+\epsilon)}\Bigg\{
   1+\epsilon\Bigg[
   \frac{\pi
   ^2}{6}-6
   -2 \text{Li}_2\left(\frac{1}{1-r^2}\right)
   \\\nonumber&
   -\ln ^2(r^2-1)
   +2\ln (r) (2\ln (r)+2)
   \Bigg]
   +\frac{\epsilon^2}6\Bigg[
   c_0+30 \text{Li}_3\left(\frac{1}{r^2}\right)
\\\nonumber&
+72 \text{Li}_3\left(\frac{1}{1-r^2}\right)
-120 \left(\text{Li}_3\left(\frac{1}{1-r}\right)-\text{Li}_3\left(\frac{1}{r+1}\right)\right)
\\\nonumber&
+12 \text{Li}_2\left(\frac{1}{r^2}\right) (\ln (r)-5)
+16 \pi ^2 \coth ^{-1}\left(1-2 r^2\right)
+8\ln ^3(r-1)-16 \ln ^3(r)
\\\nonumber&
+8 \ln ^3(r+1)
-36 \ln (r+1) \ln ^2(r-1)
+36 \ln ^2(r)\ln (r-1)
\\\nonumber&
-36 \ln ^2(r+1) \ln (r-1)-192 \ln ^2(r)+36 \ln ^2(r) \ln (r+1)    
\\\nonumber&
+120\ln (r) \ln (r-1)
+4 \pi ^2 \ln (r)-144 \ln (r)+120 \ln (r) \ln (r+1)
    \\&
    +63 \zeta_3+13 \pi ^2
+24-30 \pi ^2 \ln (2)
   \Bigg]
   \Bigg\}\;,
\end{align}
\endgroup
where the variable $r$ is defined in Eq.~(\ref{eq:var_r}).

\subsubsection{Massive-massive contribution: \texorpdfstring{$\tilde{\cal S}_{jj}$}{Sjj}}

We now examine the case in which only one massive final-state emitter is involved and we consider $\tilde{\cal S}_{jj}$, with $j=3,\,4$.
By plugging in Eq.~(\ref{eq:S_tilde}) the condition $p_i=p_j$, we obtain
\begin{align}
\label{eq:sjj}\nonumber
\tilde{\mathcal S}_{jj}&=\tilde{\mathcal S}_{jj}^{m=0}+2 m^2 \tilde{\mathcal S}_{jj}^{m\neq0}=  \frac {m^4}{\pjqu\pjqd\pjk^2}+\frac {4 m^2}{k^2 \pjqu \pjqd}
\\&
=2\left(\frac{m^4}{\pjk^3}+\frac {4 m^2}{k^2\pjk}\right)\frac 1{\pjqu}\;,
\end{align}
where $k=k_1+k_2$.
This more compact expression for $\tilde{\mathcal S}_{jj}$ allows us to obtain a simplified version of Eqs.~(\ref{eq:sumangulararm=0}), (\ref{eq:sumangulararm!=0})
\begin{equation}
\int_{(12)} \tilde{\mathcal S}_{jj}=2\left(\frac{m^4}{\pjk^3}+\frac {4 m^2}{k^2\pjk}\right) \frac {(k^2)^{-\epsilon} 2^{-2+2\epsilon}}{\pjk}\mathcal A_{1,0}\;,
\end{equation}
and, by using the result of the angular integral $\mathcal A_{1,0}$ from Ref.~\cite{Somogyi:2011ir}, we can write
\begin{equation}
\label{eq_sjj1}
\int_{(12)} \tilde{\mathcal S}_{jj}=\frac{(k^2)^{-1-\epsilon} m^2}{\pjk^2}\frac{2\pi}{1-2\epsilon}\left(\frac{m^2 k^2}{\pjk^2}+4\right) {}_2F_1\left(\frac 12,1,\frac 32 ; \vec n_j^{\,2}\right)\, .
\end{equation}
The integration of Eq.~(\ref{eq_sjj1}) can be carried out in a similar way as the integration of $\tilde{\cal S}_{ij}$ has been performed in Sect.~\ref{sub:mass-massless}.
Also in this case it is convenient to split the integrand into its singular and regular part, immediately expanding in $\epsilon$ the latter.
The final result reads
\begingroup
\allowdisplaybreaks
\begin{align}\nonumber
    \langle\int d^D k \,e^{i \vec b\cdot \vec k_T}&\int_{(12)} \tilde{\cal S}_{jj}\rangle_{\text{av.}}
=\left(\frac{b^2}{4}\right)^{2\epsilon}\pi^{2-\epsilon}\frac{ \Gamma (-2 \epsilon)}{\Gamma (\epsilon+1)}
\Bigg\{
\frac 2{\epsilon^2}
+\frac 4\epsilon \left(\ln \left(r^2\right)-1\right)
\\\nonumber&
+ \frac{1}{3} \left[-24 \text{Li}_2\left(1-r^2\right)-24 \ln \left(r^2\right)-5 \pi ^2+30\right]
\\\nonumber&
+\epsilon \Bigg[
32 \ln \left(\frac{2 r}{r+1}\right) \text{Li}_2\left(\frac{1}{2} \left(1+\frac{1}{r}\right)\right)-\frac{8}{3} \left(7+12\ln \left(\frac2{r-1}\right)\right) \text{Li}_2\left(\frac{1-r}{2}\right)
\\\nonumber&
+\left(\frac{254}{3}+64 \ln (r+1)\right)\text{Li}_2(1-r)
+2 \left(4+\frac{1}{r}+8 \ln \left(r\frac{r+1}{(r-1)^2}\right)\right) \text{Li}_2\left(\frac{1}{r^2}\right)
\\\nonumber&
+\left(-\frac{182}{3}-\frac{8}{r}+32 \ln \left(\frac{(r-1)^2}{2r^2(r+1)^2}\right)\right)\text{Li}_2\left(\frac{1}{r}\right)
+32 \ln \left(\frac{2 r}{r-1}\right) \text{Li}_2\left(\frac{r-1}{2 r}\right)
\\\nonumber&
+32 \ln \left(\frac{r-1}{2 r}\right) \text{Li}_2\left(\frac{r-1}{r}\right)
+8 \left(-3+4\ln \left(2r^2\frac{r+1}{r-1}\right)\right) \text{Li}_2(-r)
\\\nonumber&
-64 \ln (r-1) \text{Li}_2\left(\frac{1}{r+1}\right)
+\frac{8}{3} \left(7+12 \ln \left(\frac2{r+1}\right)\right) \text{Li}_2\left(\frac{2}{r+1}\right)
\\\nonumber&
+32 \ln \left(\frac{2 r}{r+1}\right)\text{Li}_2\left(\frac{r}{r+1}\right)
+32 \text{Li}_3\left(\frac{1}{2} \left(1+\frac{1}{r}\right)\right)
-32 \text{Li}_3\left(\frac{1-r}{2}\right)
\\\nonumber&
-64 \text{Li}_3(1-r)
+8 \text{Li}_3\left(\frac{1}{r^2}\right)
-64\text{Li}_3\left(\frac{1}{r}\right)
+32 \text{Li}_3\left(\frac{r-1}{2 r}\right)
-32 \text{Li}_3\left(\frac{r-1}{r}\right)
\\\nonumber&
-64 \text{Li}_3(-r)
-64 \text{Li}_3\left(\frac{1}{r+1}\right)
-32 \text{Li}_3\left(\frac{2}{r+1}\right)
-64\text{Li}_3\left(\frac{r-1}{r+1}\right)
\\\nonumber&
-32 \text{Li}_3\left(\frac{r}{r+1}\right)
+64 \text{Li}_3\left(\frac{r+1}{1-r}\right)
-8 \text{Li}_3\left(1-r^2\right)
+\frac{1}{3 r}\Bigg(
-32 r \ln ^3(r-1)
\\\nonumber&
+96 r \ln ^3(r)
+r \ln ^2(r) (77+96 \ln(2)-144 \ln (r+1))+96 r \ln ^2(r-1)
\\\nonumber&
 \times\ln (r+1)
-2 \ln (r-1) (4 r \left(5 \pi ^2-7 \ln (2)\right)+\ln (r) (-6+r (-67
\\\nonumber&
+48 \ln (2))
+72 r \ln (r))+4 r (7-48 \ln (r)) \ln (r+1)+96 r \ln ^2(r+1))
\\\nonumber&
+12 \ln (r)\left(2 r \left(9+8 \ln ^2(2)\right)-\ln (r+1) (1+2 r (5+8\ln (2))-8 r \ln (r+1))\right)
\\\nonumber&
+r (-60+\pi ^2 (27+16 \ln (2))+4 \ln ^2(2) (-7+8\ln (2))+2 \left(-24+\pi ^2\right) \ln \left(r^2\right)
\\\nonumber&
-8 \ln^3\left(r^2\right)+4 \ln (r+1) \left(6 \left(\pi ^2-4 \ln ^2(2)\right)+(7+36 \ln (2)) \ln (r+1)\right)
\\&
+12 \ln ^2\left(r^2\right) \left(-2+\ln \left(r^2-1\right)\right)+276 \zeta (3))
\Bigg)
\Bigg]
\Bigg\}\;.
\end{align}
\endgroup

\subsubsection{Massive-massive contribution: \texorpdfstring{$\tilde{\cal S}_{34}$}{S34}}
\label{sec:S34}
We finally analyse the case of the interference between the two final-state emitters.
Our task is to integrate both the mass-independent and mass-dependent expressions in Eq.~(\ref{eq:sumangulararm=0}) and Eq.~(\ref{eq:sumangulararm!=0}) for $i=3$, $j=4$.

\paragraph{Mass-independent part}
Let us start our analysis with the mass-independent contribution.
We need to integrate the dimensionless angular function $\f_{34}^{gg}(\vec n_3\cdot \vec n_4, \vec n_3^2, \vec n_4^2)$ defined in Eq.~(\ref{eq:sumangulararm=0}).

To evaluate the angular integrals contained therein, we relate them to the imaginary part of a massive box diagram via the optical theorem\footnote{See e.g. the discussion in Appendix A of Ref.~\cite{vanNeerven:1985xr}.}.
All order results for the box diagram with a single mass, equivalent to fix $a_i=a_j$ in Eq.~(\ref{eq:angular}), are presented in Ref.~\cite{Kniehl:2009pv}, while results with two different masses but only at the lowest order in $\epsilon$ can be found in Ref.~\cite{Ellis:2007qk}.
We extended the latter expression to all orders in $\epsilon$, obtaining the angular integral $\mathcal A_{1,1}^\pm$
\begin{align}
\label{eq:A11_us}\nonumber
  \mathcal A_{1,1}^\pm=&\int_0^\pi d \theta \int_0^\pi d \phi \, \frac{\sin^{D-3}\theta \sin^{D-4}\phi}
 {(1-a_3 \cos\theta)(1\pm a_4 \cos\chi \cos\theta\pm a_4 \sin\chi \sin\theta \cos\phi)}\\\nonumber
 =&
 \frac{2\pi}{2\epsilon-1}
 \frac{ \vec n_3^2+\vec n_3\cdot \vec n_4}{\sqrt{1-\vec n_3^2}  \left(\vec n_3^2+\vec n_4^2+(\vec n_3\cdot \vec n_4)^2+2
   \vec n_3\cdot \vec n_4-\vec n_3^2 \vec n_4^2\right)}\times
   \\\nonumber&
   \times F_1\left(\frac{1}{2}-\epsilon
   ;1,\frac{1}{2};\frac{3}{2}-\epsilon ;\frac{\vec n_3\cdot \vec n_4^2-\vec n_3^2
   \vec n_4^2}{(\vec n_3\cdot \vec n_4)^2+2 \vec n_3\cdot \vec n_4+\vec n_3^2-\vec n_3^2
   \vec n_4^2+\vec n_4^2},1-\frac{\vec n_3^2}{\vec n_3^2-1}\right)
 \\
 &+(3\leftrightarrow 4)\;.
\end{align}
By following the same strategy applied in the previous sections, we isolate in the angular function $\f_{34}^{gg}(\vec n_3\cdot \vec n_4, \vec n_3^{\,2}, \vec n_4^{\,2})$ a term $\sigma_{34}^{m=0}$ that will give rise to singularities upon integration over $k^2$ and a regular term $\rho_{34}^{m=0}$ that vanishes in the $k^2 \to 0$ limit and that can be directly expanded in $\epsilon$
\begin{equation} \label{eq:fsplit}
\f_{34}^{gg}(\vec n_3\cdot \vec n_4, \vec n_3^{\,2}, \vec n_4^{\,2})
=
-\frac{\pi}{\epsilon}
\left[
\sigma_{34}^{m=0} +
\epsilon\, \rho_{34}^{m=0}
+ (p_3 \leftrightarrow p_4)
\right]\;.
\end{equation}
By using Eq.~(\ref{eq:A11_us}), the $k^2$-singular part can be written in the following way
\begin{equation}\label{eq:fsing}
\sigma_{34}^{m=0}
=
2-2 \, h(\epsilon) \left(
\frac{1-\vec n_3^2     }{4}
\right)^{-\epsilon}
\left[
1- 2\epsilon\, \frac{(1-\vec n_3\cdot \vec n_4) \chi_{34}}{\vec n_3^2     - \vec n_3\cdot \vec n_4}
\,_2F_1(1,\tfrac{1}{2}-\epsilon;\tfrac{3}{2};\chi_{34})
\right]\;,
\end{equation}
where the function $h(\epsilon)$ is defined as
\begin{equation}
  \label{eq:hep}
h(\epsilon) = \pi^{-1/2}\,4^{-\epsilon}\, \Gamma(\tfrac{1}{2}-\epsilon)\,\Gamma(1+\epsilon)\;,
\end{equation}
and the $k^2$-independent coefficient $\chi_{34}$ is given by
\begin{equation}
\label{eq:def_chi34}
\chi_{34}
=
\frac{(\vec n_3^2     - \vec n_3\cdot \vec n_4 )^2}{(\vec n_3^2 - \vec n_4^2 )^2 + (\vec n_3\cdot \vec n_4 )^2 - \vec n_3^2     \, \vec n_4^2  }\;,
\end{equation}
and fulfils $0 \leq \chi_{34} \leq 1$.
We observe that the factor $(1-\vec n_3\cdot \vec n_4)/(\vec n_3^2     - \vec n_3\cdot \vec n_4)$ is also independent on $k^2$.

Let us start by considering the integration of the singular contribution.
By inserting Eqs.~(\ref{eq:fsplit}), (\ref{eq:fsing}) in Eq.~(\ref{eq:sumangulararm=0}) we obtain a sum of integrals with the following structure
\begin{equation}
I^{gg}_{34}[f_\alpha]=\int d^D k\,  e^{i \vec b \cdot \vec k_T} \frac{(k^2)^{-1-\epsilon}(p_3\cdot p_4)}{(p_3\cdot k)(p_4\cdot k)} f_\alpha(\vec n_3, \vec n_4)\;,
\end{equation}
with three possible functions $f_\alpha(\vec n_3, \vec n_4)$ ($\alpha=1,2,3$)
\begin{align}
\label{eq:f_i}
f_1(\vec n_3, \vec n_4)&=1\;,\\
f_2(\vec n_3, \vec n_4)&=\left(\frac{1-\vec n_3^2}4\right)^{-\epsilon}\;,\\
f_3(\vec n_3, \vec n_4)&=-4\pi \,h(\epsilon)\,\left(\frac{1-\vec n_3^2}4\right)^{-\epsilon}\frac{(1-\vec n_3\cdot \vec n_4) \chi_{34}}{\vec n_3^2     - \vec n_3\cdot \vec n_4}
\,_2F_1(1,\tfrac{1}{2}-\epsilon;\tfrac{3}{2};\chi_{34})\;.
\end{align}
 With those definitions, we have
 \begin{equation}
   \label{eq:sigmasingm0}
\int d^D k\,  e^{i \vec b \cdot \vec k_T}\, \frac{(k^2)^{-1-\epsilon}(p_3\cdot p_4)}{(p_3\cdot k)(p_4\cdot k)}\left(-\frac \pi \epsilon\,\sigma_{34}^{m=0} \right) = -\frac{2\pi}{\epsilon}I^{gg}_{34}[f_1]+\frac {2\pi}{\epsilon} h(\epsilon) I^{gg}_{34}[f_2]+I^{gg}_{34}[f_3]\;.
\end{equation}

We start by considering $I^{gg}_{34}[f_1]$
\begin{equation}
I^{gg}_{34}[f_1]=\int d^D k \,\frac{p_3\cdot p_4}{(p_3\cdot k)(p_4\cdot k)}\,(k^2)^{-1-\epsilon}\,  e^{i \vec b \cdot \vec k_T}\;.
\end{equation}
This integral is exactly the same appearing in Eq.~(\ref{eq:i34qspace}) for the case of soft $q{\bar q}$ emission,
and the result up to ${\cal O}(\epsilon)$ was already presented in Eq.~(\ref{eq:iqq34_result}).
In the present case, however, we need it up to ${\cal O}(\epsilon^2)$, since in Eq.~(\ref{eq:sigmasingm0}) $I^{gg}_{34}[f_1]$ already appears with an overall factor $1/\epsilon$.
We find
\begin{align}
    \braket{I^{gg}_{34}[f_1]}_{\text{av.}}=&
\pi ^{1-\epsilon }\, \Gamma (1-\epsilon ) \Gamma (-2 \epsilon )\left(\frac{b^2}{4}\right)^{2\epsilon} \frac{1+\beta^2}{2\beta}
\left\{-\frac{1}{\epsilon} L_0 - 2 L_1+ \epsilon ( 2 P_2 - L_2 )
\right.\\\nonumber&\left.
-\frac 23 \epsilon^2\left(L_3-6P_3-3Q_3\right)
+ {\cal O}(\epsilon^3) \right\}\;.
\end{align}
The functions $L_n$, $P_n$ have been defined in Eqs.~(\ref{eq:ln})--(\ref{eq:pn}). Here we also introduced the function $Q_n$, defined as
\begin{align}
&Q_n(\beta) =
\int_{-\beta}^{\beta} \frac{dz}{1 - z^2} \text{Li}_n\left(
-\frac{z^2 \tan ^2(\theta )}{z^2-\sec ^2(\theta )}
\right)\;.
\end{align}
The explicit expressions of the functions $L_0$, $L_1$ and $P_2$ are already provided in Eqs.~(\ref{eq:L0_result})--(\ref{eq:P2_result}) while $L_2$ can be found in Eq.~(\ref{eq:L2_result}). The functions $L_3$, $P_3$ and $Q_3$, that can be obtained in a similar way. 

We can use a similar strategy to evaluate $I^{gg}_{34}[f_2]$
\begin{equation}
I^{gg}_{34}[f_2]=\left(\frac{m^2}4\right)^{-\epsilon}\int d^D k\,  e^{i \vec b \cdot \vec k_T} \frac{(k^2)^{-1-2\epsilon}(p_3\cdot p_4)}{(p_3\cdot k)^{1-2\epsilon}(p_4\cdot k)}\;.
\end{equation}
In this case, the generalisation of Feynman parametrisation needs to be applied
\begin{equation}\label{eq:genfeyn}
\frac 1{A^m B^n}=\frac{\Gamma(m+n)}{\Gamma(m)\Gamma(n)}\,\int_0^1dx\, \frac{x^{m-1}(1-x)^{n-1}}{(x A+(1-x) B)^{m+n}}\;,
\end{equation}
thereby obtaining
\begin{equation}
  I^{gg}_{34}[f_2]=(p_3\cdot p_4) \left(\frac{m^2}4\right)^{-\epsilon} (1-2\epsilon)\int_0^1 dx\, x^{-2\epsilon}\, \int d^Dk\,  e^{i \vec b \cdot \vec k_T} \frac{(k^2)^{-1-2\epsilon}}{(p(x)\cdot k)^{2-2\epsilon}}\, ,
\end{equation}
with $p(x)=xp_3+(1-x)p_4$.
The azimuthally averaged integral $\langle I^{gg}_{34}[f_2]\rangle_{\rm av.}$ can be evaluated as
\begin{equation}
\langle I^{gg}_{34}[f_2]\rangle_{\rm av.}=(p_3\cdot p_4) \left(\frac{m^2}4\right)^{-\epsilon} (1-2\epsilon) \int_0^1 \frac {dx}{(p^2(x))^{1-\epsilon}}x^{-2\epsilon}\,\langle I_2^{\text{aux}}(x)\rangle_{\rm av.}\,,
\end{equation}
with
\begin{align}\label{eq:iaux2}\nonumber
\braket{I_2^{\text{aux}}(x)}_{\text{av.}}=&\braket{\int d^D k \, e^{i \vec b \cdot \vec k_T} \frac{(k^2)^{-1-\epsilon}}{(p(x)\cdot k)^2}\, p^2(x)\,\left(\frac{k^2 p^2(x)}{(p(x)\cdot k)^2}\right)^{-\epsilon}}_{\text{av.}}\\
=&\left(\frac{b^2}4\right)^{2\epsilon}\frac{\pi^{1-\epsilon}\,2^{-2-2\epsilon}}{\epsilon^2 (1-2\epsilon)}\Gamma(1-2\epsilon)\Gamma(1-\epsilon)\left(1+\frac{p_T^2(x)}{p^2(x)}\right)^{2\epsilon}\;.
\end{align}
The leftover integral over the Feynman parameter can be performed in a standard way and the solution expressed in terms of multiple polylogarithms.

The last integral we need for the evaluation of the singular part is
\begin{align}
I^{gg}_{34}[f_3]=-4\pi h(\epsilon)\int d^D k\,  &e^{i \vec b \cdot \vec k_T} \frac{(k^2)^{-1-\epsilon}(p_3\cdot p_4)}{(p_3\cdot k)(p_4\cdot k)} 
\nonumber\\&
\times
\left(\frac{1-\vec n_i^2}4\right)^{-\epsilon}\frac{(1-\vec n_3\cdot \vec n_4) \chi_{34}}{\vec n_3^2     - \vec n_3\cdot \vec n_4}
\,_2F_1(1,\tfrac{1}{2}-\epsilon;\tfrac{3}{2};\chi_{34})\;,
\end{align}
with $h(\epsilon)$ and $\chi_{34}$ defined in Eqs.~(\ref{eq:hep}) and (\ref{eq:def_chi34}), respectively.

In order to simplify the expression of the integrand we define the auxiliary momentum
\begin{equation}
  \label{eq:ell}
\ell^\mu=\frac 1v \left(p_3^\mu-\frac{m^2}{(p_3\cdot p_4)} p_4^\mu\right)\;,
\end{equation}
with $v$ defined as in Eq.~(\ref{eq:var_v}). 
By using the definition of $\ell^\mu$ and an integral representation for the hypergeometric function, we can rewrite $I^{gg}_{34}[f_3]$ as
\begin{align}\label{eq:f3l}\nonumber
I^{gg}_{34}[f_3]=&-\int d^D k\,e^{i \vec b \cdot \vec k_T}\frac {2\pi(p_3\cdot p_4)}{v}\,\frac {(k^2)^{-1-2\epsilon}}{(p_4\cdot k)}\left(\frac{m^2}{(p_3\cdot k)^2}\right)^{-\epsilon}\frac 1{\ell\cdot k}
\\\nonumber&
\times
\int_0^1\,d t\,(t)^{-\frac 12 -\epsilon}(1-t)^{\epsilon}\frac{\chi_{34}}{1-t\,\chi_{34}}\;.
\\\nonumber
=&2\pi (m^2)^{-\epsilon} \frac{(p_3\cdot p_4)}{v^2} \int_0^1\,d t\, (t)^{-\frac 12 -\epsilon}(1-t)^\epsilon
\int\,d^D k (k^2)^{-1-2\epsilon}\frac{e^{i \vec b\cdot \vec k_T}}{(p_3\cdot k)^{1-2\epsilon}}
\\&
\times\Bigg(
\left(\frac 1{1-\frac t{v^2}}\right)\left(-\frac 1{(p_4\cdot k)}\right)
+\frac{m^2}{2(p_3\cdot p_4)}\sum_{\sigma=\pm1}\frac 1{1+\sigma\frac{\sqrt{t}}v}\frac1{\ell_{(\sigma)}\cdot k}
\Bigg)\;,
\end{align}
where for convenience we introduced
\begin{equation}
  \label{eq:ellpm}
\ell_{(\pm)}^\mu=p_3^\mu\pm\sqrt t \,\ell^\mu\;.
\end{equation}
By analysing the dependence on $t$ of the integrand, we observe that it is expressed as a sum of functions, some of which feature a divergence for $t=v^2$.
This singularity has no physical origin and will eventually cancel in the final result when summing together these divergent contributions.
In order to regularise it, we fix a small imaginary part for $v$ and take its finite limit to zero at the end of the computation.
The dependence of the integrand on $k$ is via a factor
\begin{equation}
I^{gg}_{34}[f_3]\propto \frac {(k^2)^{-1-2\epsilon}}{(p_3\cdot k)^{1-2\epsilon}}\,\left\{\frac 1{(p_4\cdot k)};\frac1{(\ell_{(\pm)}\cdot k)}\right\}\;.
\end{equation}
By applying the generalised Feynman parametrisation introduced in Eq.~(\ref{eq:genfeyn}), we obtain a single momentum integral equivalent to $I_2^{\text{aux}}$ in Eq.~(\ref{eq:iaux2}), the only difference being the expression of the auxiliary momentum, which now reads:
\begin{equation}
p(x)=x p_3+(1-x) P\;,
\end{equation}
with the three possibilities
\begin{equation}
P=p_4,\qquad P=\ell_{(+)}\qquad P=\ell_{(-)}\;.
\end{equation}
The integration over $k$ can thus be performed by using the partial results already obtained.
The leftover integrals over the Feynman parameter $x$ and the variable $t$ we used for the integral representation of the hypergeometric function can be performed with standard techniques, and the result can be expressed in terms of multiple polylogarithms.

We now consider the contribution to $\f^{gg}_{34}$ that vanishes in the $k^2 \to 0$ limit, $\rho_{34}^{m=0}$.
It can be written at all orders in $\epsilon$ in the following compact way
\begin{align}\nonumber
\frac{1}{\pi}\, \rho_{34}^{m=0} =& \frac{\Gamma(\tfrac{1}{2}-\epsilon) \Gamma(\epsilon)}{\sqrt{\pi}}
\left( \frac{1-\vec n_3^2     }{\vec n_3^2     }
\right)^{-\epsilon}
\frac{1+D_{34}-2\sqrt{\vec n_3^2     }}{\sqrt{\vec n_3^2     }}
+ \frac{1}{\epsilon} \left[
2-(1+D_{34}) _2F_1(\tfrac{1}{2},1;1+\epsilon,1-\vec n_3^2     )
\right] \\
&- \frac{2(1-\vec n_3\cdot \vec n_4 ) \chi_{34}}{ \vec n_3^2     - \vec n_3\cdot \vec n_4 }
\int_0^1 du \frac{(1-u)^{-\frac{1}{2}-\epsilon}}{\sqrt{1-u \chi_{34}}}
\left[
(1+ u \psi)^\epsilon- u^\epsilon(1+ \psi)^\epsilon
\right] \nonumber \\
&+ D_{34} \gamma
\int_0^1 du
\frac{(1-u)^{\frac{1}{2}-\epsilon}}{\sqrt{1-\vec n_3^2     u}(1-(1-u)\gamma)}\;,
\end{align}
where the coefficient $\chi_{34}$ is defined in Eq.~(\ref{eq:def_chi34}) and
\begingroup 
\allowdisplaybreaks
\begin{align}
\label{eq: def chi}
&D_{34}
=
\frac{(1+\vec n_3\cdot \vec n_4 )(\vec n_3^2     + \vec n_3\cdot \vec n_4)}{(\vec n_3^2 + \vec n_4^2 )^2 + (\vec n_3\cdot \vec n_4 )^2 - \vec n_3^2     \, \vec n_4^2  }\;,\\
&\gamma
=
\frac{(\vec n_3\cdot \vec n_4)^2 - \vec n_3^2     \vec n_4^2  }{(\vec n_3^2 + \vec n_4^2 )^2 + (\vec n_3\cdot \vec n_4 )^2 - \vec n_3^2     \, \vec n_4^2  }\;,\\
\label{eq: def psi}
&\psi
=
-\frac{(\vec n_3\cdot \vec n_4)^2 - \vec n_3^2     \vec n_4^2  }{(\vec n_3^2 - \vec n_4^2 )^2 + (\vec n_3\cdot \vec n_4 )^2 - \vec n_3^2     \, \vec n_4^2  }\;.
\end{align}
\endgroup
Because of its regular behaviour, $\rho_{34}^{m=0}$ can be safely expanded in $\epsilon$
\begin{equation}
\rho_{34}^{m=0}
=
\rho_{34}^{m=0,\,(0)}
+
\epsilon\, \rho_{34}^{m=0,\,(1)}
+{\cal O}(\epsilon^2)\;.
\end{equation}
Due to the complexity of the functions $\rho_{34}^{m=0,\,(0)}$ and $\rho_{34}^{m=0,\,(1)}$, we perform numerically the last steps of their integration.
The representation of the soft integrals in $q_T$-space, rather than the impact-parameter space used until now, is more convenient to this purpose, as it allows us to trivially carry out the $D-2$ dimensional integration of the transverse components of the soft momentum $k$.
The conversion to the representation in $b$-space can be obtained by applying to the final result the relation in Eq.~(\ref{eq:space_conversion}).
To be specific, the integral that we will compute is
\begin{equation}
\left\langle\int d^D k \, \delta^{(D-2)}(\vec k_T-\vec q_T) \frac{(k^2)^{-1-\epsilon} (p_3\cdot p_4)}{(p_3\cdot k) (p_4\cdot k)}
(\rho_{34}^{m=0,\,(0)} + \epsilon\, \rho_{34}^{m=0,\,(1)})\right\rangle_{\text{av.}} \;.
\end{equation}
To perform its azimuthal average, we can fix the azimuthal angle $\phi$ such that $\vec{p}_{T,3} \cdot \vec q_T = p_{T,3} \, q_T \cos \phi$. With this choice the integral over the other angles becomes straightforward.
After integrating over $d^{D-2} k_T$, the remaining computation can be performed for instance by introducing an integral over the virtuality of $k$.
We obtain
\begin{equation}\label{eq:numerical1}
\frac{(q_T^2)^{-1-\epsilon}}{B(\tfrac{1}{2},\tfrac{1}{2}-\epsilon)}
\int_{-1}^1 d\cos \phi \int_0^\infty dx \, dy \,
\frac{1-\vec n_3\cdot \vec n_4}{2\, x\, y} x^{-1-\epsilon} (1-\cos^2\phi)^{-1/2-\epsilon}
(\rho_{34}^{m=0,\,(0)} + \epsilon\, \rho_{34}^{m=0,\,(1)})\;,
\end{equation}
where we introduced the dimensionless integration variables
\begin{equation}
\label{eq: def adim xy}
x = \frac{k^2}{q_T^2} \;, \hspace*{2cm}
y = \frac{k_-}{q_T}\;.
\end{equation}

Given that the functions $\rho_{34}^{m=0,\,(0)}$ and $\rho_{34}^{m=0,\,(1)}$ vanish by construction in the limit $x \to 0$, the integrand in Eq.~(\ref{eq:numerical1}) can be safely expanded in $\epsilon$ and integrated numerically over $x$, $y$ and $\cos\phi$. 
The result is a function of the LO phase-space, which can be reduced to the dependence on $\beta$ and $\cos\theta$ and can thus be provided in the form of a two-dimensional grid.

In order to obtain a more stable and fast numerical evaluation of Eq.~(\ref{eq:numerical1}) we isolate its contributions that can be expressed only as a function of $\beta$.
To be specific, we observe that in the $\epsilon$-expanded expression,
\begin{align}
\nonumber
\frac{(q_T^2)^{-1-\epsilon}}{B(\tfrac{1}{2},\tfrac{1}{2}-\epsilon)}&
\int_{-1}^1 d\cos \phi \int_0^\infty dx \, dy \,
\frac{1-\vec n_3\cdot \vec n_4}{2\, x^2\, y}  (1-\cos^2\phi)^{-1/2}
\Big[
\rho_{34}^{m=0,\,(0)} + \epsilon\, \rho_{34}^{m=0,\,(1)}
\\&
- \epsilon\ln\left[x(1-\cos^2\phi)\right] \rho_{34}^{m=0,\,(0)}
\Big]\;,
\end{align}
the integral of the first two terms in the square bracket is independent of $\cos\theta$, allowing us to simply compute a one-dimensional grid. In addition, these terms can be integrated by using the following identity,
\begin{align}\label{eq:numeric2}\nonumber
&\frac{1}{\pi}\int_{-1}^1 d\cos \phi \int_0^\infty dx \, dy \,
\frac{1-\vec n_3\cdot \vec n_4}{2\, x^2\, y} x^{-1-\epsilon} (1-\cos^2\phi)^{-1/2-\epsilon}
f(\vec n_3\cdot \vec n_4,\vec n_3^2     ,\vec n_4^2  )  \;
\\ &
=\int_0^1 dt \int_{-1}^1 d\cos\phi
\frac{t^2}{1-t^2} \frac{1}{1- v t \cos\phi}
\, f \left(
1-\frac{1-t^2}{1-v t \cos\phi},
t^2,
1-(1-v^2)\frac{1-t^2}{(1-v t \cos\phi)^2}
\right)\;,
\end{align}
which is valid for a generic function $f$, and that can by proven by using the relation in Eq.~(\ref{eq:theta_trick}). 
The right hand side of the equation is only a two-fold integral, leading to a further simplification of the corresponding numerical integration.

\paragraph{Mass-dependent part}
We now consider the case of the contribution coming purely from the massive case, $\tilde{\cal S}^{m\texorpdfstring{\neq}{!=}0}_{34}$.
We can approach the computation in a way similar to that used for the mass-independent part we just described, considering Eq.~(\ref{eq:sumangulararm!=0}) rather than Eq.~(\ref{eq:sumangulararm=0}). 
The factor multiplying the angular functions, however, is not anymore symmetric under the exchange $p_3 \leftrightarrow p_4$. Unlike the mass-independent contribution, we thus cannot assume that the final result respects such symmetry and we need to separately compute the integral of $\tilde{\cal S}^{m\texorpdfstring{\neq}{!=}0}_{34}$ and the one of $\tilde{\cal S}^{m\texorpdfstring{\neq}{!=}0}_{43}$.

As it is by now customary, we write the dimensionless angular function $\g_{34}^{gg}(\vec n_3 \cdot \vec n_4, \vec n_3^{\,2}, \vec n_4^{\,2})$ defined in Eq.~(\ref{eq:sumangulararm!=0}) as the sum of a singular and a regular contribution
\begin{equation} \label{eq:gsplit}
\g^{gg}_{34}(\vec n_3\cdot \vec n_4, \vec n_3^{\,2}, \vec n_4^{\,2})
=
-\frac{\pi}{\epsilon}
\left[
\sigma_{34}^{m\neq 0} +
\epsilon\, \rho_{34}^{m \neq 0}
+ (p_3 \leftrightarrow p_4)
\right]\;.
\end{equation}

The singular part, which generates singularities after integration over $k^2$, can be written as
\begin{align}
  \label{eq:singularm!=0}
\nonumber \sigma_{34}^{m\neq 0}=
  &-\frac{(1-\vec n_3^2     )^{-\epsilon } \Gamma \left(\frac{1}{2}-\epsilon \right) \Gamma
     (1+\epsilon)}{\sqrt{\pi }}
     \left(1+\epsilon\,\frac{2 \chi_{34} (1-\vec n_3 \cdot \vec n_4)  \, _2F_1\left(1,\frac{1}{2}-\epsilon
   ;\frac{3}{2};\chi_{34}\right)}{\vec n_3^2     - \vec n_3 \cdot \vec n_4}\right)\\
   &+\frac{(1-\vec n_4^2 )^{-\epsilon } \Gamma \left(\frac{1}{2}-\epsilon \right) \Gamma(1+\epsilon)}{\sqrt{\pi }}\left(1+ \epsilon\,\frac{2  \chi_{34} (1-\vec n_3 \cdot \vec n_4) \, _2F_1\left(1,\frac{1}{2}-\epsilon;\frac{3}{2}; \chi_{34}\right)}{\vec n_3 \cdot \vec n_4-\vec n_4^2 }
   \right)\;,
\end{align}
while the regular part can be directly expanded in $\epsilon$ and we can symbolically write
\begin{equation}
\rho_{34}^{m\neq 0}
 =
\rho_{34}^{m\neq 0\,(0)}
 +
\epsilon\, \rho_{34}^{m \neq 0\,(1)}
+{\cal O}(\epsilon^2)\;,
\end{equation}
where higher order terms can be neglected in our computation.

Let us start with the integration of the singular part in Eq.~(\ref{eq:singularm!=0}). 
Its computation requires the evaluation of integrals in the form
\begin{equation}
I^{gg}_j[f_\alpha]=\int d^D k\,  e^{i \vec b \cdot \vec k_T} \frac{(k^2)^{-1-\epsilon}}{(p_j\cdot k)^2} f_\alpha(\vec n_3, \vec n_4)\;,
\end{equation}
with $j=3,4$ and where the possible functions $f_\alpha(\vec n_3, \vec n_4)$ are the same already introduced in the mass-independent case in Eq.~(\ref{eq:f_i}).
With this definition we have
\begin{align}\nonumber
\sum_{j=3,4}\int d^D k\,  e^{i \vec b \cdot \vec k_T}\, \frac{(k^2)^{-1-\epsilon}}{\pjk^2}& \left(-\frac\pi\epsilon (\sigma_{34}^{m\neq 0}+\sigma_{43}^{m\neq 0})\right)=\Bigg(-\frac\pi\epsilon I^{gg}_3[f_1]+\frac\pi\epsilon h(\epsilon)I^{gg}_3[f_2]+\frac 12 I^{gg}_3[f_3]\\
&+\frac\pi\epsilon I^{gg}_4[f_1]-\frac\pi\epsilon h(\epsilon)I^{gg}_4[f_2]+\frac 12 I^{gg}_4[f_3]\Bigg)+(p_3\leftrightarrow p_4)\;.
\end{align}
We start from $I^{gg}_j[f_1]$,
\begin{equation}
  I^{gg}_j[f_1]=\int d^D k\,   \frac{1}{(k^2)^{1+\epsilon}}\frac{e^{i \vec b \cdot \vec k_T}}{(p_j\cdot k)^2}\;.
\end{equation}
This integral has already been computed in Sect.~\ref{sub:qq} and we have $m^2 I^{gg}_j[f_1]=I_{jj}^{q{\bar q}}$,
where $I_{jj}^{q{\bar q}}$ is defined in Eq.~(\ref{eq:ijjqspace}).
The result for $\braket{I_{jj}^{q\bar q}(\vec b)}_{\text{av.}}$ was reported in Eq.~(\ref{eq:iqqjj_result}).

We now turn our attention to $I^{gg}_j[f_2]$. We first consider $I^{gg}_3(f_2)$.
The integral to compute is
\begin{equation}
  I^{gg}_3[f_2]=\int d^D k\,  e^{i \vec b \cdot \vec k_T} \frac{(k^2)^{-1-2\epsilon}}{(p_3\cdot k)^2}\left(\frac{m^2}{4 (p_3\cdot k)^2}\right)^{-\epsilon}\;,
\end{equation}
which has the same structure as $I_2^{\text{aux}}$, introduced in Eq.~(\ref{eq:iaux2}). We can thus take the result of $\langle I^{gg}_3[f_2]\rangle_{\rm av.}$ from Eq.~(\ref{eq:iaux2}) with the replacement $p(x)\to p_3$
\begin{align}
    \braket{I_3^{gg}[f_2]}_{\text{av.}}=\frac{1}{m^2}\left(\frac{b^2}4\right)^{2\epsilon}\frac{\pi^{1-\epsilon}\,2^{-2-2\epsilon}}{\epsilon^2 (1-2\epsilon)}\Gamma(1-2\epsilon)\Gamma(1-\epsilon)\left(1+\frac{p_{3,T}^2}{m^2}\right)^{2\epsilon}\;.
\end{align}
The integral $I^{gg}_4[f_2]$, on the other hand, is not proportional to $I_2^{\text{aux}}(x)$
\begin{equation}
  I^{gg}_4[f_2]=\int d^D k\,  e^{i \vec b \cdot \vec k_T} \frac{(k^2)^{-1-2\epsilon}}{(p_4\cdot k)^2}\left(\frac{m^2}{4 (p_3\cdot k)^2}\right)^{-\epsilon}\;,
\end{equation}
but the structure of $I_2^{\text{aux}}(x)$ can be recovered by applying the generalised Feynman parametrisation of Eq.~(\ref{eq:genfeyn}):
\begin{align}
\label{eq:I4g2_comp}\nonumber
\braket{I_4^{gg}[f_2]}_{\text{av.}}=&-2^{1+2\epsilon}\epsilon\,(1-2\epsilon)\int_0^1 dx\, x\,(1-x)^{-1-2\epsilon}\, \braket{\int d^Dk\,  e^{i \vec b \cdot \vec k_T} \frac{(k^2)^{-1-2\epsilon}}{(p(x)\cdot k)^{2-2\epsilon}}}_{\text{av.}}\\
=&-2^{1+2\epsilon}\epsilon\,(1-2\epsilon) \int_0^1 \frac {dx}{(p^2(x))^{1-\epsilon}}x\,(1-x)^{-1-2\epsilon}\,\braket{I_2^{\text{aux}}(x)}_{\text{av.}}\nonumber\\\nonumber
  =&-\frac{1}{ 4\,\beta m^2\,\epsilon} \pi ^{1-\epsilon} \left(\frac{b^2}4\right)^{2 \epsilon}
  \left(\frac{\tau}2\right)^{1-\epsilon}\Gamma (1-2 \epsilon) \Gamma (1-\epsilon)
 \\&\times
  \int_{-\beta}^{\beta}\,dy\, \left(1+\frac y\beta\right) \left(1-\frac
   y\beta\right)^{-1-2 \epsilon} \left(\frac1{1-y^2}\right)^{1-\epsilon} \left(\frac {B\tau}{1-\tau} \frac{
   y^2}{1-y^2}+1\right)^{2 \epsilon}\;,
\end{align}
where in the last step we used the known result of $\braket{I_2^{\text{aux}}(x)}_{\text{av.}}$ from Eq.~(\ref{eq:iaux2}).
At this stage, the integrand cannot yet be expanded in $\epsilon$, since the integral does not converge in the limit $\epsilon\to0$ due to a singularity in $y=\beta$.
In order to perform the expansion, we need to isolate the singular behaviour and subtract it. 
We consider the following auxiliary integral:
\begin{align}
\label{eq:auxiliary_sing_int}
\nonumber
  \int_{-\beta}^{\beta}\,dy\,&\left(1-y^2\right)^{2 \epsilon -1}\left(1-\frac y\beta\right)^{-2
   \epsilon -1} \left(1+\frac y\beta\right)^{1-2 \epsilon }
   =
   \\\nonumber
   =&2\sqrt\pi \Gamma(-2\epsilon)\frac{\beta}{1-\beta^2}\left(\frac 1{\Gamma\left(\frac 12 -2\epsilon\right)}{}_2F_1\left(\frac 12 , -2\epsilon, \frac 12 -2\epsilon, \beta^2\right)
   \right.\\&\left.
   + \epsilon\frac{1+\beta^2}{\Gamma\left(\frac 32-2\epsilon\right)}{}_2F_1\left(\frac 12 , 1-2\epsilon, \frac 32 -2\epsilon, \beta^2\right)\right)
   \;.
\end{align}
We observe that the integrand in Eq.~(\ref{eq:auxiliary_sing_int}) has the same behaviour as $I_4[f_2]$ in the $y\to \beta$ limit, while having a simpler structure that allows for a straightforward analytic integration.
We can thus add the r.h.s. of Eq.~(\ref{eq:auxiliary_sing_int}) to Eq.~(\ref{eq:I4g2_comp}), while subtracting the l.h.s. at the integrand level in order to obtain a regular expression that can be safely expanded.
The resulting integral can be evaluated separately at each order in $\epsilon$ in terms of multiple polylogarithms.

Let us finally consider the contribution of the function $f_3$.
By following the same procedure already applied for the evaluation of $I^{gg}_{34}[f_3]$ in the mass-independent case, we define the momentum $\ell$ as in Eq.~(\ref{eq:ell}) and by using Eq.~(\ref{eq:f3l}) we have:
\begin{align}
  \label{eq:Iig3_1}
\braket{I^{gg}_3[f_3]}_{\text{av.}}=-\frac {\pi}{v}\braket{\int d^D k\,e^{i \vec b \cdot \vec k_T}\,\frac {(k^2)^{-1-2\epsilon}}{(p_3\cdot k)}\left(\frac{m^2}{(p_3\cdot k)^2}\right)^{-\epsilon}\frac 1{\ell\cdot k}\int_0^1\,d t\,t^{-\frac 12 -\epsilon}(1-t)^{\epsilon}\frac{\chi_{34}}{1-t\,\chi_{34}}}_{\text{av.}}\;,
\end{align}
which, once we substitute in it the definition of $\chi_{34}$ as in Eq.~(\ref{eq:def_chi34}), gives us an expression that only depends explicitly on two momenta, $p_3$ and $\ell$. By applying partial fractioning and defining $\ell_\pm$ as in Eq.~(\ref{eq:ellpm}) we obtain:
\begin{align}
\label{eq:I3f3_comp}\nonumber
  \braket{I_3^{gg}[f_3]}_{\text{av.}}=-\frac {\pi}{v}\,\langle\int d^D k\,e^{i \vec b \cdot \vec k_T}\,\frac{(m^2)^{-\epsilon}}2
  &\int_0^1\,d t\, t^{-1 -\epsilon} (1-t)^{\epsilon}\Bigg(\frac{(k^2)^{-1-2\epsilon}}{(p_3\cdot k)^{1-2\epsilon}(\ell_-\cdot k)}
  \\ &
  -\frac{(k^2)^{-1-2\epsilon}}{(p_3\cdot k)^{1-2\epsilon}(\ell_+\cdot k)}\Bigg)\rangle_{\text{av.}}\;.
\end{align}
By applying the generalisation of Feynman parametrisation introduced in Eq.~(\ref{eq:genfeyn}) we can reduce the dependence of the denominators of the integrand to a single momentum, and retrieve the structure of $I_2^{\text{aux}}(x)$ as defined in Eq.~(\ref{eq:iaux2}).
The leftover integral over the Feynman parameter $x$ and the variable $t$ can be computed in terms of multiple polylogarithms with a standard procedure.

The evaluation of $I^{gg}_4[f_3]$ can be performed by following the same steps, 
but the differences in the integrand make the procedure of partial fractioning a bit more involved.
After introducing the momentum $\ell$ 
and applying partial fractioning for a first time, we obtain an expression similar to Eq.~(\ref{eq:I3f3_comp}):
\begin{align}\nonumber
  \braket{I^{gg}_4[f_3]}_\text{av.}=-\frac {\pi}{v}\,&\langle\int d^D k\,e^{i \vec b \cdot \vec k_T}\,\frac{(m^2)^{-\epsilon}}{2 (p_4\cdot k)^2}\\
  &\int_0^1\,d t\, t^{-1 -\epsilon} (1-t)^{\epsilon}\left(\frac{(k^2)^{-1-2\epsilon}}{(p_3\cdot k)^{-1-2\epsilon}(\ell_-\cdot k)}-\frac{(k^2)^{-1-2\epsilon}}{(p_3\cdot k)^{-1-2\epsilon}(\ell_+\cdot k)}\right)\rangle_\text{av.}\;,
\end{align}
but, in this case, we can not yet apply Feynman parametrisation, since each denominator involves three different products of the momenta.
We can circumvent this problem by performing an additional partial fractioning:
\begin{align}
\label{eq:I4f3comp}\nonumber
  \braket{I^{gg}_4[f_3]}_\text{av.}=&-\frac \pi{v^2} (m^2)^{-\epsilon}\langle\int d^D k\,e^{i \vec b \cdot \vec k_T}\,\int_0^1\,d t\, \frac{t^{-\frac 12-\epsilon}\, (1-t)^\epsilon}{\left(1-\frac t{v^2}\right)}\left\{
  \frac{(k^2)^{-1-2\epsilon}}{(p_3\cdot k)^{-2\epsilon}(p_4\cdot k)^2}
  \right.\\\nonumber&\left.
  +\frac{m^2}{p_3\cdot p_4}
  \left[
  -\frac{\left(1+\frac t{v^2}\right)}{\left(1-\frac t{v^2}\right)}\frac{(k^2)^{-1-2\epsilon}}{(p_3\cdot k)^{1-2\epsilon}(p_4\cdot k)}
  +\frac{m^2}{2p_3\cdot p_4}\,\frac{\sqrt t}{v}
  \right.\right.
  \\&\left.\left.
  \times\left(\frac{\left(1+\frac t{v^2}\right)}{\left(1-\frac t{v^2}\right)}\frac{(k^2)^{-1-2\epsilon}}{(p_3\cdot k)^{1-2\epsilon}(\ell_-\cdot k)}-\frac{\left(1-\frac t{v^2}\right)}{\left(1+\frac t{v^2}\right)}\frac{(k^2)^{-1-2\epsilon}}{(p_3\cdot k)^{1-2\epsilon}(\ell_+\cdot k)}\right)\right]\right\}\rangle_\text{av.}\;.
\end{align}
It is now possible to apply Feynman parametrisation to Eq.~(\ref{eq:I4f3comp}), obtaining integrals over the momentum $k$ that can be written in terms of $I_2^{\text{aux}}$.
By using the known result of $\braket{I_2^{\text{aux}}}$ provided in Eq.~(\ref{eq:iaux2}) to perform the integration over $k$, we are left with the final two integrals over the Feynman parameter and the variable $t$, that can be performed with a standard procedure.

Let us finally analyse the regular part of $\g_{34}^{gg}(\vec n_3\cdot \vec n_4, \vec n_3^2, \vec n_4^2)$, that can be safely be expanded in $\epsilon$.
We need to evaluate the following integral:
\begin{align}
\langle{\sum_{j=3,4}\int d^D k\,  e^{i \vec b \cdot \vec k_T}\, \frac{(k^2)^{-1-\epsilon}}{\pjk^2}
  (\rho_{34}^{m\neq 0}+\rho_{43}^{m\neq 0})}\rangle_{\text{av.}}\;.
\end{align}
The explicit all-orders expression of the integrand reads:
\begingroup\allowdisplaybreaks
\begin{align}\nonumber
    \frac 1\pi\sum_{j=3,4}\frac{1}{p_j\cdot k}&\left( \rho_{34}^{m\neq0}+ \rho_{43}^{m\neq0}\right)=
    (1-\vec n_3^2)\left(-\frac{2-5\epsilon}{\epsilon(1-2\epsilon)}-\frac1{1-2\epsilon}\frac{\vec n_3\cdot \vec n_4}{\vec n_3^2}\right)
    \\\nonumber&
    +(2-\vec n_3^2-\vec n_4^2)\left(\frac 1\epsilon +\frac{\Gamma(1-2\epsilon)\Gamma(\epsilon)}{\Gamma(1-\epsilon)}\left(\frac{1-\vec n_3^2}4\right)^{-\epsilon}\left(\frac{D_{34}}{n_3^{1/2-\epsilon}}-1\right)\right)
    \\\nonumber&
    +\frac{1}{\sqrt{\pi}}\Gamma\left(\frac 12-\epsilon\right)\Gamma\left(\epsilon\right)(1-\vec n_3^2)^{1-\epsilon}\left(\frac{\vec n_3^2+\vec n_3\cdot\vec n_4}{2(\vec n_3^2)^{3/2-\epsilon}}-\frac{6\vec n_3^2}{2(\vec n_3^2)^{3/2-\epsilon}}+2\right)
    \\\nonumber&
    +\frac 1\epsilon\left(3(1-\vec n_3^2)+D_{34}(2-\vec n_3^2-\vec n_4^2)\right){}_2F_1\left(\frac 12 ,1,1+\epsilon;1-\vec n_3^2\right)
    \\\nonumber&
    - \frac{1-\sqrt{\vec n_3^2}}{\epsilon\vec n_3^2}(\vec n_3^2+\vec n_3\cdot \vec n_4){}_2F_1\left(1,1-\epsilon,1+\epsilon;\frac2{1+\sqrt{n_3^2}}-1\right)
    \\\nonumber&
    -\frac{1-\vec n_3\cdot \vec n_4}{\vec n_3^2-\vec n_3\cdot \vec n_4}(2-\vec n_3^2-\vec n_4^2)\chi_{34}
    \int_0^1 \,du\,\frac{(1-u)^{-\frac 12 -\epsilon}}{\sqrt{1-\chi_{34}u}}[(1+u\psi)^\epsilon-u^\epsilon(1+\psi)^\epsilon]
    \\&
    +D_{34}\frac{\gamma}{1-\gamma}(2-\vec n_3^2-\vec n_4^2)\int_0^1\,du\,\frac{(1-u)^{\frac 12-\epsilon}}{\sqrt{1-\vec n_3^2 u}}\left(1+u \frac \gamma{1-\gamma}\right)^{-1}\;,
\end{align}
\endgroup
where the variable $\chi_{34}$ has been defined in Eq.~(\ref{eq:def_chi34}) while $D_{34}$, $\gamma$ and $\psi$ are given in Eqs.~(\ref{eq: def chi})--(\ref{eq: def psi}).
We introduce the following notation for the $\epsilon$-expansion of the integrand:
\begin{equation}
\rho_{ij}^{m\neq0}
=
\rho_{ij}^{m\neq0,\,(0)}
+
\epsilon\, \rho_{ij}^{m\neq0,\,(1)}
+{\cal O}(\epsilon^2)\;.
\end{equation}

As for the case of the mass-independent contribution, due to the complexity of the functions involved in the integrand, we perform numerically the last steps of this computation.
We follow the same steps as for the integration of $\rho_{34}^{m=0}$, by considering the $q_T$-space representation of the integral and by fixing the azimuthal angle $\phi$ such that $\vec{p}_{T,3}\cdot \vec q_T=p_{T,3}\, q_T \cos\phi$.
After switching to the dimensionless variables $x$, $y$ already introduced in Eq.~(\ref{eq: def adim xy}) and inserting an integral over the virtuality of the momentum, we obtain:
\begin{align}
\label{eq: def massive regulars}\nonumber
\langle\sum_{j=3,4}\int d^D k \, &\delta^{(D-2)}(\vec k_T-\vec q_T)  \frac{(k^2)^{-1-\epsilon} }{(p_j\cdot k)^2}(\rho_{34}^{m=0,\,(0)} + \epsilon\, \rho_{34}^{m=0,\,(1)})\rangle_{\text{av.}}=
\\\nonumber
=&\frac{q_T^{-1-\epsilon}}{B(\frac 12, \frac 12-\epsilon)}
\int_{-1}^1 \,d\cos\phi\,\int_0^\infty dx\,dy\,\frac{1}{2x^2 y}(1-\cos^2\phi)^{-1/2}
\\&
\times\sum_{j=3,4}\frac{1}{p_j\cdot k}
\Big[
\rho_{34}^{m\neq0,\,(0)} + \epsilon\, \rho_{34}^{m\neq0,\,(1)}
- \epsilon\ln\left[x(1-\cos^2\phi)\right] \rho_{34}^{m\neq0,\,(0)}
\Big]
\;.
\end{align}
We can observe that also in this case the integral of the first two terms in the square bracket only depends on $\beta$ and can be thus evaluated on a one-dimensional grid.
We can also reduce the $3$-fold integrals of Eq.~(\ref{eq: def massive regulars}) in $2$-fold ones by replacing the exponential by a $\theta$-function in the corresponding $b$-space representation, in a similar fashion as it was done in Eq.~(\ref{eq:numeric2}).
Adapting it to the present functions we obtain:
\begin{align}\nonumber
&\frac{1}{\pi}\int_{-1}^1 d\cos \phi \int_0^\infty dx \, dy \,
\frac{1}{2\, x^2\, y} x^{-1-\epsilon} (1-\cos^2\phi)^{-1/2-\epsilon}
\rho(\vec n_3\cdot \vec n_4,\vec n_3^2     ,\vec n_4^2  )  \;
\\&
=\int_0^1 dt \int_{-1}^1 d\cos\phi
\frac{t^2}{1-t^2}
\, \rho \left(
1-\frac{1-t^2}{1-v t \cos\phi},
t^2,
1-(1-v^2)\frac{1-t^2}{(1-v t \cos\phi)^2}
\right)\;.
\end{align}

We have now summarized in detail the technical aspects of our calculation, and presented explicit partial results in the cases the expressions were obtained in a compact analytic form. Our complete final results are collected and implemented in a numerical code, which is described in the next Section.

\section{Numerical results}
\label{sec:code}

\begin{figure}
\begin{center}
\includegraphics[width=0.66\textwidth]{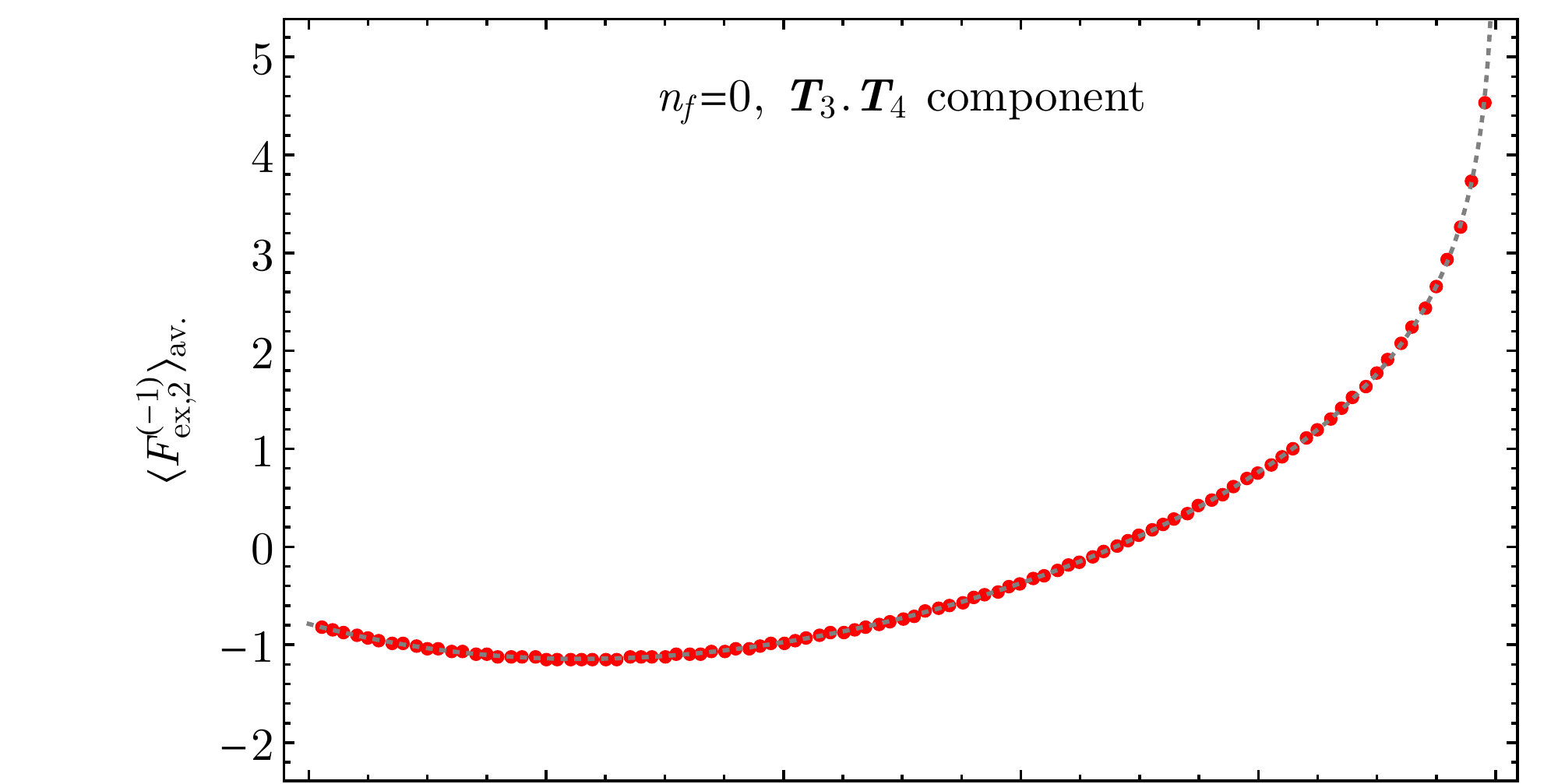}
\\
\includegraphics[width=0.66\textwidth]{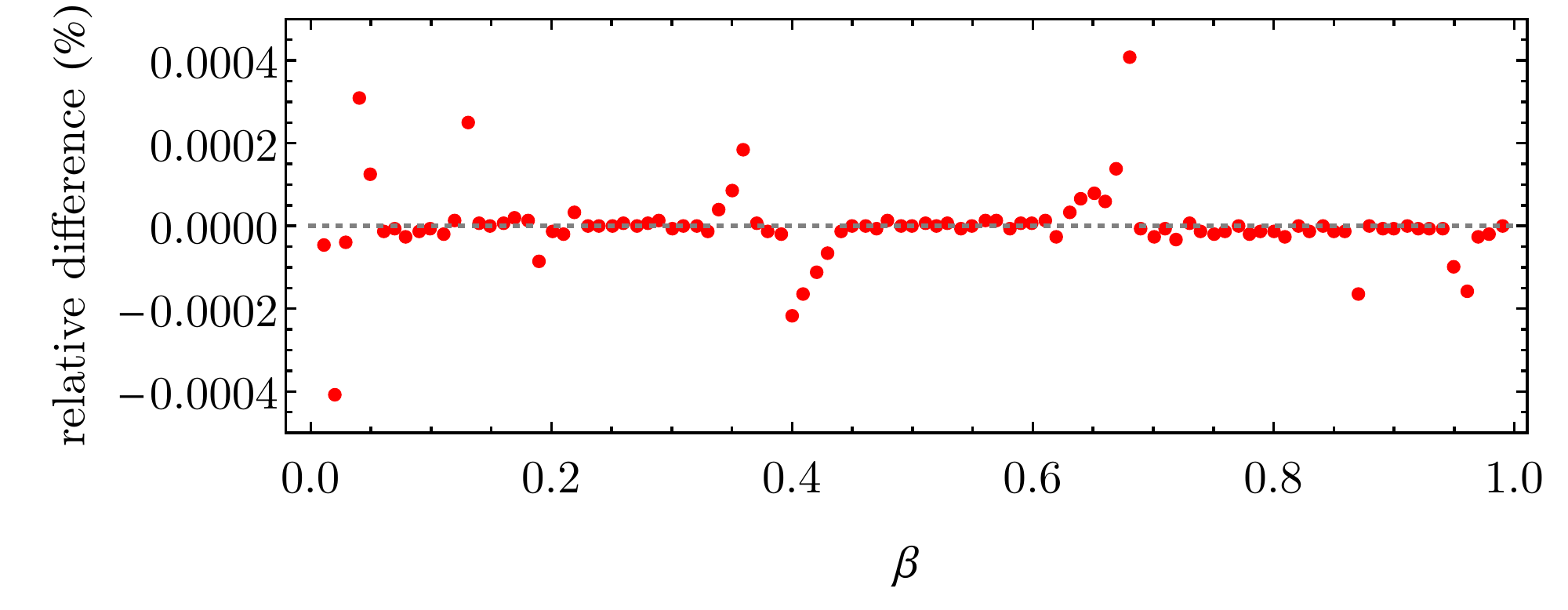}
\end{center}
\vspace*{-0.7 cm}
\caption{Numerical results for the coefficient of the $1/\epsilon$ pole of the contribution proportional to ${\mathbf T}_3 \cdot {\mathbf T}_4$ in ${\mathbf F}_{\text{ex},2}$ (red points), compared to the expected analytical result (gray curve) from Eq.~(\ref{eq: Fex21 check}). The lower plot shows the relative difference (in percentage) between the two.}
  \label{fig:pole_Fex2}
\end{figure}

In Sect.~\ref{sec:resum+soft} we described in detail the ingredients entering the transverse-momentum resummation formalism for heavy-quark production.
To the purpose of the application to the $q_T$-subtraction framework, the key role is played by the coefficient ${\cal H}^{Q{\bar Q}}$ defined in Eq.~(\ref{eq:HQQ}), which depends on the subtracted matrix element $\widetilde{\cal M}$ via the master formula (\ref{eq:HD}), while $\widetilde{\cal M}$ can be obtained through Eq.~(\ref{eq:mtilde new}).
All the ingredients entering in these equations are finite, since the cancellation of the IR poles has been carried out at the operator level as described in Eq.~(\ref{eq:KK}).
The cancellation is guaranteed by the relation with the subtracted soft anomalous dimension $\mathbf{\Gamma}_{\rm sub}$ in Eqs.~(\ref{eq: Fex1 check})--(\ref{eq: Fex21 check}): we were able to verify analytically this cancellation for all the contributions, with the exception of the $n_f$-independent part of the term proportional to the colour factor ${\mathbf T}_3 \cdot {\mathbf T}_4$.
This term depends only on the variable $\beta$.
As described in Section~\ref{sec:S34}, part of this term was evaluated numerically, and, therefore, only a numerical check of the cancellation is possible.
In Figure~\ref{fig:pole_Fex2} we compare the coefficient of the $1/\epsilon$ pole as a function of $\beta$ computed analytically with Eq.~(\ref{eq: Fex21 check}) against our numerical result.
The lower plot shows the relative difference between the two:
The relative difference is below the $0.0005\%$ in all the regions of the phase-space, showing a perfect agreement with the prediction and providing a strong cross-check of our computation.

Having discussed the cancellation of the IR singularities, we now
consider the ingredients needed for the implementation of the function ${\cal H}^{Q{\bar Q}}$.
In the $q_T$-subtraction formalism, a final average over the azimuthal degree of freedom of $\vec b$ is required (see Eq.~(\ref{eq:HQQ})), and since the operator $\mathbf{D}$ is defined in such a way that $\braket{\mathbf{D}}_\text{av.}=1$, it gives no contribution in our computation, except when interfering with the azimuthally dependent part of the $C$ coefficients in Eq.~(\ref{eq:HQQ}). Therefore, as a new perturbative ingredient at NNLO we just need to evaluate the subtracted amplitude
$\widetilde{\cal M}$ through Eq.~(\ref{eq:mtilde new}) at second order.
At this order the subtracted amplitude $\mathbf{Z}^{-1}\ket{\cal M}$ appearing in Eq.~(\ref{eq:mtilde new}) is provided by the numerical grids in Ref.~\cite{Baernreuther:2013caa}. The operator $e^{V_c^{\text{fin}}}$ at the same order is already known from the implementation of $q_T$-subtraction for a colourless final state at NNLO \cite{Catani:2013tia} (see Eq.~(\ref{eq:Vcfin})).

We are left with the coefficient $\mathbf h$, whose perturbative expansion is given in Eq.~(\ref{eq:h}).
The term involving the commutator produces contributions proportional to three-parton correlators, which vanish when evaluated on the Born $c{\bar c}\to Q{\bar Q}$ amplitude.
Therefore we can simply write
\begin{align}
    \label{eq:h_final}\nonumber
    &\mathbf{h}(\as)=
    1+\frac{\as}{2\pi}\braket{\mathbf{F}_{{\rm ex},1}^{(0)}}_{\rm av.}
    +\left(\frac{\as}{2\pi}\right)^2\Big\{\braket{(\mathbf{F}_{{\rm ex},1}^{(0)})^2}_{\rm av.}-\frac 12 \left(\braket{\mathbf{F}_{{\rm ex},1}^{(0)}}_{\rm av.}\right)^2
    \\&\phantom{\mathbf{h}(\as)=}
    +\braket{\mathbf{F}_{{\rm ex},2}^{(0)}}_{\rm av.}
    -2\pi\beta_0\braket{\mathbf{F}_{{\rm ex},1}^{(1)}}_{\rm av.}
    \Big\}+\mathcal{O}(\as^3)\, .
\end{align}
The two last terms in the ${\cal O}(\as^2)$ contribution involve the product of two colour charges; we have chosen to write the results in the numerical implementation in terms of the colour structures ${\mathbf T}_3 \cdot {\mathbf T}_4$ and ${\mathbf T}_i \cdot {\mathbf T}_j$ with $i = 1, 2$ and $j = 3, 4$.

The results for the ${\mathbf T}_i \cdot {\mathbf T}_j$ structure are obtained in a fully analytical way, and the explicit expression, which can be obtained from the results in the previous Sections, is implemented in the numerical code.
The results corresponding to the colour structure ${\mathbf T}_3 \cdot {\mathbf T}_4$ have contributions from the integral of the soft correlators ${\cal S}_{34}^{m=0}$ and ${\cal S}_{34}^{m\neq 0}$ of Eq.~(\ref{eq:sij0}) and (\ref{eq:sijm}), which are partially obtained numerically in the form of a two-dimensional grid.

The numerical integration is performed using the implementation of global adaptive strategies available in {\sc Mathematica}.
For the terms independent of $\theta$, the integral is evaluated for a grid in the variable $\beta$ from 0 to 1, in steps of $0.001$ in the range $(0;0.8)$ and a smaller step of $0.0001$ in the high-energy region $(0.8;1)$, in which the variation of the function is larger.
For the remaining term, which depends both on $\beta$ and $\cos\theta$, the integral is evaluated for a total number of 5000 phase-space points in the range $\beta \in (0;1)$, $\cos\theta \in (0;1)$ (the result is symmetric under the exchange $\cos\theta\to -\cos\theta$), which were obtained from the NNLO parton level generator {\sc Matrix} \cite{Grazzini:2017mhc} after the optimisation for the integration of the LO $t\bar t$ cross section.

Given that a numerical interpolation of the grid is already needed, and also due to the fact that the numerical evaluation of the analytical terms entering the ${\mathbf T}_3 \cdot {\mathbf T}_4$ structure of $\langle {\mathbf F}_{\text{ex},2}^{(0)} \rangle_{\text{av.}}$ is computationally very expensive, we decided to encode all the contributions proportional to ${\mathbf T}_3 \cdot {\mathbf T}_4$ in the terms $\langle {\mathbf F}_{\text{ex},2}^{(0)} \rangle_{\text{av.}}
-2 \pi \beta_0\, \langle {\mathbf F}_{\text{ex},1}^{(1)} \rangle_{\text{av.}}$ in a two-dimensional grid composed by the same phase-space points used for the numerical integration of the aforementioned pieces of ${\cal S}_{34}^{m=0}$ and ${\cal S}_{34}^{m\neq 0}$.
The different pieces entering the final result are defined in three independent grids and combined afterwards, in order to have a fully flexible
implementation in the number of light-quark flavours $n_f$.
The numerical evaluation of the multiple polylogarithms appearing in some of our analytic expressions, needed for the construction of the grids, is performed using {\sc GiNaC}~\cite{Bauer:2000cp,Vollinga:2004sn}.

In addition to the contributions described above, the result for the azimuthal average of the square of the NLO result, i.e. the term $\langle ({\mathbf F}_{\text{ex},1}^{(0)})^2 \rangle_{\text{av.}}$, is also obtained numerically, by simply starting from the known result for ${\mathbf F}_{\text{ex},1}^{(0)}$ and computing the azimuthal average of its square, again in the same set of phase-space points used before.
In this case, the results are grouped in three different colour structures, $({\mathbf T}_3 \cdot {\mathbf T}_4)^2$, $C_F\; {\mathbf T}_3 \cdot {\mathbf T}_4$ and $C_F^2$.

The grids described above are afterwards fitted using a spline approximation~\cite{dierckx1995curve}.
Given that we do not expect our results for each phase-space point to have a large deviation from the correct value, as the uncertainties of the numerical integration are at the per mille level, the parameters of the spline fitting are chosen such that the fit is very close to the original points.
In addition, and in order to improve the quality of the fit, the grids are divided by appropriate factors depending on $\beta$ and $\cos\theta$ before performing the fit, which were checked to generate surfaces with smaller variations and therefore easier to fit.
A concrete example of this procedure is given by the way to handle the threshold region: all the grids had a divergent logarithmic behaviour in the $\beta\to 1$ limit.
We thus divided all the points by a factor $(1+\log^2(1-\beta^2)^n)$, with the value of $n$ chosen in order to get a regular grid in such limit, and this factor was added back after the fitting.
Also, in order to work with more evenly distributed points, we worked with the variables $\beta^2$ (instead of $\beta$) and $\cos\theta$.

We have studied the self-consistency of the fit by comparing the results obtained with it to the original values on the grids used to construct it.
We observed that, for the majority of the points ($93.9\%$), the difference is below the per mille level, while the points that agree better than $1\%$ almost cover the full phase-space ($98.9\%$).
The largest relative differences show up in the grids corresponding to $\langle ({\mathbf F}_{\text{ex},1}^{(0)})^2 \rangle_{\text{av.}}$, in the regions in which simultaneously $\beta$ and $|\cos\theta|$ are close to $1$, the reason being the sudden variation of the fitted function in that area, and its value being very close to zero.

In order to see if the error coming from the fitting of the grids has an impact in the computation of a physical quantity, we checked the difference between the original grid and the fit once combined with all the other ingredients entering the coefficient ${\cal H}^{Q{\bar Q}}$.
This involves, among other things, the evaluation of lower-order (colour-correlated) matrix elements, the finite part of the two-loop amplitudes, plus all the soft contributions that were obtained and encoded analytically.
We performed this check for the specific case of top-quark pair production, using {\sc OpenLoops}~\cite{Buccioni:2019sur} for the evaluation of tree-level and one-loop amplitudes, and the results from Ref.~\cite{Baernreuther:2013caa} for the two-loop corrections.
We observed that the point-wise difference is always below $0.25\%$, and that, from the total of points, only a handful of them present a deviation larger than $0.05\%$, indicating that the accuracy obtained through the fit is more than enough to reproduce the original results.

The checks described above only tested the accuracy of the fit on the very same points used to generate it: it is also important to perform some checks on the rest of the phase-space.
To this end, we reduced the number of points used to perform the fit by a factor of $2$ and checked how the accuracy of the final result is affected, finding results similar to the ones described above, thereby confirming the reliability of our implementation.

\begin{figure}[t]
\begin{center}
\includegraphics[width=0.49\textwidth]{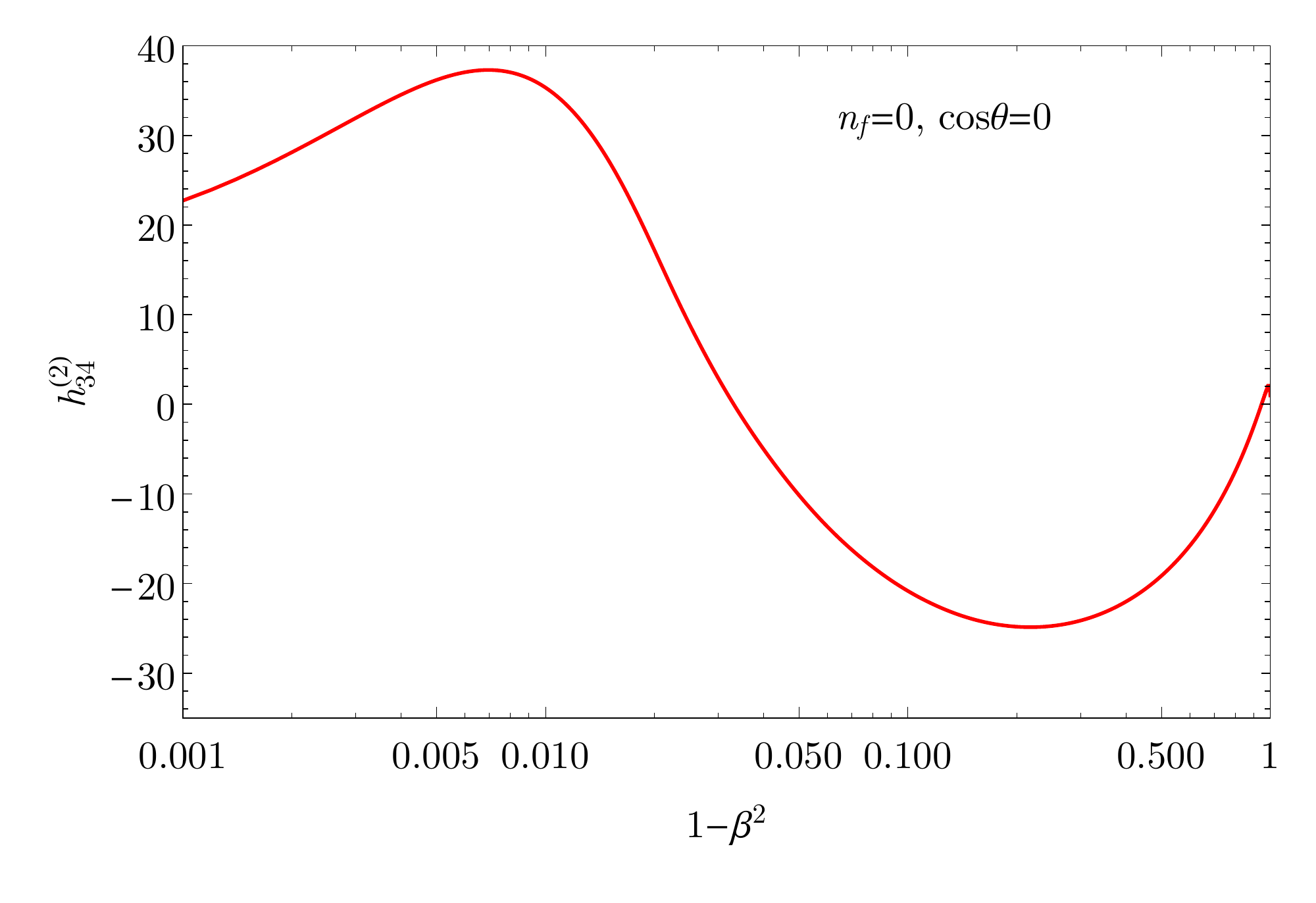}
\includegraphics[width=0.49\textwidth]{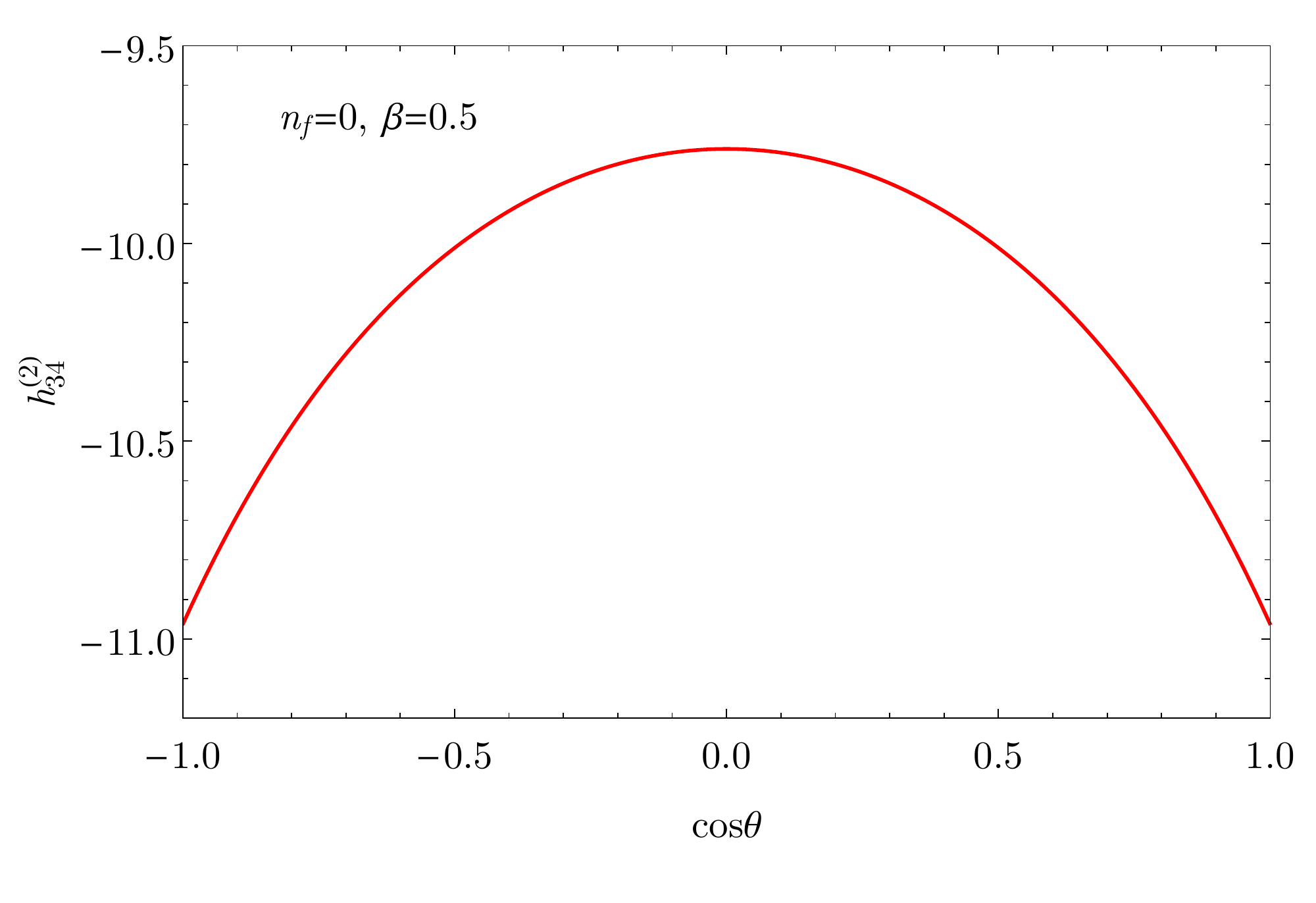}\\
\includegraphics[width=0.49\textwidth]{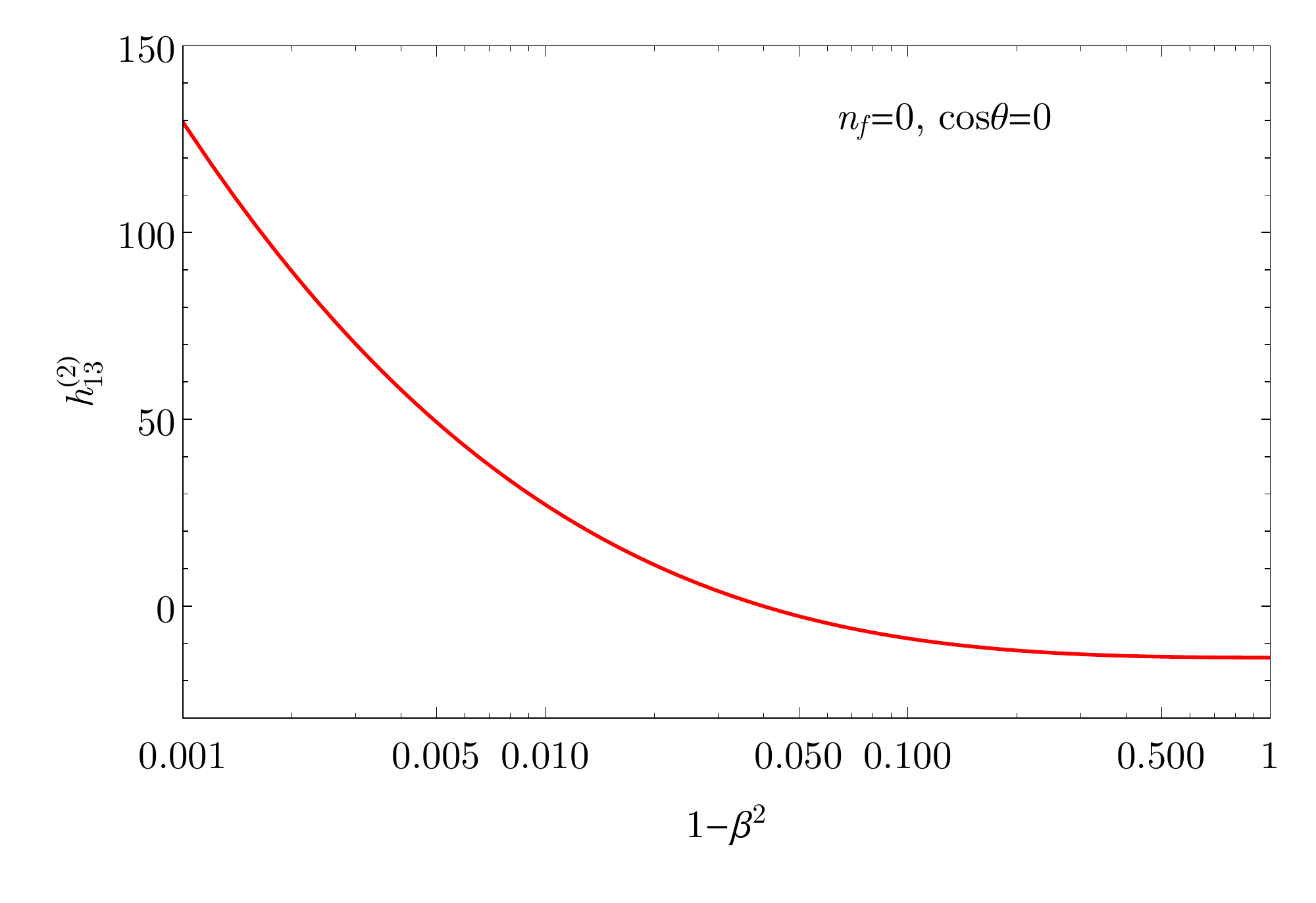}
\includegraphics[width=0.49\textwidth]{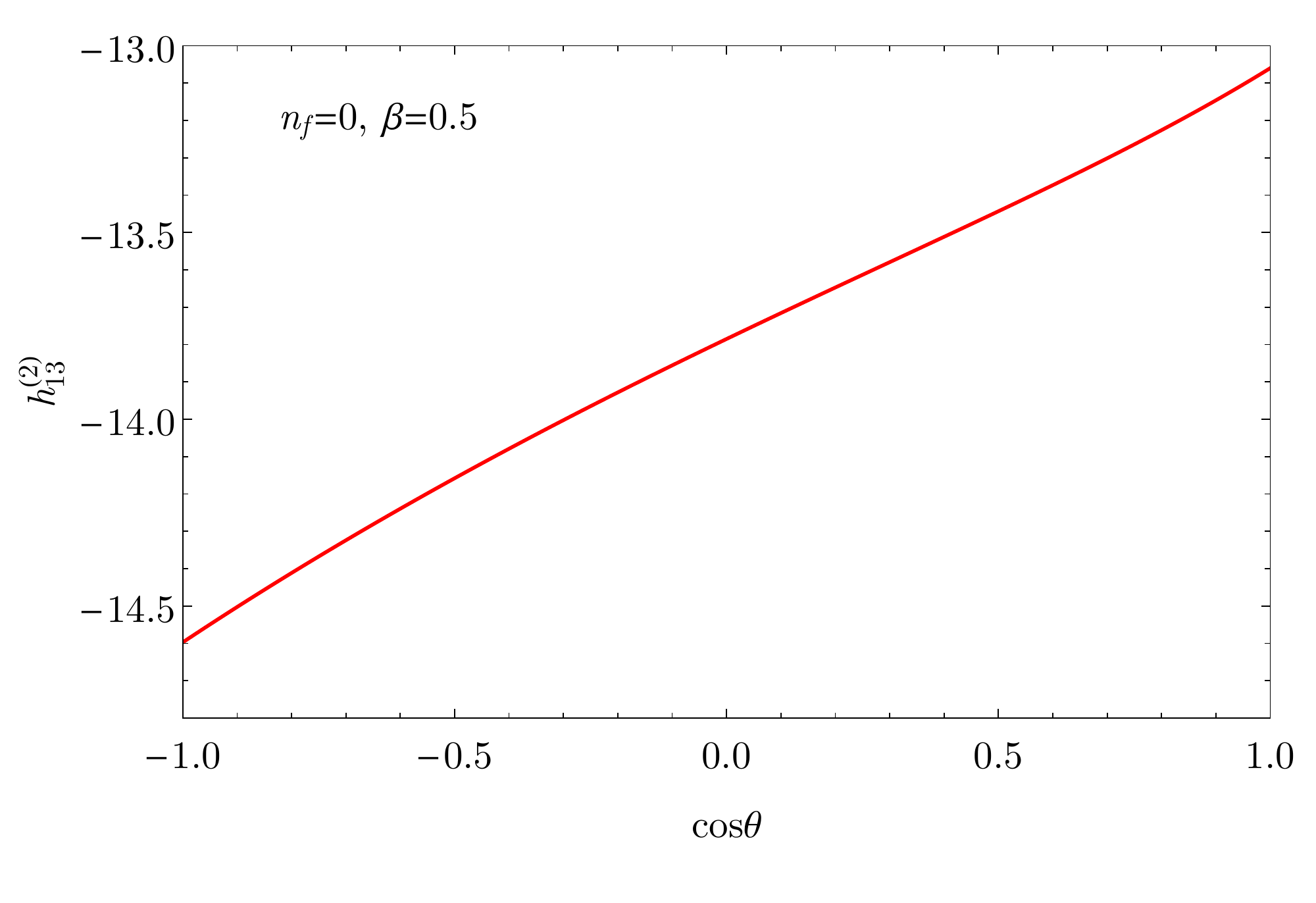}
\end{center}
\vspace*{-0.5 cm}
\caption{Contributions to the second order coefficient of $\mathbf h$ proportional to ${\mathbf T}_3 \cdot {\mathbf T}_4$ (upper panels) and ${\mathbf T}_1 \cdot {\mathbf T}_3$ (lower panels). The results correspond to $n_f=0$. The left panels show the $\beta$ dependence for a fixed value of $\cos\theta=0$, while the $\cos\theta$ dependence is shown in the right panels for $\beta=0.5$.}
  \label{fig:h2_TT}
\end{figure}

\begin{figure}
\begin{center}
\includegraphics[width=0.49\textwidth]{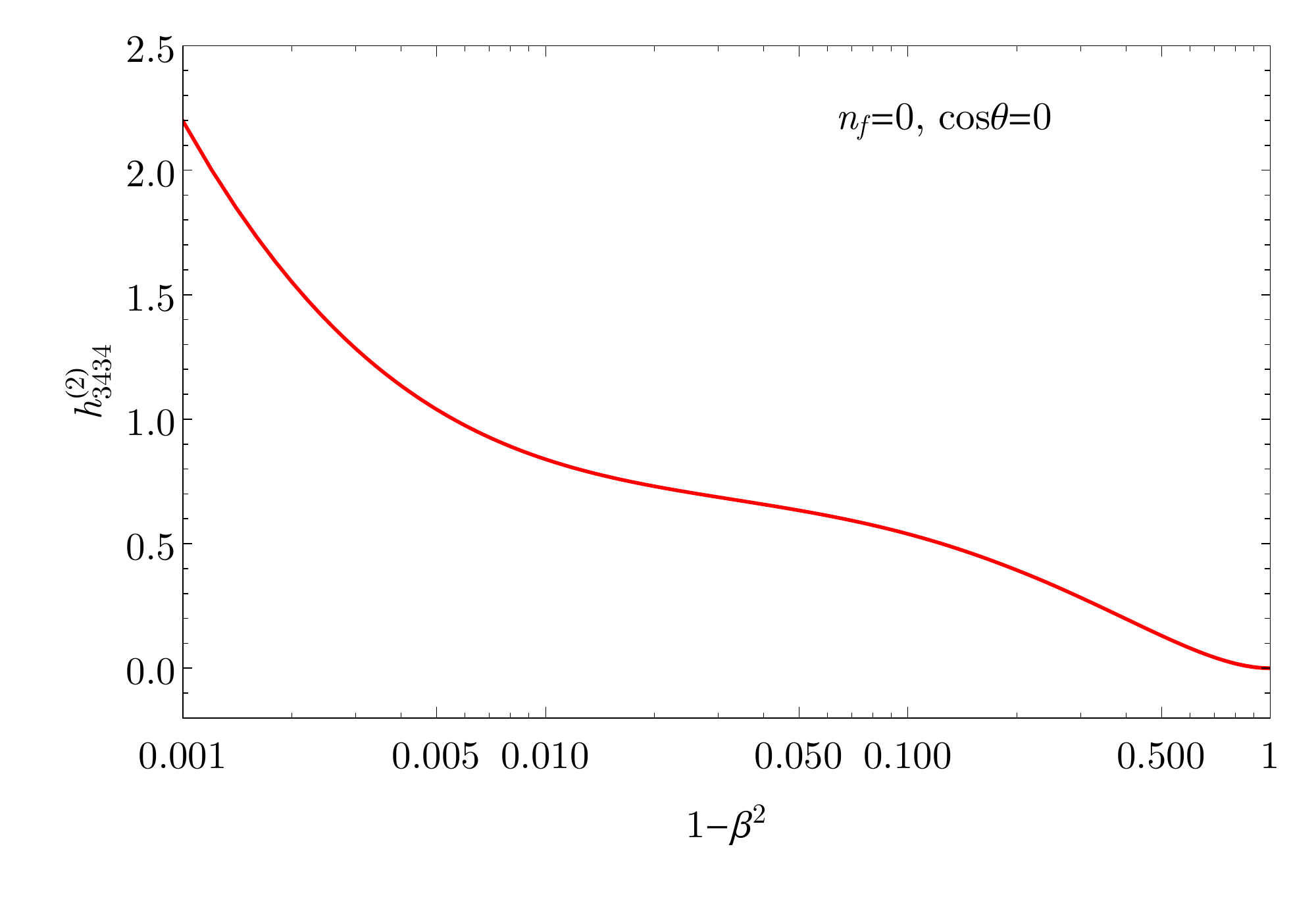}
\includegraphics[width=0.49\textwidth]{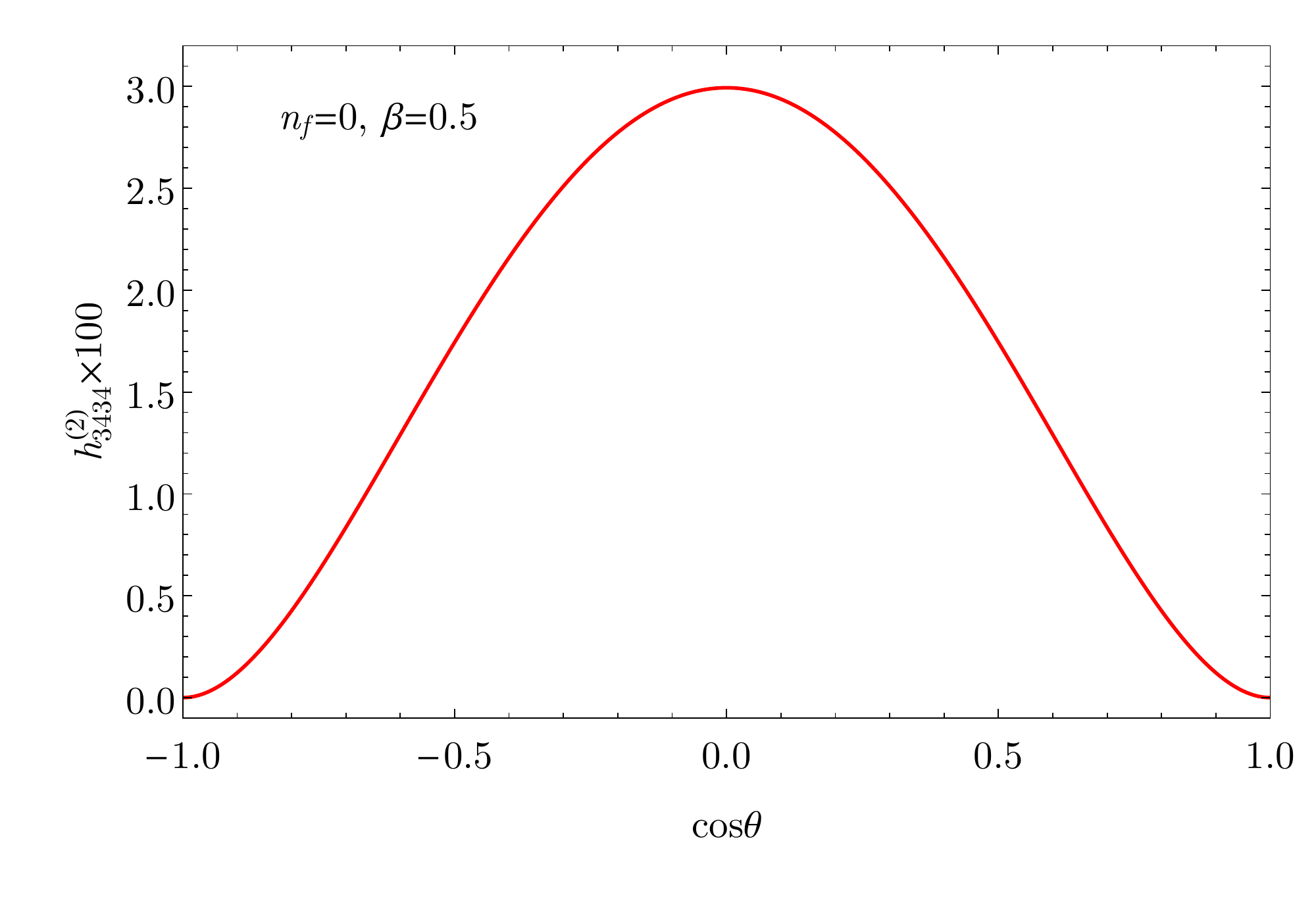}\\
\includegraphics[width=0.49\textwidth]{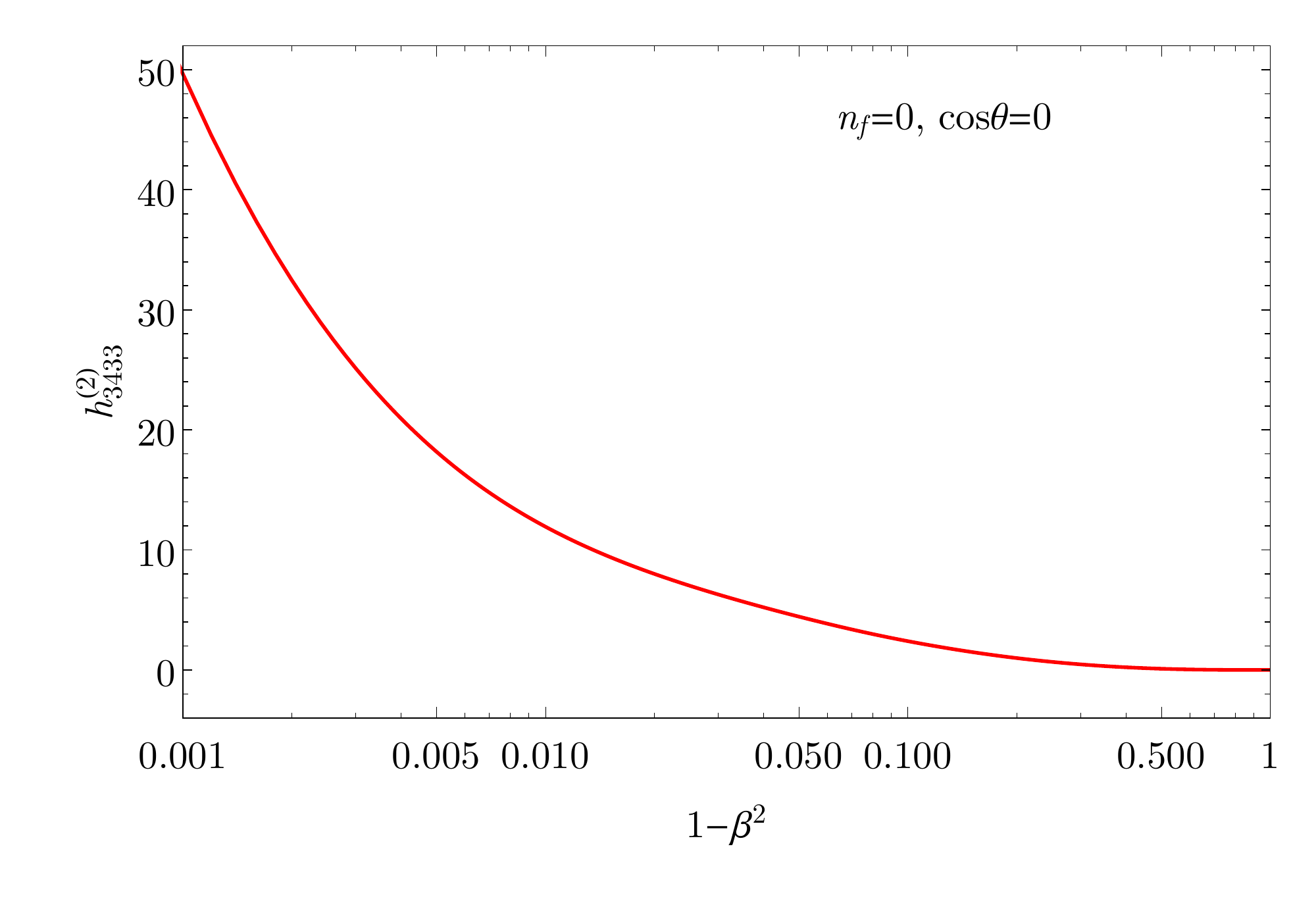}
\includegraphics[width=0.49\textwidth]{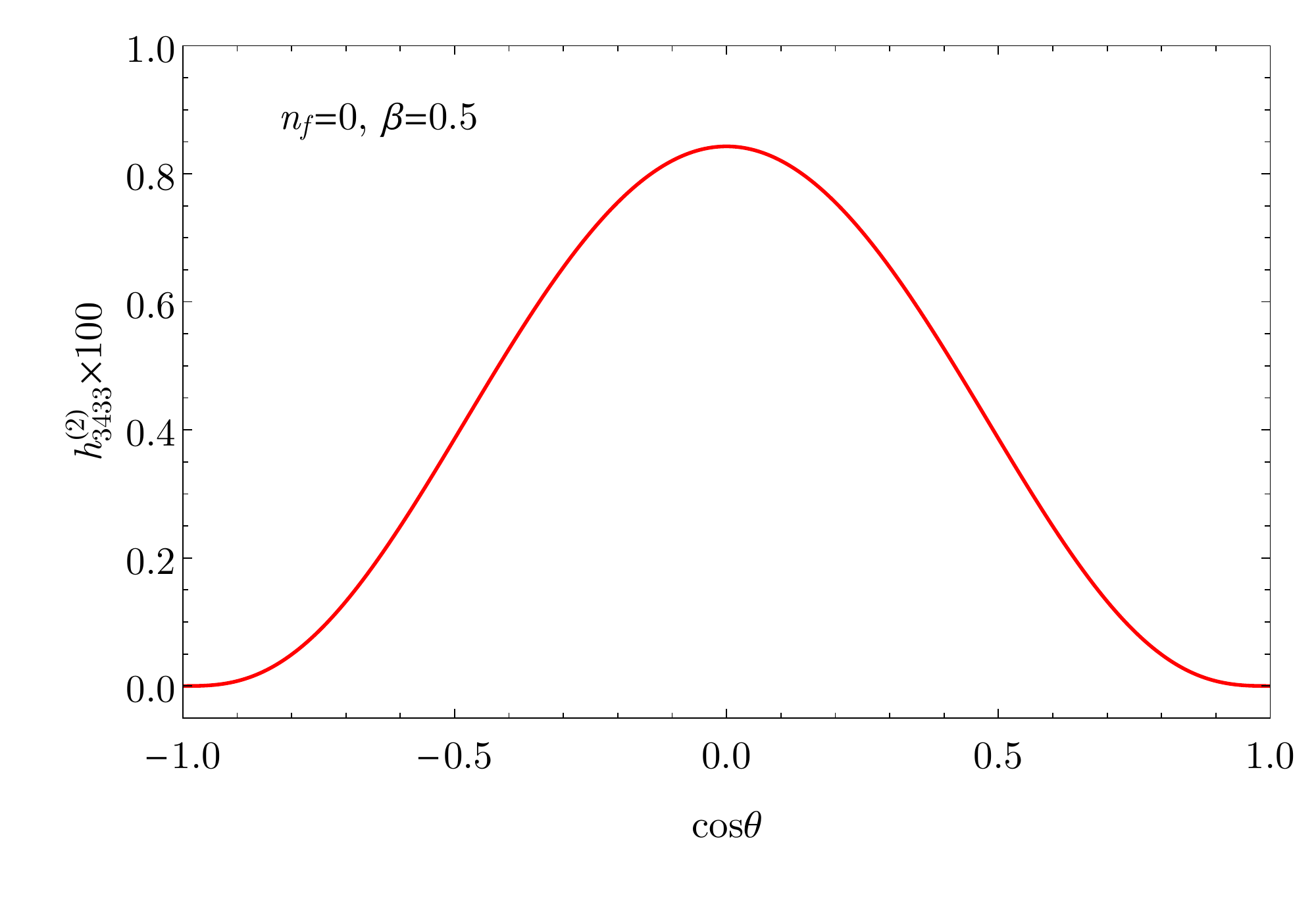}\\
\includegraphics[width=0.49\textwidth]{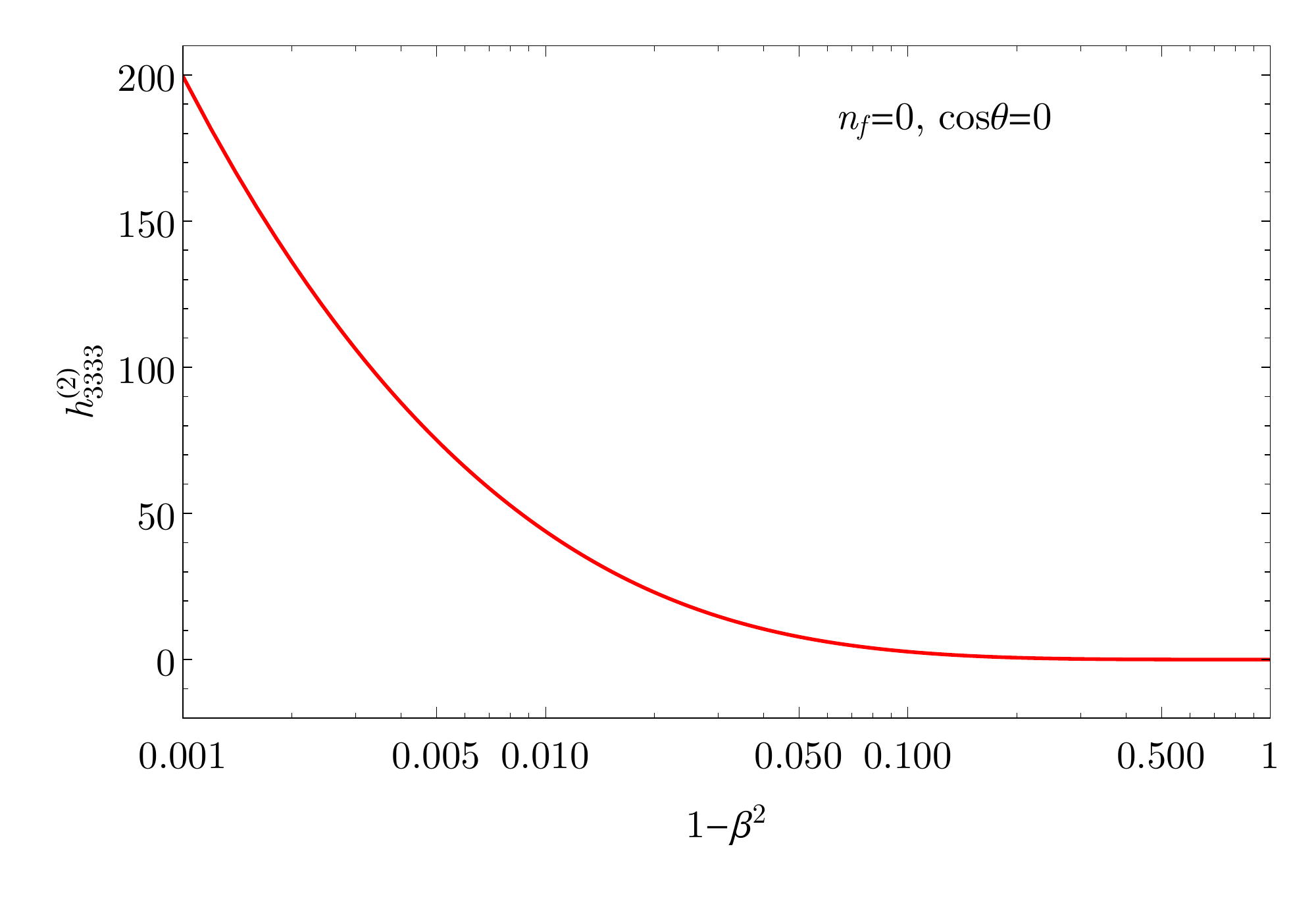}
\includegraphics[width=0.49\textwidth]{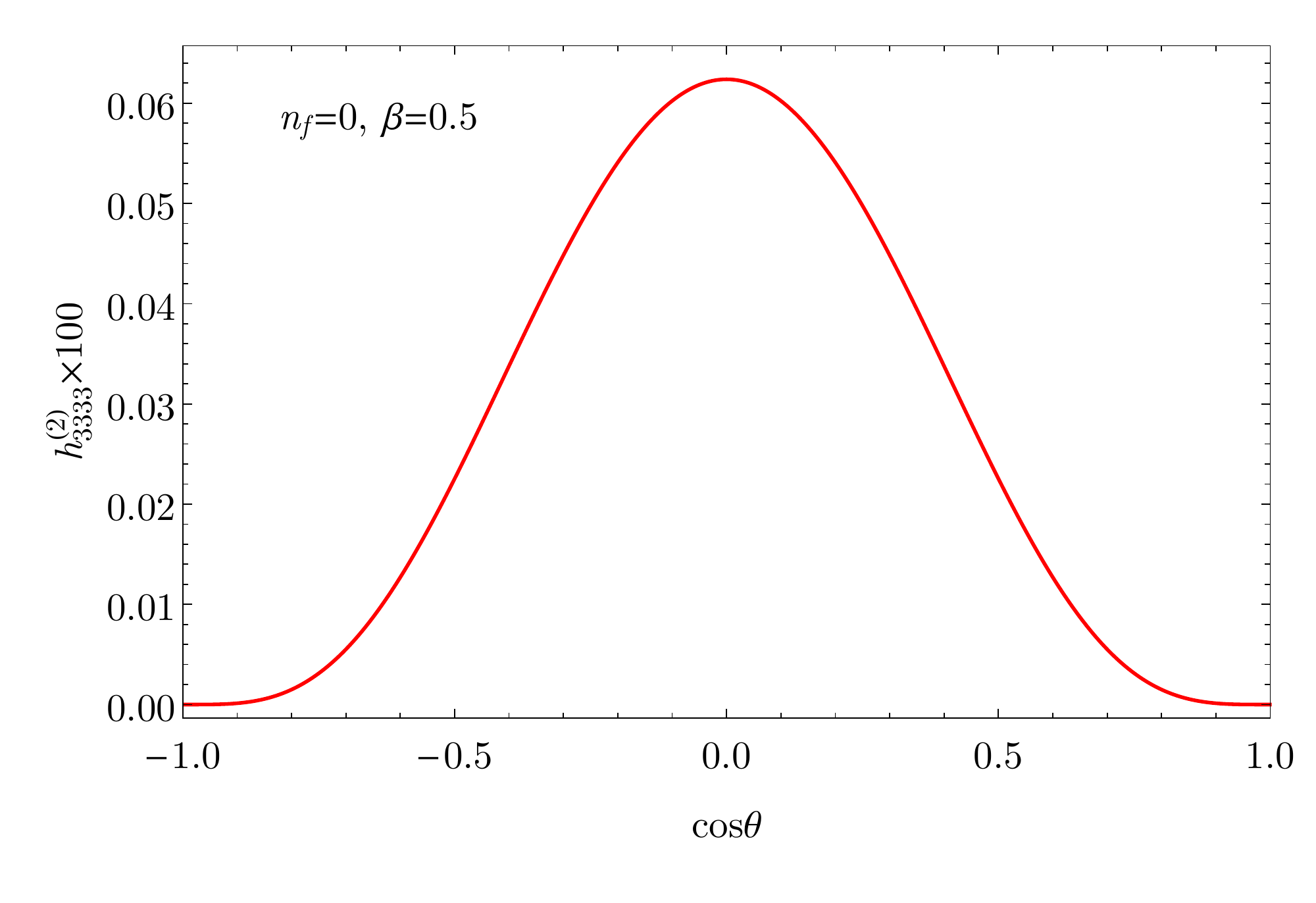}
\end{center}
\vspace*{-0.5 cm}
\caption{Same as in Fig.~\ref{fig:h2_TT} for the contributions proportional to ${\mathbf T}_3 \cdot {\mathbf T}_4 \; {\mathbf T}_3 \cdot {\mathbf T}_4$ (upper panels), $C_F\, {\mathbf T}_3 \cdot {\mathbf T}_4$ (middle panels) and $C_F^2$ (lower panels).}
  \label{fig:h2_TTTT}
\end{figure}

We illustrate our final results in Figs.~\ref{fig:h2_TT} and \ref{fig:h2_TTTT}. 
As described in the text, we split our results into the different colour structures appearing in $\mathbf{h}$, specifically
\begin{align}\label{eq:final}\nonumber
    \mathbf{h}(\as)\, =& \,1 + \frac{\as}{2\pi} \left( h^{(1)}_{34}\, {\mathbf T}_3 \cdot {\mathbf T}_4 + h^{(1)}_{33}\, C_F \right) \\[1ex]
     &+ \left(\frac{\as}{2\pi}\right)^2 \Big( h^{(2)}_{34}\, {\mathbf T}_3 \cdot {\mathbf T}_4 
    + h^{(2)}_{13}\, {\mathbf T}_1 \cdot {\mathbf T}_3 + h^{(2)}_{14}\, {\mathbf T}_1 \cdot {\mathbf T}_4
    + h^{(2)}_{23}\, {\mathbf T}_2 \cdot {\mathbf T}_3 + h^{(2)}_{24}\, {\mathbf T}_2 \cdot {\mathbf T}_4   
    \nonumber \\
    &+\; h^{(2)}_{3434}\, {\mathbf T}_3 \cdot {\mathbf T}_4 \; {\mathbf T}_3 \cdot {\mathbf T}_4
    + h^{(2)}_{3433}\, {\mathbf T}_3 \cdot {\mathbf T}_4 \; C_F 
    + h^{(2)}_{3333}\, C_F^2 \Big) + {\cal O}(\as^3)
    \;.
\end{align}
We note that this particular choice of colour structures is not unique, and different choices can be made which are related by colour conservation.
Results for $h^{(2)}_{34}$ and $h^{(2)}_{13}$ are given in Fig.~\ref{fig:h2_TT}, while $h^{(2)}_{3434}$, $h^{(2)}_{3433}$ and $h^{(2)}_{3333}$ are 
presented in Fig.~\ref{fig:h2_TTTT}. In both cases, the results correspond to $n_f=0$.
The numerical code used to evaluate all the terms in Eq.~(\ref{eq:final}) is included as supplemental material of this paper, allowing for the evaluation of our final results for arbitrary values of $\beta$, $\cos\theta$ and $n_f$.

\section{Summary}
\label{sec:summa}
This paper has been devoted to the evaluation of the soft-parton contributions that are relevant when a heavy-quark pair is produced at small transverse momenta in hadronic collisions.

When a colourless system (vector boson(s), Higgs boson(s) and so forth) is produced in hadron collisions only soft and collinear radiation from the initial-state colliding partons plays a role.
When a heavy-quark pair is produced, the coloured heavy quarks can emit in turn soft radiation (soft gluons and light quark-antiquark pairs), which gives an additional contribution to the structure of the singular contributions at small transverse momenta. We have evaluated such soft-parton contributions to NNLO in QCD perturbation theory.

Our computation has been carried out by using a semi-numerical approach, and evaluating all the relevant integrals in impact parameter space.
We have explicitly considered only the contributions that are relevant to apply the $q_T$ subtraction formalism to
this process.
After having introduced our framework in Sect.~\ref{sec:resum+soft}, in Sect.~\ref{sec:details} we have provided the details of our calculation, by first starting from the NLO in Sect.~\ref{sub:nlo}, which had already been obtained in Ref.~\cite{Catani:2014qha}. We then moved to the evaluation of the integrals from soft-gluon emission at one-loop order in Sect.~\ref{sub:1L}, soft light-quark pairs in Sect.~\ref{sub:qq}, and finally double gluon emission in Sect.~\ref{sub:gg}. We have provided all the relevant details of the computation by highlighting the difficulties that had to be overcome. 
At NNLO the most challenging contributions are those from the double-real emission, and in particular, those from double gluon radiation. These contributions need first to be integrated over the angles of the emitted partons by keeping their total momentum $k$ fixed. Then, the remaining integrals have been evaluated by splitting them into a singular and a regular part as $k^2\to 0$. For some of the contributions, the latter has been evaluated numerically.
After checking the cancellation of the $\epsilon$ poles, the complete results for the final remainders are provided through a numerical code that is attached to the arXiv distribution of the paper.

Together with the results already available in the literature, the soft-parton contributions presented in this paper
complete the evaluation at NNLO of the azimuthally-averaged transverse momentum resummation formula
for the production of heavy-quark pairs. In particular, the results
can straightforwardly be implemented to carry out fully differential NNLO calculations for
the production of a pair of heavy quarks with arbitrary mass by using the $q_T$ subtraction formalism.

\section*{Acknowledgments}
This work is supported in part by the Swiss National Science Foundation (SNF) under contract 200020$\_$188464.

\clearpage
\typeout{} 
\bibliography{biblio}

\end{document}